\documentclass[preprint,12pt]{elsarticle}

\usepackage{amssymb}
\usepackage{epsfig}
\usepackage{graphicx}
\usepackage{dcolumn}
\usepackage{bm}
\usepackage{times}
\usepackage{xcolor}
\usepackage[percent]{overpic}
\usepackage{cancel}
\usepackage{amsmath}
\usepackage{multirow}
\usepackage{slashbox}
\usepackage{adjustbox}
\usepackage{blindtext}
\usepackage{setspace}

\begin{document}

\begin{frontmatter}



\title{Coevolution spreading in complex networks}


\author{Wei Wang$^{1,2}$, Quan-Hui Liu$^{2,3,4}$, Junhao Liang$^{5}$, Yanqing Hu$^{6,7}$, and Tao Zhou$^{2,3*}$}
\cortext[cor1]{Corresponding Author: zhutou@ustc.edu}

\address{1. Cybersecurity Research Institute, Sichuan University,
Chengdu 610065, China}
\address{2. Big Data Research Center, University of Electronic
Science and Technology of China, Chengdu 610054, China}
\address{3. Comple$\chi$ Lab, University of Electronic
Science and Technology of China, Chengdu 610054, China}
\address{4. College of Computer Science, Sichuan University, Chengdu 610065, China}
\address{5. School of Mathematics, Sun Yat-Sen University, Guangzhou 510275, China }
\address{6. School of Data and Computer Science, Sun Yat-sen University, Guangzhou 510006, China}
\address{7. Southern Marine Science and Engineering Guangdong Laboratory, Zhuhai, 519082, China}

\begin{abstract}
The propagations of diseases, behaviors and information in real systems are rarely
independent of each other, but they are coevolving with strong interactions. To uncover
the dynamical mechanisms, the evolving spatiotemporal patterns and critical phenomena of networked coevolution spreading are extremely important, which provide
theoretical foundations for us to control epidemic
spreading, predict collective
behaviors in social systems, and so on. The coevolution spreading dynamics in complex
networks has thus attracted much attention in many disciplines. In this review, we introduce
recent progress in the study of coevolution spreading dynamics, emphasizing
the contributions
from the perspectives of statistical mechanics and network science. The theoretical
methods, critical phenomena, phase transitions, interacting mechanisms, and effects of network
topology for four representative types of coevolution spreading
mechanisms, including
the coevolution of biological contagions, social contagions, epidemic-awareness, and
epidemic-resources, are presented in detail, and the challenges
in this field as well as open issues for future studies
are also discussed.

\end{abstract}


\end{frontmatter}


\tableofcontents

\section{Introduction} \label{sec:intro}

Propagations in many real-world systems can be theoretically
described by spreading dynamics, with infectious disease,
computer viruses, information, innovation, financial risk, and
many others being treated as ``epidemics''~\cite{
keeling2008modeling,centola2018behavior,
lehmanncomplex,anderson1992infectious,hethcote2000mathematics}. When investigating
spreading dynamics, scientists wish to understand certain important aspects,
such as what are the dynamical mechanisms behind the phenomena, whether there will be an outbreak, how many individuals
will be infected, when will an individual be infected, and how to effectively
predict and contain the spread. Addressing these problems provides many beneficial aspects for our society. For governments,
the situations of epidemics could be apperceived, and
effective containment measures could be provided~\cite{brockmann2013hidden}. For
e-commerce, certain personalized recommendation strategies could be designed
to promote the diffusion of products (e.g., popular clothing)~\cite{lu2012recommender}.
For economics, financial risks may be
perceived at an early stage and thus global
economic crises could be evaded to some
extent~\cite{helbing2013globally}.

To address these problems, scientists have already made
great efforts since the first mathematical approach to
study the spread of an infectious disease by Bernoulli
in 1760~\cite{bernoulli1760essai}. Historically,
the single spreading dynamics was placed in a well-mixed population,
without any differences among individuals~\cite{
anderson1992infectious,kermack1991contributions}. Following this idea, researchers can theoretically predict
the outbreak size, critical threshold, and
associated critical phenomena. However, in reality, an individual only has contacts
with a limited number of other individuals. This
constraint can be characterized by a network $G(V,E)$,
where $V$ and $E$ are sets of nodes and edges, with nodes
representing individuals and edges denoting interactions
between individuals. Accordingly, scientists studied
the single spreading dynamics on oversimplified networks (e.g.,
regular networks and Erd\"{o}s--R\'{e}nyi (ER)
\cite{er1960random} random networks), and
analytically obtained the outbreak threshold
and epidemic prevalence~\cite{marro2005nonequilibrium}.

The above analytical results are usually far from
empirical observations, because real networks are much
more complex than the oversimplified models. For
example, the node degree (i.e., the number of edges of
a node) can vary over a few orders of magnitude, exhibiting
a highly heterogeneous nature~\cite{barabasi1999emergence}.
Other frequently observed
features that cannot be well captured by the oversimplified
models include the small-world property~\cite{watts1998collective},
community structure~\cite{newman2006modularity},
multiplexity~\cite{kivela2014multilayer}, spatiality~\cite{barthelemy2011spatial},
temporality~\cite{holme2012temporal}, and so on.
In a pioneering work~\cite{pastor2001bepidemic},
Pastor-Satorras and Vespignani studied
a single infectious virus on networks
with power-law degree distribution (named
as scale-free (SF) networks), and
revealed that there is no epidemic threshold
for a specific range of the degree exponent.
Following this work, researchers found the spreading
dynamics, such as for a global infectious disease, can be better predicted by accounting
for more topological features of real networks
\cite{hufnagel2004forecast,belik2011natural}.
Many reviews and books have already summarized
the state-of-the-art progress in single
dynamics~\cite{albert2002statistical,
castellano2009statistical,wang2016statistical,
pastor2015epidemic,
de2018fundamentals,lu2016vital,wang2016unification,de2016physics}.

The propagations of diseases, behaviors, and information in
the real world are rarely independent of each other; rather,
they are coevolving
with strong interactions. Coevolution spreading exists widely, with
important practical relevance~\cite{bauch2013social}. For example,
HIV results in the lower immunity of virus
carriers, who are therefore more susceptible to infectious diseases
such as tuberculosis and hepatitis~\cite{ferguson2003ecological,
abu2006dual}. The propagation of disease-related information in social media could
largely suppress the spreading of the corresponding epidemic disease~\cite{funk2009spread}. In an extremely important case,
in the early stage of the spreading of
the severe acute respiratory syndrome (SARS) in China, an unofficial message entitled
``There is a fatal flu in Guangzhou''
was sent to tens of millions of individuals~\cite{tai2007media}. As a result, individuals
adopted simple but effective actions (e.g., staying  home or wearing
face masks) to protect themselves from being infected by SARS,
which greatly decreased the final number of infected individuals.
Scientists have already made efforts to uncover and understand the interaction
mechanisms, spatiotemporal evolution patterns, critical phenomena, and phase
transitions of networked coevolution spreading. The complex interactions during the coevolving dynamics lead to
rich phase transition phenomena and novel physics, such as coexistence
thresholds caused by competitive interactions~\cite{newman2005threshold} and
discontinuous phase transition caused by synergetic interactions
\cite{cai2015avalanche}. In addition, the multiscale structure of networks
remarkably affects not only the
value of thresholds, but also the critical exponents~\cite{noh2005asymmetrically}
and the type of phase transitions~\cite{hebert2015complex}
of spreading dynamics. Therefore, the coevolution spreading dynamics on complex networks has attracted
increasing attention in recent years.

There are three reasons for us to write this review. First, a large number of
papers about coevolution spreading dynamics have emerged recently, but
there is still a lack of a comprehensive review to systematically organize
these results, discuss major challenges at the current stage, and point out
open issues for future studies. Second, the early literature used different expressions to describe essentially the same mathematical problems and methods. It is thus urgent to unify the problem description and the symbolic
system. Third, researchers have already
tried to find certain
potential applications, but these application possibilities are scattered in
disparate fields, and lack integration. Accordingly, this review
will be helpful to researchers already in the field,
those who intend to enter the field, and those who wish to apply the related
findings, and it will also contribute to the further development of the field.

In what follows, we will introduce the progress of studying the coevolution spreading
on complex networks, including theoretical methods, critical phenomena,
phase transitions, interaction mechanisms, effects of
network topology, and so on. Four
representative types of networked coevolution spreading are
considered,
including the
coevolution of biological contagions, social contagions,
awareness--epidemic
and resources--epidemic. In Section~\ref{bio}, we
introduce the coevolution spreading
of biological contagions, in which each contact between susceptible and
infected nodes may trigger the transmission of the infection. We
mainly focus on the most representative biological contagions,
i.e., epidemic spreading. However, in the spreading processes for political information, innovative products, and new drugs,
a single contact is insufficient to eliminate the risk of adoption
for susceptible individuals, and thus multiple contacts are
necessary.
The coevolution spreading of social
contagions is presented in Section~\ref{soc}, which focuses on the spreading dynamics involving
the above social reinforcement effects. To contain the
epidemic, certain coevolution spreading strategies are developed.
In Section~\ref{ei} and Section \ref{res}, we respectively
introduce the coevolution of awareness and an epidemic, and the
coevolution of resources and an epidemic. Finally, in Section
\ref{con} we sketch the landscape of this emerging
field, summarize representative progress, and make discuss the outlooks
of the current challenges and future open issues of the
field.

\section{Coevolution of biological contagions} \label{bio}

Empirically, epidemic spreading, virus spreading, and information
diffusion are usually modeled as biological contagions, where a single activated source
can be sufficient for infection transmission. Susceptible-infected-susceptible (SIS) and susceptible-infected-recovered (SIR) are the most
representative models for biological contagions in networks. For the reversible SIS model, a node
can be in the susceptible or infected state. At each time step,
each infected node tries to transmit the infection to every susceptible
neighbor with rate $\beta$, and then returns to the susceptible state with rate
$\gamma$. For the irreversible SIR model, each infected node also tries to infect every susceptible neighbor with rate $\beta$, but with the difference that the infected node then
becomes recovered with rate $\gamma$. The recovered node does not participate in
the remaining spreading process. The effective transmission rate is denoted as $\lambda=\beta/\gamma$. Other well-known models for biological contagions include the
the susceptible-infected (SI) model, susceptible-infected-recovered-susceptible
(SIRS) model, contact process, and so on. Unless specifically stated otherwise,
this section focuses on SIS and SIR models (others are similar).

\subsection{Single biological contagions}

As some reviews have
systemically reported the progress of single contagions on complex networks \cite{castellano2009statistical,pastor2015epidemic,Zhou2006epidemic,dorogovtsev2008critical,
zhang2016dynamics,wang2014spatial,kiss2017mathematics}, we only briefly emphasize two
aspects in the following subsections: (i) the mainstream theoretical
approaches and results, and (ii) the effects of network topology.

\subsubsection{Theoretical approaches}

In 2001, Pastor-Satorras and Vespignani first studied epidemic
spreading on SF networks~\cite{pastor2001epidemic}.
They analyzed the survival probability $P_s(t)$ of the virus
data reported by the Virus Bulletin covering 50 months, and found
$P_s(t)\sim {\rm exp}(-t/\tau)$, where $\tau$ is the characteristic
lifetime of the virus strain. Within the traditional framework for well-mixed populations of homogeneous networks (e.g., random networks and regular lattices), such a
long lifetime suggested that the effective transmission rate was much larger than the epidemic
threshold. However, the average fraction of the infected population was very small, which in contrast suggested a small value of the effective transmission rate in the traditional framework.
To understand this seemingly paradoxical phenomenon, Pastor-Satorras and Vespignani
studied the SIS model on Barab\'asi--Albert (BA) networks with power-law degree
distribution $P(k)\sim k^{-3}$ (see Ref. \cite{barabasi1999emergence} for the construction of BA networks). Using the heterogeneous mean-field
theory, the spreading dynamics is described as
\begin{equation} \label{h_men_field}
\frac{d\rho_k(t)}{dt}= -\rho_k(t)+\lambda k [1-\rho_k(t)]\Theta(t),
\end{equation}
where $\rho_k$ is the density of infected nodes with degree $k$, $\Theta(t)=\frac{1}{\langle k\rangle}\sum_{k} P(k)k\rho_k(t)$
represents the probability that a randomly selected edge points to
an infected node, and $\lambda$ is the effective transmission rate. To simplify the analysis, Pastor-Satorras and Vespignani set $\beta=\lambda$ and $\gamma=1$, and thus the first term of Eq.~(\ref{h_men_field}) implies that all the infected nodes with degree $k$ will become susceptible in the next time step, and the second
term represents the fraction of susceptible nodes with
degree $k$ that will be infected by neighbors in this time step. In the steady state,
$d\rho_k(t)/dt=0$, and the stationary density is
\begin{equation}\label{steady}
\rho_k=\frac{k\lambda\Theta}{1+k\lambda\Theta}.
\end{equation}
From Eq.~(\ref{steady}), one can conclude that the nodes with larger degrees are of higher probability to be infected. Linearizing around the initial
conditions $\rho_k(0)\rightarrow0$, the epidemic threshold can be obtained as
\begin{equation} \label{h_thereshold}
\lambda_c=\frac{\langle k\rangle}{\langle k^2\rangle},
\end{equation}
where $\langle k\rangle$ and $\langle k^2\rangle$ are the first and
second moments of the degree distribution $P(k)$, respectively.
When $\lambda\leq\lambda_c$, there is no global epidemic;
otherwise, i.e., when $\lambda>\lambda_c$, the global epidemic
is possible.


The heterogeneous mean-field theory can accurately capture
spreading dynamics on annealed networks; however, it cannot predict the threshold of quenched networks well, because these usually contain very complicated local structures that cannot be characterized only by the degree distribution. For example, BA networks \cite{barabasi1999emergence} and random Apollonian networks (RANs) \cite{zhou2005maximal} are of the same degree distribution but exhibit very different structural features, and thus different epidemic behaviors. In addition, because Eq.~(\ref{h_men_field})
assumes that the states of neighbors are independent, the dynamical correlations
among the states of neighbors are neglected, resulting in deviations from real dynamics. To address
the above shortcomings in the heterogeneous mean-field theory, scientists recently proposed some advanced approaches as follows (see also a recent review \cite{wang2016unification} about
theoretical approaches in networked spreading, as well as the related references therein).

By mapping the transmission
probability to the bond occupancy probability, bond percolation is widely used to analyze epidemic
spreading on networks for both the SIR and SIS
models \cite{newman2002spread}, and to identify influential nodes
in epidemic dynamics~\cite{lu2016vital,hu2018local}.
The final fraction of recovered nodes goes through a second-order
phase transition at the epidemic threshold $\lambda_c=
\langle k\rangle/(\langle k^2\rangle-\langle k\rangle)$, which
is determined by the network structure. Close to $\lambda_c$, the final epidemic
size $R$ behaves as $R\sim (\lambda-\lambda_c)^{\alpha_e}$,
where the critical exponent $\alpha_e=1$ for ER networks and SF networks
with power-law exponent $\gamma_D\geq 4$, and
$\alpha_e=1/(\gamma_D-3)$ when $3<\gamma_D<4$.

To capture the effects caused by the quenched topology of an undirected network, the quenched mean-field theory~\cite{goltsev2012localization},
discrete-time Markov chain approach \cite{gomez2010discrete}, and
$N$-intertwined approach \cite{van2009virus} directly explore the
adjacency matrix $A$, where $A_{ij} = 1$ when nodes $i$ and $j$ are
connected, and $A_{ij}=0$ otherwise. Through these different approaches, they arrived at the same epidemic threshold
$\lambda_c=1/{\Lambda_A},
$
where $\Lambda_A$ is the largest eigenvalue of $A$. For uncorrelated SF networks
with a power-law degree distribution $P(k)\sim k^{-\gamma_D}$, $\lambda_c \propto \langle k
\rangle/\langle k^2\rangle$ when $\gamma_D<2.5$, suggesting the same
threshold as in the heterogeneous mean-field theory.
When $\gamma_D>2.5$, $\lambda_c \propto 1/\sqrt{k_{\rm max}}$, which
indicates that there is no epidemic threshold in the thermodynamic
limit \cite{chung2003spectra}, where $k_{\rm max}$ is the maximal
degree of the network. Further analyses show that the eigenvector corresponding to $\Lambda_A$ is localized when
$\gamma_D>2.5$, which means that only hubs and their neighbors are infected,
and consequently the epidemic grows very slowly and may die out owing
to fluctuations.

Although the above approaches more accurately predict the epidemic
threshold than the heterogeneous mean-field theory, they still cannot accurately
capture the dynamical correlations among the states
of neighbors. For example, if there is only one seed node, the dynamical
correlations are obvious because the contagion path has the
same source, and the epidemic
transmission events to one node coming from two neighbors
may be correlated \cite{altarelli2014containing}. To overcome the weaknesses of the quenched mean-field approach
but retain its advantages, i.e., to take into consideration the
full network structure, the dynamic message-passing approach was
 proposed by Karrer and Newman \cite{karrer2010message}
to study the SIR model, and generalized later by Shrestha \emph{et al.} to
describe the SIS model \cite{shrestha2015message}. By disallowing
a node in the ``cavity'' state from transmitting
an infection to its neighbors but allowing it to be infected by them,
the ``echo chamber'' \cite{castellano2018relevance} (i.e., where a node is reinfected by a neighbor
it previously infected) is reduced in the dynamic
message-passing approach. The dynamic message-passing approach
predicts that the epidemic threshold is $\lambda_c=1/\Lambda_\mathbf{B}$,
where $\mathbf{B}$ is the non-backtracking matrix. Note that
$\mathbf{B}$ is a $2M\times2M$ nonsymmetric matrix with rows
and columns indexed by directed edges $i\leftarrow j$, where
$M$ is the number of edges. The element of $\mathbf{B}$ is
\begin{equation}
\mathbf{B}_{i\leftarrow j,k\leftarrow \ell}=\delta_{jk}(1-\delta_{i\ell}),
\end{equation}
where $\delta_{jk}=1$ if $j=k$, and $\delta_{jk}=0$ otherwise.
Recent studies showed that the dynamic message-passing approach predicts the
epidemic spreading dynamics well on uncorrelated
locally tree-like networks~\cite{
wang2016predicting}, and it has found wide applications in epidemic
containment~\cite{altarelli2014containing}, locating spreading
sources~\cite{altarelli2014bayesian}, and
network dismantling~\cite{zhao2016optimal}. However, the dynamic message-passing approach
needs $2E+N$ differential equations, which is time-consuming
for large networks. Moreover, some simplified approaches have been
developed, such as the edge-based compartmental approach, which
only uses four differential equations to describe the dynamics,
and predicts the dynamics well on configuration networks~\cite{kiss2017mathematics}.

The pairwise approximation approach is another well-known
approach to capture the dynamic correlations by considering
the evolution of pair states, instead of states of individual nodes \cite{eames2002modeling}.
It requires $k_{\rm max }^2$ equations to describe the dynamics, even assuming that
nodes of the same degree are statistically the same, and if it treats every node differently, $N+E$ equations are
needed. It is also too complicated for large-scale networks. Another disadvantage is that the pairwise approximation
approach usually cannot show an analytical expression of the epidemic threshold, but only a numerical value.
Moreover, other approaches were
proposed for certain specific dynamics, including the master equations
\cite{gleeson2011high} and other
generalized ones \cite{cai2016solving}.

\begin{figure}
\begin{center}
\epsfig{file=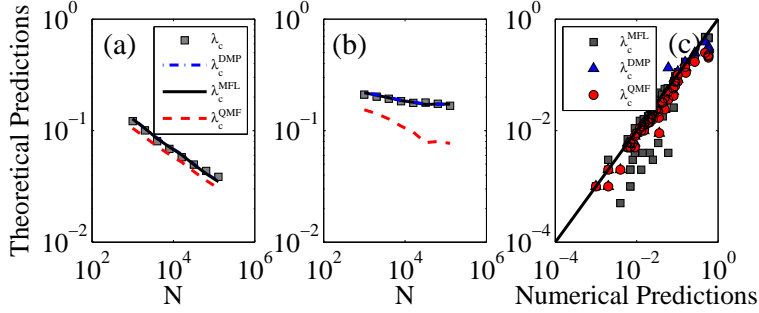,width=0.8\linewidth}
\caption{(Color online) Predicting the epidemic threshold for the
SIR model on uncorrelated SF networks and 56 real-world networks.
Theoretical predictions of $\lambda_c^{\rm MFL}$, $\lambda_c^{\rm QMF}$, and
$\lambda_c^{\rm DMP}$, and simulation results of $\lambda_c$ versus network size $N$ are extensively compared for networks with power-law exponent $\gamma_D= 2.1$
(a) and $\gamma_D = 3.5$ (b). Theoretical predictions and simulation results for the 56 real-world networks are shown in (c). Reproduced from \cite{wang2016predicting}
under CC-BY 3.0.}
\label{fig1_bio}
\end{center}
\end{figure}

Based on the SIS model, Gleeson \emph{et al.} \cite{gleeson2012accuracy} compared mean-field
predictions with numerical simulation results in 21 real-world networks, and found that the accuracy
of the mean-field theory is high when the average
degree of the nearest neighbors of a random node is sufficiently
large. Wang \emph{et al.} \cite{wang2016predicting} classified the most widely used approaches into three categories: the (i)
mean-field like (MFL), (ii) quenched mean-field
(QMF), and (iii) dynamical message passing (DMP) methods. As shown in Figs.~\ref{fig1_bio}(a) and (b), the MFL and DMP methods yield identical
predictions of epidemic thresholds for the SIR model in uncorrelated configuration networks. As for
the 56 real-world networks, as shown in Fig.~\ref{fig1_bio}(c), the epidemic thresholds obtained by
the DMP method is more accurate, because it incorporates the full network topology information and some
dynamical correlations.

\subsubsection{Effects of network topology}

A very important and challenging issue is to reveal the effects of complex topologies on networked epidemic spreading. Here, we briefly review the major progress on this issue, including the effects of degree heterogeneity, degree--degree correlation, clustering, finite network size, community structure, weight distribution, multilayer structure, and time-varying structure.

Early studies on the effects of degree heterogeneity showed that the final
epidemic size (i.e., the epidemic prevalence) scales as $R\sim(\lambda-\lambda_c)^{\alpha_e}$ near
the critical point, with the critical exponent $\alpha_e=1$ for homogeneous networks, such as ER
networks and Watts--Strogatz (WS) networks \cite{watts1998collective}. By comparison, for
Barab\'{a}si--Abert (BA) networks~\cite{barabasi1999emergence}, $R\sim {\rm e}^{-2/\lambda
\langle k\rangle}$, indicating the absence of an epidemic threshold. 
%
Degree--degree correlation is an important feature of networks~\cite{newman2002assortative}. In
an assortative network (i.e., one with positive degree--degree correlation), large-degree nodes tend to connect with large-degree nodes and small-degree nodes tend to connect with small-degree nodes, whereas in a disassortative network (i.e., one with negative degree--degree correlation), large-degree nodes tend to connect with small-degree nodes and vice versa. Bogu\~{n}\'{a} \emph{et al.} \cite{boguna2003absence} found that the epidemic threshold vanishes in the thermodynamic limit of SF networks with power-law exponent $2<\gamma_D\leq3$, whether the two-point
degree correlations are assortative or disassortative mixing patterns. Specifically, the
epidemic threshold is $\lambda_c=1/\Lambda_m$, where
$\Lambda_m$ is the largest eigenvalue of the connectivity
matrix $C_{kk^\prime}=kP(k^\prime|k)$ and $P(k^\prime|k)$
is the probability that an edge belonging to a node of degree $k$ connects
to a node of degree $k^\prime$. 

Clustering is a widely observed characteristic of disparate networks~\cite{newman2001clustering}. Egu\'{i}luz
and Klemm~\cite{eguiluz2002epidemic} built a highly clustered SF network model and obtained a
finite threshold $\lambda_c=1/(\langle k\rangle-1)$. Miller~\cite{miller2009percolation} also claimed that
clustering reduces the epidemic size and increases the
epidemic threshold, and network clustering is an important factor in
controlling the growth rate of epidemic spreading~\cite{miller2009spread}. However, the effects of clustering
seem to be dependent on the underlying network models. For example, studies on random SF networks with high clustering coefficients showed that high clustering cannot restore a finite epidemic threshold~\cite{serrano2006percolation}. In a solvable model, Newman theoretically proved that higher clustering leads to an even lower epidemic threshold, because redundant paths introduced by triangles in the network provide more opportunities for the susceptible nodes to be infected~\cite{newman2009random}.
Wang \emph{et al.}~\cite{wang2018critical}
revealed that there is a double transition
when epidemics spread on networks with
cliques.

Most theoretical analyses are under the thermodynamic limit (i.e., assuming the network size is
infinite) whereas real-world networks are of finite sizes. 
No\"{e}l proposed an accurate theoretical framework to address the time
evolution of epidemic dynamics on finite-size
networks \cite{noel2009time}. Ferreira \emph{et al.}. proposed a susceptibility
method $\chi=N(\langle \rho^2\rangle-\langle \rho\rangle^2)/
\langle\rho\rangle$ to locate the network-size-dependent epidemic
threshold $\lambda_c(N)$ for the SIS model \cite{ferreira2012epidemic}, where $\langle \cdot\rangle$ is the average of the assemble, and $\rho$
is the final epidemic outbreak size. They
found $\lambda_c(N)-\lambda_c(\infty)\sim N^{-1/\overline{\nu}}$,
where $\overline{\nu}$ is the critical exponent. Other methods
were developed to determine the epidemic thresholds in finite-size networks,
such as variability \cite{shu2016recovery},
lifespan \cite{mata2015multiple}, and finite-size scaling methods \cite{hong2007finite}.

A community is a mesoscale measurement of networks topology~\cite{fortunato2016community}. Generally speaking, nodes within a community are densely connected, whereas nodes between communities are sparsely connected. Liu and Hu~\cite{liu2005epidemic} studied the epidemic spreading on simplified-community
networks and found that the epidemic threshold fulfills $\lambda_c(p/q)-
\lambda_c(\infty)\sim q/p$, where $p$
and $q$ respectively stand for the connecting probability of links
within a community and between communities, and $\lambda_c(\infty)$
is the outbreak threshold when there are only edges in the communities.
To date, the effects of community structure
on spreading dynamics are controversial. Chen \emph{et al.} \cite{chen2012epidemic} found that an
overlapping community structure promotes epidemic
prevalence. However, Huang and Li \cite{huang2007epidemic}
claimed that strong community structure suppresses epidemic prevalence. 

In simple networks, edges are binary (i.e., edges either do or do not exist), whereas in many real networks,
interacting strengths between different node pairs are significantly different, and thus edges are associated
with weights to represent their strengths~\cite{barrat2004architecture}. By treating a simple network as a
special weighted network with each edge associated with weight 1, it is obvious that real weighted networks
are always of more heterogeneous weight distributions than that of a simple network. Indeed, scientists have demonstrated that the
heterogeneity of the weight distribution markedly affects the
epidemic dynamics, including both the epidemic threshold
and epidemic prevalence~\cite{yan2005epidemic,yang2012epidemic,
wang2014epidemic}. Among the earliest works,
Yan \emph{et al.}~\cite{yan2005epidemic} showed that nodes with larger strengths
(a node's strength is defined by the sum of the weights of its associated edges)
are preferentially infected. To accurately predict the epidemic
spreading on weighted networks, Wang \emph{et al.}~\cite{wang2014epidemic} developed
an edge-weight-based compartmental approach. Their approach shows remarkable agreement with numerical results.


Certain real-world systems are better characterized by multilayer
networks (also known as multiplex networks, networks of networks, and interdependent networks in the literature)~\cite{
boccaletti2014structure,gao2012networks,bianconi2018multilayer}. A multilayer network consists of a
few subnetworks (usually two or three subnetworks, each of which is called a layer), where each has
its own organizing rules and functions, different from the others, and nodes in different layers
may have strong interactions that can be described by cross-layer edges.
Saumell-Mendiola \emph{et al.} \cite{saumell2012epidemic}
studied the SIS model on interconnected networks. Through a generalized heterogeneous mean-field theory, they
found that the global endemic state may occur, even though the epidemics
cannot outbreak on each network separately. On the contrary,
for the SIR model on interconnected networks,
the epidemic occurs on both subnetworks when
the coupling is strong enough; otherwise, a mixed phase exists
\cite{dickison2012epidemics}. 
Arruda \emph{et al.} \cite{de2017disease} found
epidemic spreading on multilayer networks shows a localization phenomenon. Recently, Liu \emph{et al.}~\cite{liu2018measurability} constructed two multiplex contact networks from high-resolution sociodemographic data in Italian and Dutch populations, and showed that the classical concept of the basic reproduction number is untenable
in realistic populations, owing to the multiplex and clustered contact structure of the populations. 

In some real systems, network topologies are time-varying, which can be described
by temporal networks \cite{holme2012temporal}. Perra \emph{et al.} \cite{perra2012activity} proposed an activity-driven
network to model temporal networks, where each node $i$ is assigned an activity potential $x_i$
independently drawn from a given probability distribution $F(x)$, and
is active with probability $a_i\Delta t=\eta x_i\Delta t$ in each time step,
where $\eta$ is a constant. Each active node generates
$m$ edges to connect with $m$ randomly selected nodes.
At the next time step, all existent edges are
deleted and the newly active nodes generate edges to form a new network. For the SIS model on the proposed temporal
networks, the epidemic threshold is $\lambda_c
=2\langle a\rangle/(\langle a\rangle+\sqrt{\langle a^2\rangle})$,
where $\langle a\rangle$ and $\langle a^2\rangle$ are the first
and second moments of $a_i$, respectively. The results
indicated that temporal networks are more robust to
epidemic spreading than integrated static networks. Further analysis of this activity-driven model showed that memory inhibits the spreading of the SIR model, whereas it promotes the spreading of the SIS model~\cite{sun2015contrasting}. Liu \emph{et al.} \cite{liu2018epidemic} studied the SIS spreading process on time-varying multiplex networks, and found that strong multiplexity (i.e., the fraction of overlapping nodes) significantly reduces the epidemic threshold. 
Starnini \emph{et al.} \cite{starnini2017equivalence} revealed that the
non-Markovian spreading dynamics can be captured by
the effective infection rate.

\subsection{Coevolution of two biological contagions}

In many real-world scenarios, epidemics spread simultaneously and
interact with each other~\cite{vasco2007tracking}.
In this section, we first review two successive contagions,
then review some representative models
of coevolution epidemic spreading
dynamics.

\subsubsection{Successive contagions}

When two biological contagions spread in the
same population, the first contagion may affect the
latter one. For example, the hosts may be
killed or be provided permanent immunity by the first contagion \cite{castillo1996competitive,
andreasen1997dynamics}, such that the latter one cannot infect them.

Newman studied two epidemics spreading on the same network \cite{newman2005threshold}. The first and second epidemics are both described by the standard SIR model, but with different transmissibility
probabilities given by $\lambda_1$ and $\lambda_2$, respectively. The recovery probabilities for both models are simply set as 1. Using the bond percolation approach, the threshold of the first epidemic is
\begin{equation}\label{threshold_bio}
\lambda_c^1=\frac{\langle k\rangle}{\langle k^2\rangle-\langle k\rangle}.
\end{equation}
$u$ denotes the probability that a node is not
infected by a neighbor at the end of the epidemic.
For this to occur, either the infection does not transmit through
an edge with probability $1-\lambda_1$, or
the infection is transmitted through an edge but the endpoint of
this edge is not infected with probability $\lambda_1 G_1(u)$,
where $G_1(x)=\sum_{k} Q(k)x^k$ is the generating function of
the excess degree distribution $Q(k)=(k+1)P(k+1)/\langle
k\rangle$. Thus, $u$ satisfies the following equation:
$u=1-\lambda_1+\lambda_1 G_1(u).
$
 For a randomly selected node with degree $k$, it is not
infected with probability $u^k$. Therefore, the
prevalence of the first epidemic is
$R=1-G_0(u),
$
where $G_0(x)
=\sum_k P(k)x^k$ is the generating function of $P(k)$.

\begin{figure}
\begin{center}
\epsfig{file=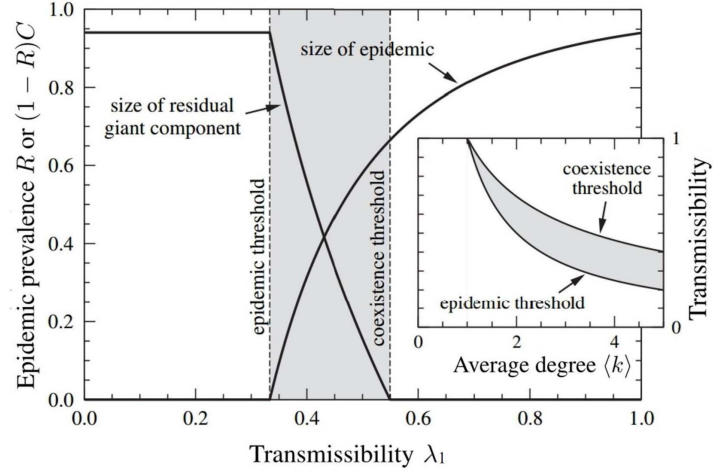,width=0.7\linewidth}
\caption{(Color online) Epidemic prevalence of
first epidemic $R$ or giant connected cluster $(1-
R)C$ on residual network versus transmissibility
$\lambda$ on ER networks with average degree $\langle
k\rangle=3$. The parameter $C$ is obtained by numerically solving
$C=1-F_0(v)$ and $v=F_1(v)$. The inset shows the two
epidemic thresholds as functions of the average degree $\langle
k\rangle$. The shaded areas denote that both
epidemics can spread. Reproduced from \cite{newman2005threshold}.}
\label{fig3_bio}
\end{center}
\end{figure}

At the ending of the first epidemic, the topology
of the residual network (i.e., after deleting the nodes
infected by the first epidemic) has obviously changed.
The second epidemic cannot transmit the infection to
nodes that are infected by the first epidemic, i.e., the second
epidemic can only spread on the residual network. For
a node $i$ that is not infected by the first epidemic on the
residual network, it connects to $m$ other uninfected
nodes with probability $P_u(m)= G_0(x)^{-1}
\sum_{k=m}P(k){m\choose k}[G_1(u)]^m[u-G_1(u)]^{k-m}$,
and its generating function can be expressed as
\begin{equation}\label{residual}
F_0(x)=\frac{1}{G_0(u)}G_0(u+(x-1)G_1(u)).
\end{equation}
Following the bond percolation theory, the
threshold of the second epidemic should fulfill the
condition $F_1^\prime(1)=1$, where $F_1(x)=F_0^\prime
(x)/F_0^\prime(1)$. Once $F_1^\prime(1)>1$, the system
undergoes an additional phase transition, and Newman called
it a coexistence transition. The threshold $\lambda_c^2$
of the second epidemic is thus named the coexistence threshold.
The corresponding results on ER networks are shown in
Fig.~\ref{fig3_bio}.

Newman and Ferrario \cite{newman2013interacting} further considered a different situation,
in which the second epidemic can only infect the nodes
that are have been infected by the first epidemic.
Their model can be used to describe the case in which
one disease increases the chance of infection by another.
For instance, once a person is infected by syphilis
and HSV-2, he/she is more likely to be infected by HIV
\cite{freeman2006herpes}. With a similar method to
Ref. \cite{newman2005threshold}, the second epidemic threshold is
\begin{equation}\label{second_threshold}
\lambda_c^2=\frac{2}{\tau+\sqrt{\tau^2-4\Delta}},
\end{equation}
where $\tau=G_1^\prime(1)-(1-2\lambda_1)G_1^\prime(1-u\lambda_1)$,
$\Delta=\lambda_1^2G_1^\prime(1)G_1^\prime(1-u\lambda_1)$, and
$\lambda_1$ is the transmission probability of the first epidemic.
When $\lambda_1\rightarrow\lambda_c^1=1/G_1^\prime(1)$, one obtains
$\lambda_c^2=1$. $\lambda_c^2$ decreases monotonously with
$\lambda_1$, and $\lambda_c^1\leq \lambda_c^2\leq1$. That is to say, the
second epidemic threshold is never smaller than the first one.

Bansal and Meyers \cite{bansal2012impact} modeled two consecutive seasonal epidemics, such
as influenza, on heterogeneous networks, and the former epidemic
inflicts immunity on the latter one. At the end of the first epidemic, a fraction of $1-f$ infected
nodes by the first epidemic are immune to the second epidemic, and
the remaining $f$ infected nodes can be
infected by the second epidemic. By using bond percolation
theory, they studied both perfect immunity (i.e., $f=0$) and partial immunity (i.e., $f>0$),
and found that the immunity of the first epidemic limits the outbreak of the second epidemic.

Funk and Jansen \cite{funk2010interacting} considered two SIR epidemic dynamics consecutively
spread on an overlay network. The overlay network is constructed by two layers, denoted as $\mathcal{A}$ and
$\mathcal{B}$. The nodes in the two layers are the same, and the network size
is $N$. The overlay network is built according to degree distribution
$P(k_1,k_2)$, where $k_1$ and $k_2$ denote the degrees of a node
in the two layers. The first epidemic spreads on $\mathcal{A}$ with
effective transmission probability $\lambda_1$, and then
all the nodes infected by the first epidemic are removed from
the overlay network. The second epidemic spreads on the residual
network with effective transmission probability $\lambda_2$. The generating function $G_{0,2}^r(x)$ of the residual degree
distribution of $\mathcal{A}$ on $\mathcal{B}$ is
\begin{equation}\label{overlay}
G_{0,2}^r(x)=\frac{G_0^J(1-\lambda_1+\lambda_1u_1,
1-h_1+h_1x)}{G_0^J(1-\lambda_1+\lambda_1u_1,1)},
\end{equation}
where $G_0^J(x,y)=\sum_{k_1}\sum_{k_2}P(k_1,k_2)
x^{k_1}y^{k_2}$ is the generating function of the
degree distribution $P(k_1,k_2)$, $h_1=
\frac{1}{\langle k_2\rangle}\frac{\partial}{\partial
y}G_0^J(1-\lambda_1+\lambda_1u_1,1)$, and
$\langle k_2\rangle$ is the average degree of network
$\mathcal{B}$.
$u_1$ denotes the probability that
an edge does not connect to an infected neighbor, which
is obtained by solving $u_1=G_{1,1}(1-\lambda_1+\lambda_1
u_1)$, where $G_{1,1}(x)$ is the generating function of the
excess degree of network $\mathcal{A}$. The parameter $h_1$ is
the probability for a node arrived at following a random
edge on $\mathcal{B}$ to be infected by the first epidemic.
If there is no overlapping between networks $\mathcal{A}$
and $\mathcal{B}$, the second epidemic threshold is
\begin{equation}\label{second_threshold_overlay}
\lambda_c^2=\frac{1}{G_{0,1}(u_1,\lambda_1)
}\frac{\langle k_2\rangle}{\langle k_2^2\rangle
-\langle k_2\rangle},
\end{equation}
where $G_{0,1}(x)=\sum_{k_1}\sum_{k_2}P(k_1,k_2)x^{k_1}$.

Funk and Jansen~\cite{funk2010interacting} further
studied networks $\mathcal{A}$ and
$\mathcal{B}$ with arbitrary overlapping, and determined the
second epidemic threshold as
\begin{equation}\label{overlay_threshold}
\lambda_c^{\rm overlay}=\frac{1-\lambda_1+\lambda_1
u_1}{q_{1|2}u_1/h_1+(1-q_{1|2})(1-\lambda_1+\lambda_1
u_1)}\lambda_c^2,
\end{equation}
where $q_{1|2}$ ($q_{2|1}$) is the probability that an
edge in network $\mathcal{B}$ ($\mathcal{A}$) is also in network
$\mathcal{A}$ ($\mathcal{B}$). Obviously, $\lambda_c^{\rm overlay}$
is strictly increasing with $q_{1|2}$, which indicated that
overlapping is beneficial for suppressing the second epidemic.
Funk and Jansen finally studied a more realistic scenario with
partial immunity. They found that once the first epidemic
provides partial immunity to the second one, the second
epidemic more easily invades the population. However, when
two interacting epidemics simultaneously spread on multilayer
networks, Zhou \emph{et al.} \cite{zhou2018propagation} found that overlapped
links have no effects on the spreading dynamics,
whereas the fraction of vulnerable nodes markedly
affects the dynamics.

\subsubsection{Competing or cross-immunity contagions}

\begin{figure}
\begin{center}
\epsfig{file=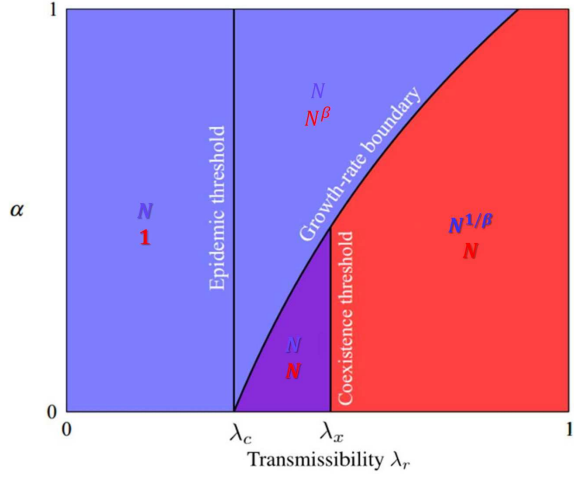,width=0.6\linewidth}
\caption{(Color online) Phase diagram of the system
with a given $\lambda_2$. The colors represent the
dominant disease, and the colored symbols represent
leading-order scaling of the expected number of
individuals infected by each epidemic. Reproduced from \cite{karrer2011competing}.}
\label{fig4_bio}
\end{center}
\end{figure}

Karrer and Newman investigated the behavior of two competing
SIR-type epidemics on the same network \cite{karrer2011competing}. The two epidemics are denoted
as red and blue epidemics with transmissibilities $\lambda_1$
and $\lambda_2$, respectively. Each node can only be
infected by one of the two epidemics. Initially,
a seed node for each epidemic is randomly selected.
It is assumed that the blue
epidemic evolves with time-step 1, and the red epidemic spreads
with time-step $0\leq \alpha \leq1$. Karrer and Newman
developed the competing percolation to
study the final state of the two epidemics theoretically.
At early time $t$, the average number of nodes infected by the
blue epidemic is $N_b={\rm e}^{t{\rm ln}R_b}$,
where $R_b=\lambda_2/\lambda_c$ is the reproductive
number for the blue epidemic, and
$\lambda_c=\langle k\rangle/(\langle k^2\rangle-
\langle k\rangle)$. Similarly, the average
number of infected nodes by the red epidemic at early time $t$
is $N_r={\rm e}^{t{{\rm ln} R_r}/\alpha}$,
where $R_r=\lambda_1/\lambda_c$ is the reproductive
number for the red epidemic. Initially, the two epidemics
increase exponentially. $\beta_0$ is defined as the ratio
of the growth rates of the two epidemic, given by
\begin{equation}\label{rate}
\beta_0=\frac{{\rm ln}(\lambda_1/\lambda_c)}{\alpha
{\rm ln}(\lambda_2/\lambda_c)}.
\end{equation}
For the case of $\beta_0>1$ the red epidemic spreads faster;
when $\beta_0<1$, the blue epidemic spreads faster. The
growth-rate boundary, i.e., $\beta_0=1$, is thus $\alpha =
{\rm ln}(\lambda_1/\lambda_c)/{\rm ln}(\lambda_2/\lambda_c)$.
In the thermodynamic
limit, the faster epidemic spreads on the network first,
and the slower one spreads on the residual network.
According to the competing percolation theory,
the phase diagram of the system is shown in Fig.~\ref{fig4_bio},
where $\lambda_x=\lambda_2{\rm ln}\langle k\rangle\lambda_2/
(\langle k\rangle\lambda_2-1)$.

Miller~\cite{miller2013cocirculation} proposed a
different competing spreading model, in which it is not necessary for the initial
seed size
to be small enough. Miller treated the fast epidemic
as completing its spread before the slow epidemic is large enough to warrant consideration,
and developed a low-dimensional edge-based compartmental approach
to describe the two SIR competing spreading dynamics.
Miller found two different situations, which depend
on the initial seed sizes $\rho_1^0$ and $\rho_2^0$ of the two epidemics.
If one epidemic has a much larger seed size than the other, the epidemic with larger seeds
infects most of the nodes, and the other epidemic cannot spread in
the residual network. When $\rho_1^0$ and $\rho_2^0$ are relatively large,
the two epidemics first grow exponentially with rates $r_1$ and $r_2$.
It is assumed that $r_1>r_2$, and $C$ is given by $C={\rm ln}\rho_2^0-(r_2/r_1){\rm ln}
\rho_1^0$. To determine the bounds of the coexistence of two epidemics,
Miller assumed that the epidemics¡¯ exponential
growth continues forever, and obtained two critical values
$C_{\rm min}$ and $C_{\rm max}$. When $C<C_{\rm min}$, epidemic 1
breaks out before epidemic 2, and vice versa for $C>C_{\rm max}$.

Considering the effects of mobility
of individuals, Poletto \emph{et al.}~\cite{poletto2013host} proposed a novel
model with two competing epidemics in a metapopulation network.
The metapopulation network is composed of $N_s$ subpopulations,
and the edges between the subpopulations are connected
according to a given degree distribution. Each subpopulation
has a given number of individuals, and each individual moves to a neighboring
subpopulation randomly. It is assumed that the two epidemics
have different infectious periods, but
the same basic reproductive number. Poletto \emph{et al.} found
that the structure of the population and the mobility
of hosts across subpopulations affect the infectious
period of the dominant epidemic.
Poletto \emph{et al.}~\cite{poletto2015characterising} further
considered a different situation, in which the two epidemics
have different basic reproductive numbers and infectious
periods. Poletto \emph{et al.}
revealed that mobility can either have no effect
on the competition dynamics or play an important role
in shaping the dominant epidemic.

The scenario of two competing SIS epidemics on complex networks has
also been widely studied. Generally speaking, two epidemics
evolve according to the SIS model with infection probabilities
$\beta_1$ and $\beta_2$. The recovery probabilities
of the two epidemics are $\gamma_1$ and $\gamma_2$.
To include the competing mechanism between two epidemics,
scientists assumed that each susceptible node infected
by one epidemic decreases its probability of being
infected by the other one. For the case of two completely
competing SIS epidemics on complex
networks, i.e., each susceptible can only be infected by
one of the two epidemics, Prakash \emph{et al.} found that
the stronger epidemic completely suppresses the other
\cite{prakash2012winner}. Once the two competing SIS
epidemics have partial immunizing functions, there is a
coexistence region of the two epidemics \cite{beutel2012interacting}.
Bovenkamp \emph{et al.} studied two competing SIS epidemics on
a complete network \cite{van2014domination}, and revealed that only
one epidemic exists when the transmission probability
is above the epidemic threshold, which is markedly different
from the observations of competing SIR epidemics.
Disallowing epidemic extension by allowing one node to
become infected automatically when there are no infected nodes,
the dominant and dominated epidemics alternate when
the two epidemics are identical. The domination period of
an epidemic depends on its initially infected nodes. Yang \emph{et al.}
developed the criteria for the extinction of both epidemics
and for the survival of only one epidemic when two competing epidemics have
general infection rates~\cite{yang2018bi,yang2018competition}.

Wang \emph{et al.} proposed a competing SIS model to describe
idea-spreading dynamics
\cite{wang2012dynamics}. Assuming that the effective
transmission rate of the first and second ideas are
$\lambda_1=\beta_1/\gamma_1$ and $\lambda_2=\beta_2/\gamma_2$,
respectively, if node $i$ is only surrounded by idea 1 (or 2),
it will be infected with rate $n_1\beta_1$ (or $n_2\beta_2$),
where $n_1$ (or $n_2$) is the number of neighbors infected
by idea 1 (or 2). If node $i$ is exposed to both ideas,
the infection probabilities are $\alpha_c\beta_1$ and $\eta_c\beta_2$
for ideas 1 and 2, respectively, where $0\leq\alpha_c,\eta_c\leq1$.
Through a generalized heterogeneous mean-field theory,
Wang \emph{et al.} found that the system has a coexistence region of the two
ideas on SF networks, and this region depends on whether the ideas have
exclusive or nonexclusive influences.

\begin{figure}
\begin{center}
\epsfig{file=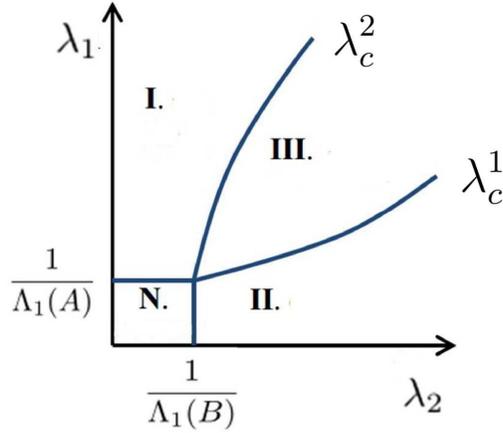,width=0.5\linewidth}
\caption{(Color online) Phase diagram of the $S
I_1SI_2S$ model on multilayer networks. The plane
is divided into the epidemic-free region N, epidemic 1
absolute-dominance region I, epidemic 2 absolute-dominance
region II, and coexistence region III.
Reproduced from \cite{sahneh2014competitive}.}
\label{fig5_bio}
\end{center}
\end{figure}

Different epidemics may transmit on distinct
networks. Considering this factor, Sahneh and Scoglio
proposed a competitive epidemic spreading model $SI_1SI_2S$ over arbitrary multilayer networks \cite{sahneh2014competitive}.
In this model, each node can be in one of three states: susceptible,
$I_1$ (i.e., infected by epidemic 1), and $I_2$
(i.e., infected by epidemic 2).
The two epidemics spread on networks $\mathcal{A}$ and $\mathcal{B}$
with effective transmission rates $\lambda_1=\beta_1/\gamma_1$
and $\lambda_2=\beta_2/\gamma_2$, respectively, where $\beta_1$ ($\beta_2$)
is the transmission rate of epidemic 1 (epidemic 2), and
$\gamma_1$ ($\gamma_2$) is the recovery rate of epidemic 1
(epidemic 2). Note that a node cannot be infected by two
epidemics simultaneously. Using the first-order mean-field
approximation, the evolutions of the fractions of nodes infected
by epidemics 1 and 2 are, respectively,
\begin{equation}\label{coevultion_sisis}
\frac{d\rho_{1,i}}{dt}=\beta_1(1-\rho_{1,i}-\rho_{2,i})
\sum_{j=1}^N A_{ij}\rho_{1,j}-\gamma_1 \rho_{1,i},
\end{equation}
and
\begin{equation}\label{coevultion_sisis2}
\frac{d\rho_{2,i}}{dt}=\beta_2(1-\rho_{1,i}-\rho_{2,i})
\sum_{j=1}^N B_{ij}\rho_{2,j}-\gamma_2 \rho_{2,i},
\end{equation}
where $i\in\{1,2,\cdots N\}$, $\rho_{1,i}$ ($\rho_{2,i}$)
represents the probability that node $i$ is in the infected
state of epidemic 1 (epidemic 2) in network $\mathcal{A}$
(network $\mathcal{B}$), and $A$ ($B$) is the adjacent
matrix of network $\mathcal{A}$ ($\mathcal{B}$).
In the steady state, 
$\rho_{1,i}^\ast$ and $\rho_{2,i}^\ast$ denote the
equilibrium probabilities of node $i$ being
infected by epidemics 1 and 2, respectively.
Through a bifurcation analysis of the model,
the system has four regions when two competing
epidemics spread on multilayer networks.
These are the epidemic-free, absolute dominance of epidemic 1,
absolute dominance of epidemic 2, and coexistence
regions, as shown in Fig.~\ref{fig5_bio}.
When the two layers are identical,
there is no coexistence region.
For the epidemic-free region, i.e., $\rho_{1,i}^\ast
=\rho_{2,i}^\ast=0$, the system is stable when $\lambda_1\leq
1/\Lambda_1(A)$ and $\lambda_2\leq1/\Lambda_1(B)$, where
$\Lambda_1(A)$ and $\Lambda_1(B)$ are the
leading eigenvalues of the adjacent matrices
$A$ and $B$, respectively. For a given $\lambda_2$,
the survival threshold $\lambda_{c}^1$ of epidemic 1 is
the critical value at which the coexistence equilibrium
emerges, say
\begin{equation}\label{threshold_1}
\lambda_{ c}^1=\frac{1}{\Lambda({\rm diag}\{1-y_i\})A},
\end{equation}
where $y_i$ is the stable value of $\rho_{2,i}^\ast$
at the epidemic 2 absolute-dominance equilibrium, i.e.,
$\rho_{1,i}=0$ and $\rho_{2,i}=y_i>0$. Similarly,
another critical point $\lambda_{ c}^2$ is obtained,
where coexistence equilibrium emerges for a given
$\lambda_1$:
\begin{equation}\label{threshold_2}
\lambda_{ c}^2=\frac{1}{\Lambda({\rm diag}\{1-x_i\})B},
\end{equation}
where $x_i=\rho_{1,i}>0$ and $\rho_{2,i}=0$.

\subsubsection{Cooperative contagions}

\begin{figure}
\begin{center}
\epsfig{file=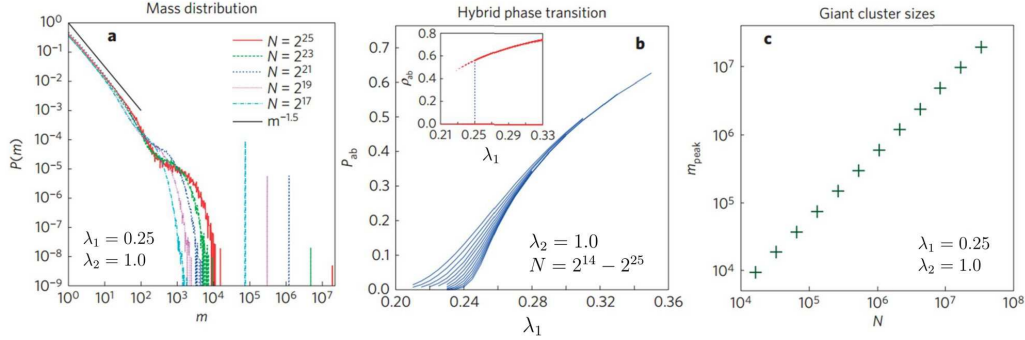,width=1\linewidth}
\caption{(Color online) Cooperative contagions on ER
networks with $\langle k\rangle=4$. (a) The mass
distribution of infected clusters at threshold $\lambda_c^1
\approx0.25$ and $\lambda_2=1.0$. (b) Phase diagram of $P_{ab}$
versus $\lambda_1$. The network sizes scale from $2^{14}$ to $2^{25}$.
The inset exhibits $\rho_{ab}$ versus $\lambda_1$. (c) The
peaks $m_{\rm peak}$ of the giant cluster size as a function
of $N$ at $\lambda_c^1$. Reproduced from \cite{cai2015avalanche}.}
\label{fig6_bio}
\end{center}
\end{figure}

Cooperation between two epidemic spreading dynamics means
that when a node is infected by one epidemic, the probability increases that this node may become infected by
the other epidemic. Such a cooperative effect is
also called a synergetic effect. For example, once a person
is infected by HIV/AIDS, his/her immune system become severely compromised, and thus the probability that he/she may become infected by
systemic lupus erythematosus (SLE), syphilis, or hepatitis increases.
Conversely, if a person infected by the latter diseases (e.g.,
SLE) is exposed to HIV, his/her infection
probability is increased.
Cai \emph{et al.} \cite{cai2015avalanche} proposed a coevolution SIR spreading model with a
cooperative effect between two epidemics, in which a node that has not yet been
infected will be infected by one of the two epidemics with
probability $\lambda_1$, and a node that has already been infected by
one epidemic will be infected by the other with probability
$\lambda_2>\lambda_1$. The two epidemics have the same
recovery probability. The recovered nodes acquire immunity
against the epidemic they had, but not against the other epidemic.
Extensive numerical simulations have been performed on ER
networks with average degree $\langle k\rangle=4$.
Two order parameters, namely the probability $\mathcal{P}$ of forming a giant
infected cluster and the fraction $\rho$ of nodes
belonging to the giant connected cluster, are used to describe the phase
transition. The system undergoes a hybrid discontinuous
transition. At the critical point,
 the system already has a finite fraction of
an infected cluster. The fraction $\rho_{ab}$ of nodes
infected by both epidemics 1 and 2 exhibits a
discontinuous phase transition for a given $\lambda_2$, as
shown in Fig.~\ref{fig6_bio}.
Furthermore, the system undergoes
a typical continuous transition on two-dimensional lattices.
Specifically, the critical behavior is $\mathcal{P}
\sim \rho_{ab}\sim (\lambda_1-\lambda_c^1)^{\alpha_e}$, where $\alpha_e\approx
5/36$. Grassberger \emph{et al.} further discussed the roles of network
topologies on the phase transition \cite{grassberger2016phase},
and revealed that loops are crucial for the emergence of
a discontinuous transition. For cooperative spreading dynamics on two-dimensional lattices with
local contacts or on BA networks, there is no discontinuous transition. However,
discontinuous transitions always appear for cooperative contagions on two-dimensional
lattices with long-range connections, on four-dimensional lattices, and on
ER networks. Chen \emph{et al.} further investigated the fundamental properties of cooperative contagion processes
of two SIS epidemics \cite{chen2017fundamental,chen2013outbreaks}. It was assumed that a node had a higher rate of infection
by one epidemic once it was already infected by another epidemic.
Mathematically, it was assumed that the spreading dynamics occurred in well-mixed populations, and the system exhibited
a discontinuous phase transition. In a recent study, Chen \emph{et al.} \cite{chenphase}
investigated a model of two interacting SIS epidemics, in which the reproduction number
is altered by the interaction introducing
a potential change in the secondary infection propensity.
When the susceptible nodes move faster than the infected nodes,
and the interaction strength is not very strong (i.e., neither
too competitive nor too cooperative), there is a nontrivial spatial
infection pattern in the system.

\begin{figure}
\begin{center}
\epsfig{file=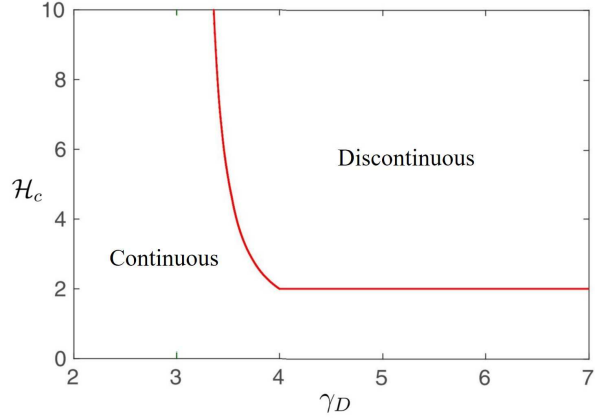,width=0.6\linewidth}
\caption{(Color online) The minimum cooperativity $\mathcal{H}_c$ needed for a
discontinuous transition versus degree
exponent $\gamma_D$. Partially reproduced from \cite{cui2017mutually}.}
\label{fig7_bio}
\end{center}
\end{figure}

Cui \emph{et al.} \cite{cui2017mutually} further analyzed two cooperative
SIR epidemics proposed in Ref.~\cite{cai2015avalanche} on uncorrelated
SF networks, and developed a generalized
heterogeneous mean-field theory to describe the cooperative spreading dynamics.
They found that the outbreak threshold is $\lambda_c^1=\langle k\rangle/(\langle k^2\rangle
-\langle k\rangle)$, which is the same as the classical epidemic outbreak
threshold. Near the critical point, the auxiliary function $\phi_\infty=
\frac{1}{\langle k\rangle}\sum_k (k-1)P(k)R_k(\infty)$ can be expressed
as
\begin{equation}\label{critical_syn}
\phi_\infty\sim (\frac{\lambda_1-\lambda_c^1}{\lambda_c^1})^{1/\alpha_e},
\end{equation}
where $R_k(\infty)$
is the probability that a node with degree $k$ is in the recovered state
when $t\rightarrow\infty$, and $1/\alpha_e$ is the critical exponent of the
system. The type of phase transition is determined by the values
of $1/\alpha_e$. (i) For the case in which the degree exponent $2<\gamma_D<3$, there is
no epidemic threshold, i.e., $\lambda_c^1=0$, and the phase transition is continuous for any
value of the cooperativity $\mathcal{H}=\lambda_2/\lambda_1$. The critical exponent is
$1/\alpha_e=(\gamma_D-2)/(3-\gamma_D)$. (ii) For the case in which $3<\gamma_D<4$, there is a
critical value $\mathcal{H}_c$, above which the phase transition is discontinuous, and
$\mathcal{H}_c$ is always larger than 2. For the continuous phase transition, the
critical exponent is $1/\alpha_e=1/(\gamma_D-3)$. When $\mathcal{H}=\mathcal{H}_c$,
$1/\alpha_e=1$. (iii) For the case in which $\gamma_D>4$, the critical exponent is
$1/\alpha_e=1$, and the phase transition is continuous when $\mathcal{H}<\mathcal{H}_c=2$; otherwise, the system
exhibits a discontinuous phase transition. The phase diagram is shown in Fig.~\ref{fig7_bio}.
Cui \emph{et al.} further demonstrated that the discontinuity decreases
with clustering through extensive numerical
simulations~\cite{cui2018effect}.

H\'{e}bert-Dufresne and Althouse proposed a synergistic coinfection model on clustered networks
\cite{hebert2015complex}, which is constructed by inducing overlapping communities. In their model, a node that has not yet been
infected will be infected by epidemics 1 and 2 with
probabilities $\lambda_1$ and $\lambda_2$, respectively. A node
that has already been infected by epidemic 1 (or 2) will be infected by the
other with probability $\lambda_1+\Delta$ (or $\lambda_2+\Delta$).
They found that epidemics spread faster on clustered networks
than on the equivalent random networks, indicating that clustering
has an opposite role on synergistic coinfection spreading dynamics to that on single
epidemic dynamics. By investigating the basic reproductive
number, H\'{e}bert-Dufresne and Althouse gave an accurate estimation
of the coupling strength for which epidemics on clustered networks spread faster
than the equivalent random networks.

Azimi-Tafreshi~\cite{azimi2016cooperative} considered the effects of multiplexity on cooperative spreading
dynamics, and described multiplex networks by
using a joint degree distribution $P(k_1,k_2,k_{12})$, where $k_1$ and $k_2$
respectively denote the number of edges in networks $\mathcal{A}$
and $\mathcal{B}$, and $k_{12}$ denotes the
number of overlapped edges. It is assumed that both epidemics
follow the SIR model, denoted as 1 and 2,
respectively. Epidemic 1 (2) spreads through
edges in network $\mathcal{A}$ ($\mathcal{B}$) with
transmission probability $\lambda_1$ ($\lambda_2$), and spreads through overlapped edges
with probability $\lambda_{12}$, and it is assumed that $\lambda_{12}>\lambda_1\lambda_2$. By using a generalized
percolation theory, the fraction of nodes infected by each
epidemic and both epidemics in the final state can be obtained. The system exhibits a
tricritical point, and the phase transition changes from continuous
to hybrid with the increase in the strength of cooperation. Zhao \emph{et al.} developed a unified
theoretical approach for coevolution spreading dynamics on multiplex networks,
which can be used to describe the competitive, cooperative, and asymmetrical
interactions between two different dynamics~\cite{zhao2014unified}.

\subsection{Coevolution of multiple biological contagions }

\begin{figure}
\begin{center}
\epsfig{file=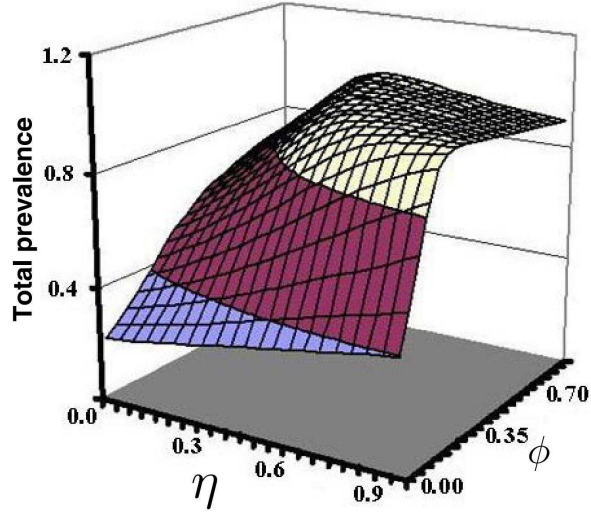,width=0.6\linewidth}
\caption{(Color online) Total prevalence versus $\eta$ and $\phi$.
The number of epidemics is $n=50$ and the effective transmission rate of
each epidemic is $e=\gamma/(\mu+\gamma)=0.3$. Reproduced from \cite{abu2008interactions}.}
\label{fig8_bio}
\end{center}
\end{figure}

In biological systems, the variation of a virus such
such as influenza~\cite{ferguson1999effect},
HIV~\cite{nowak1995antigenic}, meningitis~\cite{richardson2002mutator},
and dengue~\cite{hughes2001evolutionary} could
lead to thousands of new viruses. Understanding the
statistical mechanics, strain structure, and spreading patterns
has attracted much attention, especially in the fields of
biomedicine and statistical physics. Here, we mainly introduce
the progress related to physical science.
Abu-Raddad \emph{et al.}~\cite{abu2008interactions} proposed
a model consisting of
multiple interacting epidemics with the existence of
coinfection, cross-immunity, and arbitrary strain
diversity. It is assumed that the
system has $n$ different epidemics, denoted
as $\mathcal{H}=\{1,2,\cdots,n\}$. Nodes in the recovered and infected
states acquire immunity against the
epidemic they had, but remain susceptible
to the epidemics by which they have not been infected. Mathematically, $\mathcal{I}_\mathcal{J}^\mathcal{L}$ denotes
the set of nodes that have recovered from the set of epidemics $\mathcal{J}$ and are infected
by the set of epidemics $\mathcal{L}$. There is no intersection between $\mathcal{J}$ and
$\mathcal{L}$. The birth rate of susceptible nodes is $\iota$ and the
death rate is $\mu$. For an epidemic $l$, its transmissibility
rate is $\beta_l$, recovery rate is $\gamma_l$, and recovery
period is $\gamma_l^{-1}$. For
each epidemic, the strength of infection can be expressed
as $\Lambda^i=\beta_i\sum_{\mathcal{J}\subseteq \mathcal{H}\backslash\{i\}}
\sum_{\mathcal{L}\subseteq \mathcal{H}\backslash \{\{i\}\}\bigcup \mathcal{J}}$.
Abu-Raddad \emph{et al.} assumed the existence of cross-immunity in the
system, i.e., the susceptibility of the individuals in
$\mathcal{I}_\mathcal{J}^\mathcal{L}$ should be multiplied by
by a factor of $\sigma_{\mathcal{J},\mathcal{L}}^i$.
The evolutions of the system are
\begin{equation}\label{multi}
\begin{split}
\dot{\mathcal{I}}_\mathcal{J}^\mathcal{L}&=\iota\delta_{\mathcal{J},\emptyset}\delta_{\mathcal{L},\emptyset}
-\mu \mathcal{I}_\mathcal{J}^\mathcal{L}-\sum_{i\nsubseteq \mathcal{J}\bigcup \mathcal{L}}\Lambda^i
\sigma_{\mathcal{J},\mathcal{L}}^i\mathcal{I}_\mathcal{J}^\mathcal{L}\\
&+\sum_{l\in \mathcal{L}}\Lambda^l\sigma_{\mathcal{J},\mathcal{L}\backslash l}^l\mathcal{I}_\mathcal{J}^{\mathcal{L}\backslash \mathcal{I}}
-\sum_{l\in \mathcal{L}}\gamma_lI_\mathcal{J}^\mathcal{L}+\sum_{j\in
\mathcal{J}}v_\mathcal{J}\mathcal{I}_{\mathcal{J}\backslash j}^{\mathcal{L}\bigcup
\{j\}}.
\end{split}
\end{equation}
On the right hand of Eq.~(\ref{multi}), the third term is the
infection rate by epidemic $i$, the fourth term is the rate at which
a new node is infected by this epidemic, the fifth term stands for
the rate at which nodes recover from an epidemic in $\mathcal{L}$, and the
last term is the rate at which nodes that have recovered from
an epidemic not in $\mathcal{L}$ will be infected by an epidemic in $\mathcal{L}$.
Abu-Raddad \emph{et al.} further analyzed the final state of the spreading
dynamics. They assumed that $\sigma_{j,l}=\eta_j
\phi_l$, where $\eta_j$ is the prior cross-immunity
rate of being infected by a new epidemic after
exposure by and recovery from $j$ different
epidemics in the past, and $\phi_l$ is the cross-immunity
rate of being infected by a new epidemic for a node that
is currently infected by $l$ different epidemics.
The further assumed that $\phi_l=1$ if $l=0$, $\phi_l=\phi$ if $l>0$,
$\eta_j=1$ if $j=0$, and $\eta_j=\eta$ if $j>0$. Under the
above assumptions, the total prevalence is as shown in Fig.~\ref{fig8_bio}.
The cross-immunity against coinfection has a more prominent influence on
multiple epidemics than that of the prior exposure cross-immunity.

Gog and Grenfell~\cite{gog2002dynamics} included the effects of
Reduced transmission and polarized immunity into the dynamics
of multiple strains. Polarized immunity means that partial
cross-immunity renders certain individuals totally immunized.
Specifically, Gog and Grenfell assumed that the system has
$n$ different epidemics, denoted as $\mathcal{H}
=\{1,2,\cdots,n\}$. For epidemic $l$, its infection and
recovery rates are $\beta_l$ and $\gamma_l$,
respectively. If an individual $i$ is infected by epidemic $l$,
the probability for $i$ to be infected by another epidemic,
such as epidemic $l^\prime$, is lower than $\beta_{l^\prime}$,
as given by $(1-\sigma_{ll^\prime})\beta_{l^\prime}$, which suggests the effect
is caused by cross-immunity~\cite{abu2005characterizing,kamo2002effect}
between epidemics $l$ and $l^\prime$.
In such coevolution dynamics, Gog and Grenfell found that
the system has an epidemic cluster. If the infectious period
is short, there is only one dominant cluster, whereas for
a relatively long infectious period, many clusters coexist
and alternate with each other.

Shrestha \emph{et al.}~\cite{shrestha2011statistical}
proposed a novel model to describe the interactions among
different epidemics. In their proposed model, they assumed that
two epidemics interact with each other when the node
has ever been or is currently infected by the two epidemics.
For each epidemic, an SICR model is adopted, in which
the compartment state (C state) is used to represent
the period of convalescence, or a temporal period of
immunosuppression or strain-transcending cross-immunity.
The SIR part has the same meaning as in classical
epidemic dynamics. Furthermore, the host demography
is also included. When there are no interactions
among the epidemics, the system exhibits
damped oscillations. The interactions induce the
emergence of sustained oscillations. Shrestha \emph{et al.}
proposed a statistical
inference approach to investigate the stochastic temporal variation,
under-reporting, and over-aggregation of multiple
epidemics.

Juul and Sneppen~\cite{juul2011locally} investigated a multiple-epidemic spreading model on a two-dimensional square lattice with periodic
boundary conditions and $N=L^2$ nodes, which is a generalization of
their previous model \cite{sneppen2010minimal}. At each time step, each node $i$ will
spawn a new epidemic $l$ with probability $\sigma$. If node $i$ is
infected by more than one epidemic, it can only transmit one of
its infected epidemics to a neighbor $j$. If $j$ is not infected
and immunized against epidemic $\iota$, it will be infected by $\iota$.
After a period of $\tau_0$ steps, node $j$ recovers. If a node
has been infected by $n$ different epidemics and not yet recovered from any of them,
the probability that it tries to transmit one of its infected
epidemics to a given neighbor is
$\lambda=1-{\rm exp}(-\tau_0/4n)$, which corresponds to
the percolation probability. Numerical simulations indicated
that the exponent of cluster mass size versus diameter $L$ is
close to the fractal dimension of percolation, with diameter
1.896.

Zarei \emph{et al.}~\cite{zarei2019exact}
recently developed an exact solution for a cooperative SIR model
with $n$ epidemics. For a node $i$ that has been infected by
$\ell$ different epidemics, it is infected by the $(\ell+1)$-th
epidemic with probability $\beta_{\ell+1}$. The cooperative
strength is defined as $\mathcal{H}_{\ell}=\beta_\ell/\lambda_1$. At time $t$,
the fraction of nodes infected by $\ell$ different epidemics is $\rho_{\ell}(t)$.
The evolution of $\rho_\ell(t)$ is
\begin{equation}
\frac{d\rho_{\ell}(t)}{dt}=(\ell\beta_{\ell-1}\rho_{\ell-1}-(n-\ell)\beta_\ell\rho_\ell)X,
\end{equation}
where $X$ is the fraction of nodes transferring one specific epidemic.
The evolution of $X$ can be written as
\begin{equation}
\frac{dX}{dt}=\left(-1+\sum_{\ell=0}^{n-1}{{n-1}
\choose{\ell}}\beta_\ell\rho_\ell\right)X.
\end{equation}
For well-mixed populations, and setting $\mathcal{H}_\ell=\mathcal{H}$,
Zarei \emph{et al.} revealed that the critical condition is
\begin{equation}
\beta_1^{\rm crit,n}=1-\sqrt{\frac{2\rho_0}{2}(\mathcal{H}(n-1)-n)},
\end{equation}
where $\rho_0$ is the fraction of seeds. From the above equation,
the minimum value of $\mathcal{H}$ yields $n/(n-1)$ as the discontinuity of the system.

\subsection{Summary}

In this section, we presented the progress on coevolution spreading
biological contagions. For single biological contagions
on complex networks, the system
always exhibits a continuous phase transition, and the threshold
and critical behavior are associated with the network topologies.
To describe the spreading dynamics quantitatively, each widely
used theoretical approach has limitations and advantages. To summarize,
the DMP approach can take into account the full
topology of the network and deal with partial dynamical correlations, whereas the heterogeneous mean-field approach,
bond percolation theory, edge-based compartmental approach,
and quenched mean-field theory only perform well in
networks with specific topological properties (e.g., uncorrelated local tree-like networks).
For two successive biological contagions, the first epidemic may
provide immunity or convenience to the second epidemic, and thus
suppress or promote the second one. Generally, the coexistence
threshold of two epidemics is larger than the first epidemic
threshold. For the coevolution of two epidemics, scientists
found that competing or cross-immunity interactions can induce
the coexistence phase, and the phase diagram is affected by the
network topology. For cooperative epidemics, the
discontinuous phase transition and hysteresis loop are
included, depending on the network dimensions and
spreading dynamics. Finally, when multiple epidemics
are simultaneously spreading, epidemic clusters emerge, with
the exponent of cluster mass size versus diameter being
close to the fractal dimension of percolation. The
threshold and phase transition can be analytically solved on well-mixed
populations.

\section{Coevolution of social contagions} \label{soc}

The diffusion of news, innovations, and cultural fads,
as well as participation in health behaviors and political
protests are all examples of social contagions, in which the state
of an individual is not only impacted by interaction with peers,
but also strongly influenced by his or her psychological and cognitive factors,
as well as social affirmation~\cite{rogers2010diffusion}. In the adoption of a social behavior,
multiple confirmation, i.e., the social reinforcement effect, of the
credibility and legitimacy of the social behavior is always sought,
which has become a key differentiating factor from biological
contagions, in which a simple contact is sufficient to trigger the infection.
Thus, social contagion processes cannot be described by the
biological contagion model. To understand the underlying mechanisms
and to make the full use of social contagion, much empirical analysis
and modeling work has been devoted to this research area. Therein,
Centol's \cite{centola2010spread} health behavior experiment on an online
social network reveals that the social enforcement effect really exists,
the threshold model \cite{Granovetter:1973,watts2002simple} is a well-known model in studying social contagions, and so on.

As in the case of biological contagions, the main concern of this section regards how different types of interplay impact the coevolution of two or more social contagions, such as the cooperation between two types of behavior adoption. This section is organized as follows. The first part focuses on single social contagions, starting from the empirical studies and the fundamentals of the mathematical models. As a generalization of
the classical Watts threshold model~\cite{watts2002simple}, we also address the spreading threshold model and introduce the relevant progress. The interest in this part lies in uncovering how the dynamical mechanisms and network structures affect social contagions. The second part is devoted to the interaction of two social contagions. There, we will revisit the models established to capture successive and simultaneous social contagions on complex networks. Our main concern is how the mutual interactions impact the threshold and the type of phase transitions. In the final part, we will generalize the interactions of two social contagions to the interactions of multiple social contagions, empirically and theoretically. This topic will be quite challenging, because their interactions become more complicated, and it is also more important, as they are widely observed in natural, social, and technological systems.

\subsection{Single social contagions}

One of the key issues in network science is to understand, predict, and finally control the dynamics of social contagion processes on complex networks. As early as 1973, Granovetter showed that information
spreading through ``weak ties'' between casual acquaintances is faster than that diffusing through ``strong ties'' among close friends~\cite{Granovetter:1973}. This is because weak ties are usually edges connecting distant nodes, which can accelerate the spreading. The theory of weak ties explained the rapid spread of the HIV disease and information well; however, it cannot give the reason why it fails when using the contagion of preventative measures to stop the HIV disease~\cite{coates2008behavioural} on the same network. This is mainly because of the difference between the spread of an infectious disease and the contagion of preventative measures, where the former is a simple contagion for which one infected individual is sufficient to reproduce the infection, and the latter is a social contagion that requires multiple sources of activation, as this type of spread is usually uncertain, risky, and costly. In this subsection, we will revisit the progress of empirical analyses that reveal the potential mechanism, i.e., social reinforcement, of social behavior adoption and the established mathematical models for single social contagions on complex networks.\\

\subsubsection{Social reinforcement}

\begin{figure}
\begin{center}
\epsfig{file=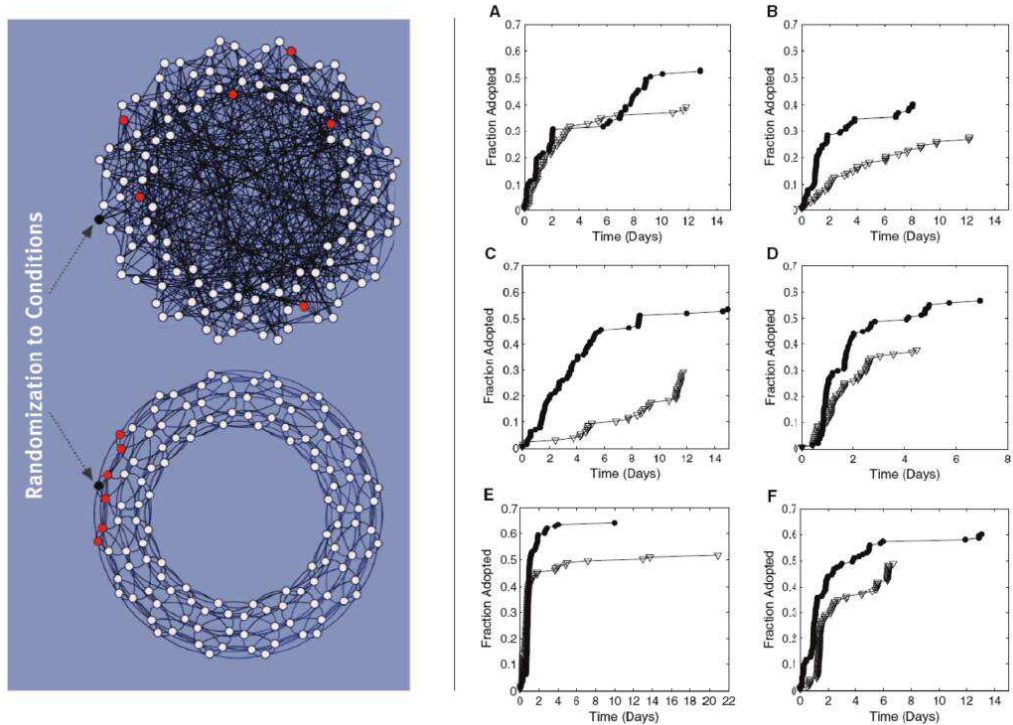,width=1\linewidth}
\caption{(Color online) The left panel shows randomization of clustered-lattice
and random-network conditions in a single trial of this study ($N=128$, $\langle k
\rangle=6$). In each condition, the black node shows the focal node of a
neighborhood to which an individual is assigned, and the red nodes
correspond to that individual's neighbors in the network. In the clustered-lattice
network, the red nodes share neighbors with each other, whereas in the random
network, they do not. White nodes indicate individuals who are not connected
to the focal node. The right panel presents the fraction of health behavior adoption
versus time through the
clustered-lattice (solid black circles) and
random (open triangles) social networks.
(A) to (F) show six independent trials for
different network size $N$ and average degree
$\langle k\rangle$. Specifically, (A) shows a trial with parameters
$N=98$, $\langle k\rangle=6$; (B--D) shows $N=128$, $\langle k\rangle=6$;
and (E,F) shows $N=144$, $\langle k\rangle=8$. The speed of the diffusion
process is evaluated by comparing the time required for the behavior to
spread to the greatest fraction reached by both conditions in each trial.
Reproduced from Ref.~\cite{centola2010spread}.}
\label{fig_empircal_sci}
\end{center}
\end{figure}

The rapid development of Internet technology has enabled large-scale social experiments on online social networks. Centola~\cite{centola2010spread} recruited 1528 participants online and tested the effects of the network structure on the contagion of health behavior. By comparing the spreading of health behavior on a regular clustered network and a random network (as shown in the left panel of Fig.~\ref{fig_empircal_sci}), Centola found that the behavior spreads farther and faster across clustered networks than random networks, because the participants can receive social reinforcement from multiple neighbors in the former network (as presented in the right panel of Fig.~\ref{fig_empircal_sci}). This is significantly different from the results of biological contagions, where a clustered structure usually suppresses the spreading. To investigate the robustness of Centola's experiment, L\"{u} \emph{et al.}~\cite{lu2011small} proposed an unknown-known-approved-exhausted model, which emphasizes the effect of social reinforcement by incorporating a mechanism by which redundant signals can increase the approval rate. They found that under certain conditions, information spreads faster and more broadly in a regular clustered network than in a random network, which to some extent supports the results of Centola's experiments. However, increasing the network size tends to favor effective spreading in a random network, which challenges the validity of the abovementioned experiment for larger-scale systems. Moreover, they found that introducing a low degree of randomness into a regular network yields the most effective information spreading. Similar to Ref.~\cite{lu2011small}, Zheng \emph{et al.}~\cite{zheng2013spreading} further claimed that increasing the network size or decreasing the average degree enlarges the difference in the final fraction of approved nodes between regular and random networks. Smoking behaviors explicitly inflict the effects on people around the smokers; as a result, they also impact the initiation and cessation of smoking~\cite{powell2005importance}. Christakis and Fowler studied the effects of peer influences on quitting behavior by analyzing a network of 12,067 people who underwent repeated assessments of their smoking behavior and social-network ties over a period of 32 years~\cite{christakis2008collective}. They found that the likelihood for a smoker to quit smoking depends on their exposure to multiple contacts with nonsmokers. The social reinforcement in encouraging smokers to abstain is also found online. Myneni \emph{et al.}~\cite{myneni2015content} found that smokers are more likely to abstain if they are exposed to several abstinent users by examining peer interactions over QuitNet---a social media platform for smokers attempting to quit. A series of other studies also demonstrated that the effects of social reinforcement from the peers were strengthened when the peers come from different social groups, exhibit the value of structural diversity, and share some key characteristics with the ego in the dynamics of social contagion~\cite{aral2017exercise}.

To popularize innovations in the most timely manner, economists and marketers have been focused on how technological innovations diffuse though a population by performing various controlled experiments. In developing countries, the adoption of new agricultural technologies is an important way to help people to escape poverty. Bandiera and Rasul studied the adoption process of a new crop for farmers in Mozambique, and found that the farmers were more likely to adopt the crop when they had a higher number of adopters among their family and friends~\cite{bandiera2006social}. In the diffusion of online innovations, Karsai~\emph{et al.} analyzed a dataset recording the adoption process of the world's largest voice over Internet protocol service, Skype, and found that the probability of adoption via social influence is linearly proportional to the fraction of adopting neighbors~\cite{karsai2014complex}. An empirical analysis in recruiting individuals to use Facebook~\cite{ugander2012structural} and Twitter~\cite{toole2012modeling} indicates that the more exposures an individual receives, the higher the probability that he will adopt the applications.

Social reinforcement effects also occur widely in other types of behavioral adoption processes, such as using menstrual cups~\cite{oster2012determinants}, adopting seeding strategies~\cite{banerjee2013diffusion}, taking a new diagnostic method~\cite{weiss2014adoption}, joining social movement activities~\cite{bond2012million}, retweeting
politics hashtags~\cite{romero2011differences}, and learning a new industry~\cite{gao2017collective}.

\subsubsection{Threshold model}

One early mathematical model established to describe the dynamics of social contagions is the threshold model~\cite{watts2002simple,granovetter1978threshold,miller2016equivalence}, based on the Markovian process without memory, where the adoption of behaviors depends only on the states of the currently active neighbors (i.e., individuals who have adopted the behavior), and an individual adopts a behavior only when the number or the fraction of his/her active neighbors is equal to or exceeds the adoption threshold. Granovetter~\cite{granovetter1978threshold} proposed the linear threshold model, in which all individuals are in the active or inactive states, and an individual becomes active if and only if the current absolute number of active neighbors is equal to or exceeds the corresponding threshold. As real-world networks are highly heterogenous, within the linear threshold framework, hub nodes are too vulnerable. To overcome this weakness, Watts takes into account the heterogeneity of individuals' number of contacts and proposes a novel threshold model (later named the Watts threshold model)~\cite{watts2002simple}. In the model, each node is initially assigned a threshold $\phi$, randomly drawn from a distribution $h(\phi)$. When the fraction of active neighbors of a node is equal to or exceeds its threshold $\phi$, it becomes active. Adopting the fraction of active neighbors instead of the absolute number is the essential difference compared with the linear threshold model. The function $h(\phi)$ can be defined arbitrarily, but satisfies the condition $\int_0^1h(\phi)d\phi=1$. For an infinite network with finite average degree, the only way that an initial seed node can grow is for at least one of its immediate neighbors to have a threshold such that $\phi \leq 1/k$, where $k$ is the degree of this neighbor. These nodes with $\phi \le 1/k$ are called vulnerable nodes, because they become active with only one active neighbor. Specifically, for a node of degree $k$, the probability $\rho_k$ that it is vulnerable is
\begin{equation}
\rho_k=\left\{
\begin{array}{rcl}
{1,} & { k=0 ,}\\
{F(1/k),} & { k>0 ,}
\end{array} \right.
\end{equation}
{\noindent}where $\mathcal{F}(\phi)=\int_0^{\phi} h(\varphi)d\varphi$. By using the method of generating functions, the critical condition on an uncorrelated configuration model is~\cite{watts2002simple}
\begin{eqnarray} \label{critical_condition}
\sum_k k(k-1)\rho_kP(k)=\langle k\rangle,
\end{eqnarray}
where $P(k)$ and $\langle k\rangle$ respectively represent the degree distribution and the average degree of the network. The critical condition can be explained as follows. When $\sum_k k(k-1)\rho_kP(k)<\langle k\rangle$, most of the vulnerable clusters in the network are small and are not connected, and thus the initial seed node cannot induce the global behavior adoption. However, for $\sum_k k(k-1)\rho_kP(k)>\langle k\rangle$, there exists a giant vulnerable cluster in the network whose size cannot be neglected when the size of network becomes infinite, and therefore a random initial seed node can trigger a global cascade. A remarkable result is that if all individuals are of the same adoption threshold, the final adoption size first grows continuously and then decreases discontinuously with the increase in the average degree. Meanwhile, a global cascade occurs more easily in a network with a more heterogeneous degree distribution. In modeling the propagation of opinions, the diffusion of innovations, and the adoption of behaviors, the Watts threshold model is well recognized as the fundamental model. Similar to epidemic spreading dynamics, researchers have explored social contagions on complex networks based on the Watts threshold model for two aspects. One is to reveal the effects of network topology and the other is to understand the mechanisms of diffusion at the individual level.

Clustering measures the edge density among the neighbors of an individual, which plays a key role in epidemic spreading. One manifest conclusion is that high clustering suppresses epidemic spreading, leading to an increase in the epidemic threshold and a decrease in the infection size~\cite{zhou2005maximal}. In investigating the effects of clustering on the adoption of behaviors, Ikeda \emph{et al.}~\cite{ikeda2010cascade} studied the Watts threshold model on a clustered network generated from a projection of bipartite graphs and compared it with a nonclustered network with the same degree distribution. Similar to the spreading of health behavior~\cite{centola2010spread, lu2011small,zheng2013spreading}, global cascades occur more easily on clustered networks than on nonclustered networks~\cite{ikeda2010cascade}. Hackett \emph{et al.}~\cite{hackett2011cascades} explored cascades on clustered networks produced by the configuration model with adjustable clustering~\cite{newman2009random}, and found that there exists a range of $\langle k \rangle$ in which increasing the clustering of the network results in an increase in the mean cascade size, whereas outside of this range it will decrease the mean cascade size \cite{miller2015complex}. Hackett and Gleeson~\cite{hackett2013cascades} further explored cascade phenomena on highly
clustered clique-based graphs~\cite{gleeson2009bond} and obtained a closed-form expression for the final fraction of active nodes within a clique of arbitrary size.

Social networks are mostly assortative, whereas technological and biological networks are usually disassortative~\cite{newman2002assortative}. Gleeson~\cite{gleeson2008cascades} studied the Watts threshold model on correlated networks and put forward an analytical approach to compute the mean cascade size. Dodds and Payne~\cite{dodds2009analysis} developed a generating function method that can not only calculate the mean cascade size, but also obtain the probability that a randomly chosen seed will trigger a global cascade. Moreover, they also validated the theoretical results on random networks with bimodal degree distributions. Payne \emph{et al.}~\cite{payne2009information} studied the Watts threshold model on degree-correlated random networks by numerical simulations. They found that increasing the positive degree--degree correlation of a network expands
the global cascade regions. Moreover, the degree--degree correlations impact the relationship between the initiator's degree and its ability to trigger a large cascade.

Galstyan and Cohen~\cite{galstyan2007cascading} studied the linear threshold model on a network composed of two loosely coupled communities. They found if the seeds are contained in one of the communities initially, the peaks of the activation dynamics in each community are well-separated in time. Curato and Lillo~\cite{curato2016optimal} used the linear threshold model to explore the optimal structure of a network consisting of two communities to maximize the asymptotic extent of the diffusion. They found that the optimal structure can be assortative, core-periphery, or even disassortative when the average degree and the fraction of initiators are constrained. In looking for a minimal fraction of initial seeds needed to trigger a global cascade, they showed that the optimal network is a very dense community linked to a much more sparsely connected periphery. The impact of community structure on the cascade processes based on the Watts threshold model was theoretically studied by Nematzadeh \emph{et al.}~\cite{nematzadeh2014optimal}. They constructed a network model with two homogeneous modules, where the internal connectivities of the two communities are the same, and one parameter, i.e., the fraction of edges between the two communities, is used to control the strength of the community structure. There exists an optimal network modularity at which the cascade size is maximized. This study was extended to the case of a network with multiple modular communities~\cite{wu2016optimal}, which yielded similar results, i.e., that modular structures facilitate the cascade and an optimal modularity exists.

Weighted networks provide meaningful representations of the strengths of interactions between entities in the real world~\cite{yang2012epidemic}. However, only a limited number of works have studied the Watts threshold model
on weighted networks. Hurd and Gleeson~\cite{hurd2013watts} studied the Watts threshold model on weighted networks, where the weight of an edge depends on the degrees (e.g., $k$ and $k^{\prime}$) of its nodes and is proportional to $(kk^{\prime})^{-\alpha_p}$. In the case of $\alpha_p>0$ ($\alpha_p<0$), the edge strength decreases (increases) with the increase in the product of the connectivities $k$ and $k^{\prime}$. The cascade window (i.e., the range of $\langle k \rangle$ for which global behavioral adoption occurs) is shifted to higher values when $\alpha_p>0 $, because highly connected nodes have relatively less influence on their neighbors. Unicomb \emph{et al.}~\cite{unicomb2018threshold} found that the heterogeneities of the weight distribution show a nonmonotonous effect on the spreading dynamics, which can accelerate or decelerate cascade processes.

In the real world, the interactions that trigger an individual to become active may come from the multiple sources~\cite{boccaletti2014structure}. Brummitt \emph{et al.}~\cite{brummitt2012multiplexity}
generalized the Watts threshold model to multiplex networks, in which a node becomes active if the fraction of its active neighbors in any channel exceeds a certain threshold, i.e., $\underset{i=1,...,r}{\rm max}\big(\frac{m_i}{k_i}\big)\geq{\phi}$, where $m_i$ and $k_i$ respectively represent the number of active neighbors and the number of neighbors in channel $i$. Compared with contagions on single networks that have the same topology but without considering multiplexity, Brummitt \emph{et al.} found that a multilayer network has a higher probability to experience a global cascade. Ya\u{g}an and Gligor~\cite{yaugan2012analysis} proposed
a content-dependent linear threshold model for social contagion in multiplex networks. In this model, each edge type $i$ is associated with a content-dependent parameter $c_i$ in $[0,\infty]$ that measures the relative bias of type $i$ in propagating this content. An inactive node becomes active if the total perceived influences, i.e., $\sum c_im_i/\sum c_ik_i$, where $m_i$ and $k_i$ respectively represent the number of active neighbors and the number of neighbors in type $i$, exceeds its threshold $\phi$. The authors showed that the content and edge types are important in characterizing a global cascade. Zhuang \emph{et al.}~\cite{zhuang2017clustering} studied the content-dependent linear threshold model on clustered multiplex networks, and found that the clustering plays a double-faceted role in cascade processes, where the clustering decreases the cascade size when the average degree
of the network is small, and facilitates cascades when the average degree is large. Along this line, Lee \emph{et al.}~\cite{lee2014threshold} studied the effect of individuals' heterogenous responses of Watts threshold model in multiplex networks. Two types of responses are introduced. In the first type, an individual becomes active if in at least one layer, a sufficiently large fraction of neighbors is active. In the second type, an individual becomes active only if the fraction of active neighbors is sufficiently large in every layer. They showed that varying the fractions of nodes following either rule facilitates or inhibits cascades. Furthermore, they found that the global cascades become discontinuous near the inhibition regime, and the cascade size grows slowly as the network density increases. Li \emph{et al.}~\cite{li2015cross} explored cross-layer cascade processes in multiplex networks. They found that multiplexity accelerates the cascade if the additional layer can provide extra short paths for rapid spreading.

Karimi and Holme~\cite{karimi2013threshold} modeled the cascade process in temporal networks, where individuals are only influenced by their contacts within a finite time window from the past to the present. The randomization of time stamps makes the cascades larger for the fractional-threshold mechanism~\cite{watts2002simple}, whereas it makes the cascades smaller in the case of the absolute-threshold mechanism~\cite{granovetter1978threshold}. Takaguchi \emph{et al.}~\cite{takaguchi2013bursty} also studied the linear threshold model on empirical temporal networks incorporated with a decaying mechanism of the exposures, and showed that burst activity patterns facilitate the contagion. However, in another study of the linear threshold model on empirical temporal networks, Bachlund \emph{et al.}~\cite{backlund2014effects} found that some networks support cascades, and some do not. Their further analysis manifested that there exists competition between the inhibition effect of burst activity patterns and the promotion effect of timing correlations between contacts on adjacent edges.

Some specific models incorporating the underlying mechanisms that affect the behavioral adoption of individuals have also been established. Melnik \emph{et al.}~\cite{melnik2013multi} purposed a multistage social contagion model accounting for the fact that individuals in different stages of the spreading process exert different levels of influence on their neighbors, which can reproduce multistage cascade phenomena. Liu \emph{et
al.}~\cite{liu2018impacts} considered a specific situation, in which individuals in the network have several opinion leaders, who affect their behaviors markedly. The impact of opinion leaders makes global cascades
occur more easily, which can not only reduce the lowest average degree of the network required for a global cascade, but also increase the highest average degree of a network for which a global cascade can occur. Kobayashi~\cite{kobayashi2015trend} generalized the Watts threshold model with a trend-driven mechanism, which introduces another type of node, global nodes whose states depend on the fraction of activated
nodes in the population. When the fraction of activated nodes in the population exceeds a threshold, a trend emerges and the global nodes become active. Kobayshi showed that global nodes accelerate cascades once a trend emerges, whereas their existence reduces the probability of a trend emerging. Accordingly, there exists a moderate fraction of global nodes that maximizes the average size of cascades. The persuasion mechanism, which can strengthen the ability of activated nodes to convince their neighbors to adopt the behavior is also considered in the Watts threshold model purposed by Huang \emph{et al.}~\cite{huang2016contagion}.
They found that this introduced mechanism can render networks more vulnerable to global cascades, especially in heterogeneous networks. Ruan \emph{et al.}~\cite{ruan2015kinetics} generalized the Watts
threshold model with mechanisms of spontaneous adoption and complete reluctance to adoption (i.e., immune or blocked nodes). They showed that the speed of spreading depends strongly on the density of blocked nodes. When the fraction of blocked nodes is small, spontaneous adopters are able to generate a large cascade. When the fraction of blocked nodes is large, because spontaneous adopters dominate the spreading, only small cascades can be generated and the spreading becomes slow. Juul and Porter~\cite{juul2018synergistic} incorporated synergistic effects into the Watts threshold model, and found that constructive synergy (i.e., the peer pressure experienced by
a node is larger than in the Watts threshold model) accelerates the contagion process and interfering synergy (i.e., the peer pressure experienced by a node is less than in the Watts threshold model) slows down the contagion process. Oh and Porter~\cite{oh2018complex} accounted for the case in which individuals wait for some period of time before they adopt a behavior. Their results indicated that heterogeneously distributed wait times
can change the adoption order of nodes, and either accelerate or decelerate the spread of adoptions.

\subsubsection{Memory effects}

\begin{figure}
\begin{center}
\epsfig{file=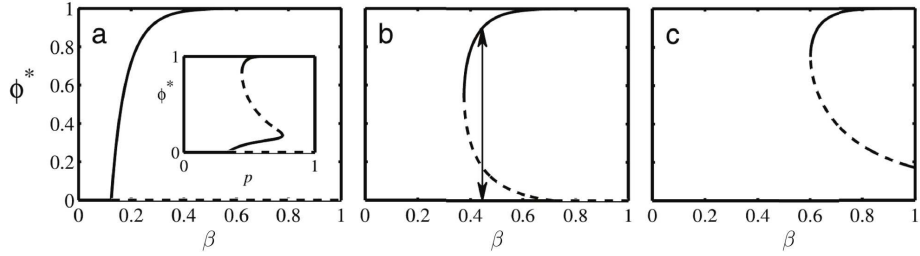,width=0.9\linewidth}
\caption{(Color online) Numerical analysis of Eq. (\ref{phi_steady}). (a) Class I: Epidemic threshold.
(b) Class II: Vanishing critical mass. (c) Class III: Pure critical mass. Dose sizes are lognormally distributed with mean 1 and variance 0.433, $T=10$, and thresholds are uniformly set at (a) $d^*=0.5$, (b) $d^*=1.6$, and (c) $d^*=3$. Inset in (a): Example of a
more complicated fixed point diagram. Here, $T=20$, dose size is set to unity, $f(d)=\delta(d-1)$ is the delta function, and $d^*=1$ with probability 0.15 and 6 with probability 0.85. Reproduced from Ref.~\cite{dodds2004universal}.}
\label{fig_single_watts}
\end{center}
\end{figure}

Social reinforcement resulted from multiple exposures of peer influence is a key feature of social contagions. In the above-discussed models, whether a node becomes active depends only on the number of current exposures from its neighbors, without memory effects. However, the historical records may be relevant; for example, recent exposures may be more important than long-ago exposures. In such cases, memory plays a significant role and the dynamics become non-Markovian. To account for memory effects, Dodds and Watts~\cite{dodds2004universal,dodds2005generalized} proposed a generalized threshold model, where each individual $i$ contacts with one other individual $j$ drawn randomly from the population. If $i$ is susceptible and $j$ is infected, with probability $\beta$, node $i$ receives a positive dose $d_i(t)$ sampled from a distribution at time step $t$. Each individual $i$ maintains a memory of doses received over the previous $T$ time steps, recorded by $D_i(t)=\sum_{t^{\prime}=t-T+1}^t d_i(t^{\prime})$. A susceptible individual $i$ becomes infected if $D_i(t)\geq d_i^*$, where $d_i^*$ is a certain threshold initially drawn from the distribution $g(d^*)$. The probability that a susceptible individual contacts $K \leq T$ infected individuals in $T$ time steps and then becomes infected is
\begin{equation} \label{P_inf_k}
P_{inf}(K)=\sum_{k=1}^{K} \binom{K}{k} {\beta}^k {(1-\beta)}^{K-k} p_k,
\end{equation}
{\noindent}where $p_k=\int_0^{\infty}$d$d^*g(d^*)\mathcal{P}(\sum\limits_{i=1}^k d_i \geq d^*)$, which means the average fraction of individuals infected after receiving $k$ positive doses in $T$ time steps, and $\mathcal{P}(\sum\limits_{i=1}^k d_i \geq d^*)$ denotes the probability that the sum of $k$ doses drawn from $f(d)$ exceeds a given $d^{*}$. Each infected individual $i$ recovers with probability $\gamma_1$ if $D_i(t)$ is smaller than $d_i^*$, and the recovered individuals become susceptible with probability $\gamma_2$ in each time step. By adjusting the parameters, such as $d_i(t)$, $d_i^*$, $\beta$, $\gamma_1$, and $\gamma_2$, this generalized threshold model can be reduced to the SIS, SIR, Watts threshold model, and so on.
Specifically, when $\gamma_1=1$, $\gamma_2=1$, and $\phi^*$ denotes the steady-state fraction of infected individuals, Eq.~(\ref{P_inf_k}) becomes
\begin{eqnarray}\label{phi_steady}
\phi^*=\sum\limits_{k=1}^{T}{(\beta\phi^*)}^k{(1-\beta\phi^*)}^{T-k}p_k.
\end{eqnarray}
{\noindent}By analyzing the fixed points of Eq.~(\ref{phi_steady}), three universal classes of equilibrium behavior are found (see Fig.~\ref{fig_single_watts}). Furthermore, when the length of the memory $T$ is fixed, the class to which a particular system belongs is only determined by $p_1$ and $p_2$, which respectively represent the probabilities that an individual will become infected as a result of one and two exposures.

\begin{figure}
\begin{center}
\epsfig{file=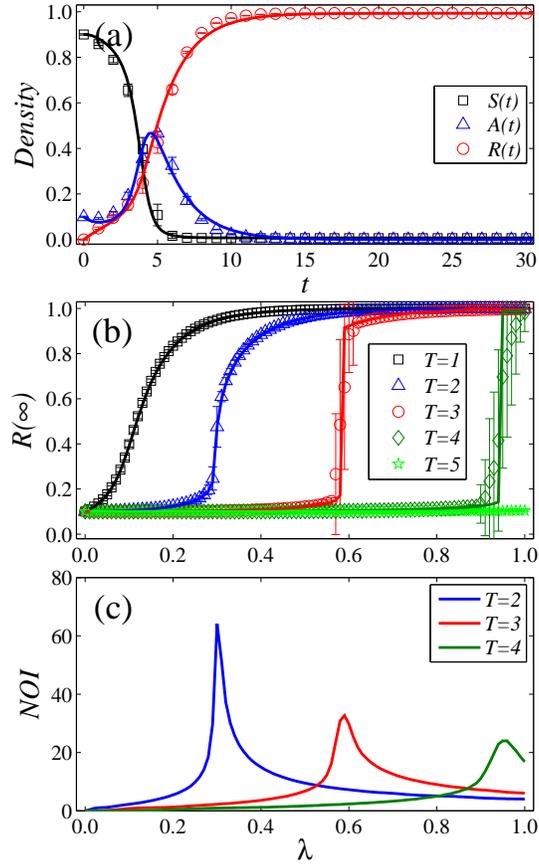,width=0.6\linewidth}
\caption{(Color online) Non-Markovian SAR model on ER networks. (a) Time evolution of the average densities of susceptible, adopted, and recovered nodes. (b) Final adoption size $R(\infty)$ versus effective transmission rate $\lambda=\beta/\mu$ for different adoption thresholds $T$. (c) Simulation results of NOI (number of iterations) as a function of $\lambda$. The lines in (a) and (b) are the theoretical predictions. The parameters in (a) are set as $\lambda=0.8$, $\rho_0=0.1$, $T=3$, and $\gamma=0.5$; in (b) and (c), $\rho_0=0.1$ and $\gamma=1.0$. Reproduced from Ref.~\cite{wang2015dynamics}.}
\label{fig_single_SAR}
\end{center}
\end{figure}

Considering the nonredundancy of received behavior information in social contagions,
Wang \emph{et al.}~\cite{wang2015dynamics} put forward a non-Markovian
susceptible-adopted-recovered (SAR) social contagion model. Initially, a fraction $\rho_0$ of individuals are randomly chosen as seeds.
At each time step, adopted individuals transmit the behavior information to their
susceptible neighbors with transmission probability $\beta$. Once an adopted
neighboring individual $v$ of individual $u$ transmits the behavior information
to $u$ successfully, the cumulative number of received pieces of information of
$u$ is increased by 1 and individual $v$ cannot transmit
the same information to $u$ again, i.e., the information transmission is nonredundant.
At time step $t$, assume a susceptible individual $u$ of degree $k$ already has
$m-1$ pieces of information, and once $u$ receives another piece of information,
the probability that it adopts the behavior is $\pi(k,m)$ whose maximum possible value
is one. The adopted individual permanently becomes recovered with probability $\gamma$.
An edge-based compartmental theory has been developed, and a variable $\theta $
was adopted to represent the probability that a random neighboring individual $v$ had not
transmitted the information to individual $u$ at the end of the spreading dynamics.
The expression of $\theta $ is
\begin{eqnarray} \label{theta_t}
 \theta =  (1-\rho_0)\frac{\sum_{k^{\prime}}
k^{\prime}P(k^{\prime})\Theta(k^{\prime},\theta )}{\langle k \rangle}
+\frac{\gamma}{\beta}(1-\theta)(1-\beta),
\end{eqnarray}
where $\Theta(k^{\prime},\theta )$ is the probability that an individual
has received $m$ pieces of nonredundant information.
{\noindent}The critical condition for the behavior to be widely adopted is
\begin{eqnarray} \label{lambda_c}
\beta_c=\frac{\gamma}{\Delta+\gamma-1},
\end{eqnarray}
where $\Delta=(1-\rho_0)[\sum_{k^{\prime}}k^{\prime}P(k^{\prime})\frac{d\Theta (k^{\prime},
\theta )}{d\theta } |_{\theta_c }]/\langle k \rangle$.
The Heaviside step
function is adopted to depict $\pi(k,m)$, as $\pi(k,m)=1$ when $m\geq T_k$.
By using a bifurcation analysis of Eq.~(\ref{theta_t}) in the steady state,
one main result is that the transition phase of the final adoption size on $\beta$
can be changed from discontinuous to continuous, as shown in Fig.~\ref{fig_single_SAR}.

Following the work of Ref.~\cite{wang2015dynamics}, researchers considered certain different situations.
Zhu \emph{et al.}~\cite{zhu2017social} investigated the SAR model on weighted networks and found that the heterogeneity of weight distributions always hinders social contagions. Su \emph{et al.}~\cite{su2018optimal} proposed a reversible social contagion model on community networks, and found an optimal community structure that can maximize spreading processes. Along this line, the SAR model was also studied in time-varying community networks~\cite{liu2017social}, in which hierarchical orders of behavior adoptions are uncovered. Shu \emph{et al.}~\cite{shu2017social} explored the social contagion process on interdependent lattice networks, and showed that the phase transition can be changed from continuous to discontinuous by increasing the fraction of dependency links. The SAR model was also studied in multiplex networks~\cite{wang2018social,zou2018social,
wu2018double,chen2018complex}. Wang \emph{et al.}~\cite{wang2018social} found that multiplex networks can promote the final adoption size. Zou \emph{et al.}~\cite{zou2018social} considered that different layers of multiplex networks have distinct reliability for social contagions. They showed that increasing the reliability of one layer can promote the contagion process. Wu \emph{et al.}~\cite{wu2018double} found a double transition, whereby a continuous transition occurs first, followed by a discontinuous transition. Chen \emph{et al.}~\cite{chen2018complex} studied the SAR model on multiplex networks and assumed that a susceptible individual becomes adopted only if the number of adopted neighbors is equal to or exceeds the adoption threshold in every layer. They found the final adoption size increases sharply with the information transmission probability when the adoption threshold is high.

Within the SAR model, there are many factors that potentially impact the behavior adoption. Wang \emph{et al.}~\cite{wang2015dynamics} took the limited contact capacity of individuals in transmitting information and suggested that enlarging the contact capacity facilitates global adoption. Considering the heterogeneity of individuals, a binary threshold model is proposed in Ref.~\cite{wang2016dynamics}, in which some individuals have a low adoption threshold (i.e., activists), and the remaining individuals have a high adoption threshold (i.e., bigots). A hierarchical adoption phenomenon is observed, where activists first adopt the behavior and then stimulate bigots to adopt the behavior. Moreover, the model also shows crossover phenomena in the phase transition. Wang \emph{et al.}~\cite{wang2018social} also explored the heterogeneous credibility of individuals in social contagion. Zhu \emph{et al.}~\cite{zhu2018dynamics,zhu2018optimal} studied local trend imitation within a social contagion model, where both tent-like adoption and gate-like adoption probabilities were analyzed. Tent-like adoption means that the adoption probability is an increasing function of $x$ (i.e., the fraction of adopted neighbors) when $x$ is smaller than a certain value, above which it becomes a decreasing function of $x$. The gate-like adoption, however, means that the behavior adoption probability is equal to 1 when $x$ is in a certain range, and it is equal to $0$ when $x$ is out of this range. For the tent-like case,
they showed that the local trend imitation capacity impacts the phase transition, where a second-order phase transition is observed when the capacity is strong, and it becomes a first-order phase transition when
the capacity is weak~\cite{zhu2018dynamics}. An optimal imitation capacity is found to maximize the final adoption size in the gate-like case~\cite{zhu2018optimal}. Su \emph{et al.}~\cite{su2017emergence} proposed a susceptible-trial-adopted-susceptible (STAS) threshold model, where individuals in the trial state accept the behavior temporarily and can transmit the information. They found that the initial conditions of the dynamics affect the final state of the dynamics, which introduces a hysteresis loop in the system. Wang \emph{et al.}~\cite{wang2017social} considered that individuals communicate with multiple channels. When an individual is active in one layer, he/she cannot use other channels to communicate, as individuals cannot simultaneously use all communication channels. Time delays occur and slow down the behavior adoption process; however, this does not affect the final adoption size. Wang \emph{et al.}~\cite{wang2018effects} proposed a model in which time delays are introduced before an individual adopts a behavior when his/her fraction of adopted neighbors is equal to or exceeds the adoption
threshold. They found that long time delays induce a microtransition, which becomes sharper when high-degree individuals have a larger probability of experiencing time delays.

\subsubsection{Generalized social contagions }

\begin{figure}
\begin{center}
\includegraphics[width=3in]{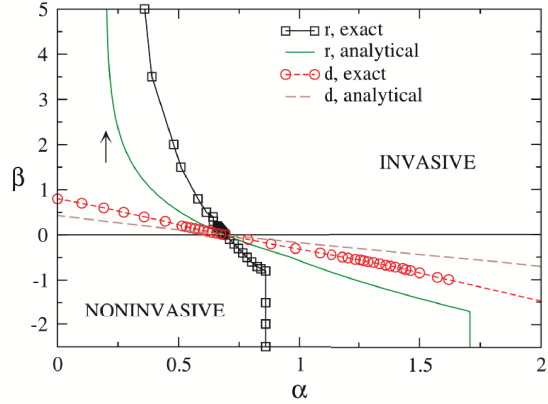}
\caption{(Color online) Phase diagram for synergistic epidemics, reproduced from Ref.~\cite{perez2011synergy}.}
\label{fig_single_synergy}
\end{center}
\end{figure}

The threshold model~\cite{granovetter1978threshold,watts2002simple} and the non-Markovian SAR model~\cite{wang2015dynamics} only take into account the synergy effects, i.e., the reinforcement effects, of susceptible individuals' direct neighbors. However, much evidence indicates that other infected individuals around a pair of susceptible-infected individuals would strengthen the transmission rate between them~\cite{ludlam2012applications}. P\'{e}rez-Reche \emph{et al.}~\cite{perez2011synergy} proposed a generalized SIR model, in which the transmission rate between an infected individual (donor) and one of its susceptible neighbors (recipient) depends on the neighborhood of this donor--recipient (d-r) pair, as
\begin{eqnarray} \label{lambda_dr}
\lambda_{d-r}(t)={\rm max}\{0,\alpha+\beta_{d-r}(t)\},
\end{eqnarray}
where $\alpha$ is a positive constant representing the basic rate of infection for an isolated $d-r$ pair without synergy effects. The rate $\beta_{d-r}(t)$ quantifies the degree of synergistic effects. In the absence of synergistic effects, $\beta_{d-r}(t)=0$ and the model reduces to the simple SIR process. Two types of synergy are introduced in the model, and the expression for $\beta_{d-r}(t)$ depends on the synergy type.
When multiple donors challenge a recipient host (i.e., r-synergy type), $\beta_{d-r}(t)=\beta[n_r(t)-1]$ with $n_r(t)$ representing the number of donors challenging a recipient host at time $t$. The rate $\beta$ measures the strength of the synergy, which is constructive for $\beta > 0$ and interfering for $\beta < 0$. The other type of synergy (i.e., d-synergy) accounts for the effect of other donors connected to a donor that is challenging a recipient host. For d-synergy type, $\beta_{d-r}(t)=\beta n_d(t)$ and $n_d(t)$ represents the number of other donors at time $t$. The rate $\beta$ is of the same meaning as for r-synergy. As shown in Fig.~\ref{fig_single_synergy}, the epidemic threshold $\alpha$ is a nonincreasing function of $\beta$ for both r-synergy and d-synergy. P\'{e}rez-Reche \emph{et al.} also showed that constructive synergy induces an exploitative behavior that results in a rapid spreading process, and interfering synergy causes a slower and sparser exploitative foraging strategy that traverses larger distances by infecting fewer hosts. Within this framework, Taraskin and P\'{e}rez-Reche~\cite{taraskin2013effects} found that increasing the connectivity of a lattice enhances the synergistic effects on spreading. Broder-Rodgers \emph{et al.}~\cite{broder2015effects} investigated the effects of both the r-synergistic mechanism and network topology (by rewiring edges in regular lattices) on the epidemic threshold of the SIR model. They found the network topology markedly impacts the spread. Under strong constructive synergy, the systems become more resilient to the spread of the epidemic after more rewiring. However, rewiring always enhances the spread of epidemics when the synergy is destructive or weakly constructive, if the local connectivity is low.

G\'{o}mez-Garde\~{n}es \emph{et al.}~\cite{gomez2015abrupt} proposed an SIRS model in which recovered individuals can be reinfected. This reinfection mechanism affects the nature of its epidemic transition, which induces an abrupt transition when the transmission rate for reinfection ($I+R\rightarrow 2I$) is larger than a critical value. G\'{o}mez-Garde\~{n}es \emph{et al.}~\cite{gomez2016explosive} investigated the synergy effects of the neighborhood of the target ignorant receiver by using the SIS model, where the transmission rate from a transmitter $j$ to an ignorant/healthy receiver $i$ is $\lambda_{j\rightarrow i}$, and its expression is as follows:
\begin{eqnarray}
\lambda_{j\rightarrow i}=\alpha \delta[n^h(i)].
\end{eqnarray}
Here, $\alpha$ accounts for the intrinsic value of the spreading phenomenon in the absence of the context, $n^h(i)$ represents the number of ignorant/healthy neighbors of the receiver $i$, and $\delta[n^h(i)]$ is a function of $n^h(i)$, which captures the effect of the context. G\'{o}mez-Garde\~{n}es \emph{et al.} analyzed both the exponential relation and the linear dependence of $\delta$ on $n^h(i)$, and found that the inhibitory mechanism (i.e., interfering synergy) leads to an explosive contagion that is not observed in the SIR model.

Liu \emph{et al.}~\cite{liu2017explosive} studied the constructive synergy of the spreader's context, where the transmission rate between a transmitter $j$ and a receiver $i$ is $p(m_j,\alpha)=1-(1-\beta)^{1+\alpha m_j}$,
with $m_j$ and $\alpha$ respectively representing the number of infected neighbors of the infected node $j$ and the strength of the synergy effect, and $\beta$ accounting for the basic transmission rate. Two theoretical
methods are established, where a master equation is used to accurately predict the simulation results and the mean-field theory is adopted to give a physical understanding. Liu \emph{et al.} obtained a critical strength
\begin{eqnarray}
\alpha_c=\frac{1}{\langle k \rangle-1}
\end{eqnarray}
by the mean-field theory when dynamical processes occur on a random regular network with average degree $\langle k \rangle$. When $\alpha>\alpha_c$ the steady-state density of the infected nodes exhibits explosive growth with respect to the basic transmission rate and a hysteresis loop emerges. Hoffmann and Bogu\~{n}\'{a}~\cite{hoffmann2018synergistic} proposed a synergistic cumulative contagion model, which takes into account the memory of past exposures and incorporates the synergy effects of multiple infectious sources. They found that the interplay of the non-Markovian feature and a complex contagion produces rich phenomena, including the loss of universality, collective memory loss, bistable regions, hysteresis loops, and excitable phases.

Rumor spreading is another type of social contagion, which rests on the basic idea that a sender continues propagating a rumor as long as it is new for the recipient. Otherwise, he/she loses interest and never spreads it. Daley and Kendall\cite{daley1964epidemics,daley1965stochastic} first proposed an ignorant--spreader--stifler model to describe the dynamical process of rumor spreading. In the model, the ignorant, spreader, and stifler states are respectively similar to the susceptible, infected, and removed states in the SIR model. In contrast to the SIR model, the recovery process (spreader becomes stifler) of the rumor model is not spontaneous. Rather, it is the consequence of individuals' interactions. In detail, both spreaders become stiflers with probability $\gamma$ once they contact, and a spreader becomes a stifler with probability $\gamma$ once he/she meets an another stifler. Instead of two spreaders becoming stiflers when they contact in the recovery process, Maki and Thompson~\cite{Maki1973Rumor} introduce a slightly distinct version, in which, when a spreader $i$ contacts another individual $j$ in the spreader state, only $i$ turns into a stifler and the state of $j$ remains unchanged. Barrat \emph{et al.}~\cite{barrat2008dynamical} studied the rumor model on the complete graph and found that the final fraction, $\rho(\infty)$, of stiflers satisfies the following self-consistent equation
\begin{equation} \label{rumor_model}
\rho(\infty)=1-e^{-(1+\beta/\gamma)\rho(\infty)},
\end{equation}
{\noindent}where, $\beta$ is the transmission rate between a spreader and an ignorant. From Eq.~(\ref{rumor_model}), it is found that the rumor becomes an outbreak for any $\beta/\gamma>0$, which is different from the finite threshold for the SIR model on homogenous networks. Moreover, when simulating the Maki--Thompson model on SF networks, Moreno \emph{et al.}\cite{Moreno2004a,Moreno2004b} found that the outbreak threshold for rumor spreading is finite. Nekovee \emph{et al.}~\cite{Nekovee2007} found that if the spontaneous recovery process is allowed with a constant rate, the rumor model behaves exactly as the SIR model, such as an infinite threshold on a SF network. These findings show that the dynamic process of rumor spreading is markedly different from the SIR model.

\subsection{Coevolution of two social contagions}

Real-world social contagions usually interact with each other successively or simultaneously, which are thus named the ecology of contagions~\cite{guilbeault2018complex}. In this section, we will revisit the literature
on the interactions between two social contagion processes.

\subsubsection{Successive contagions}

\begin{figure}
\begin{center}
\includegraphics[width=4.5in]{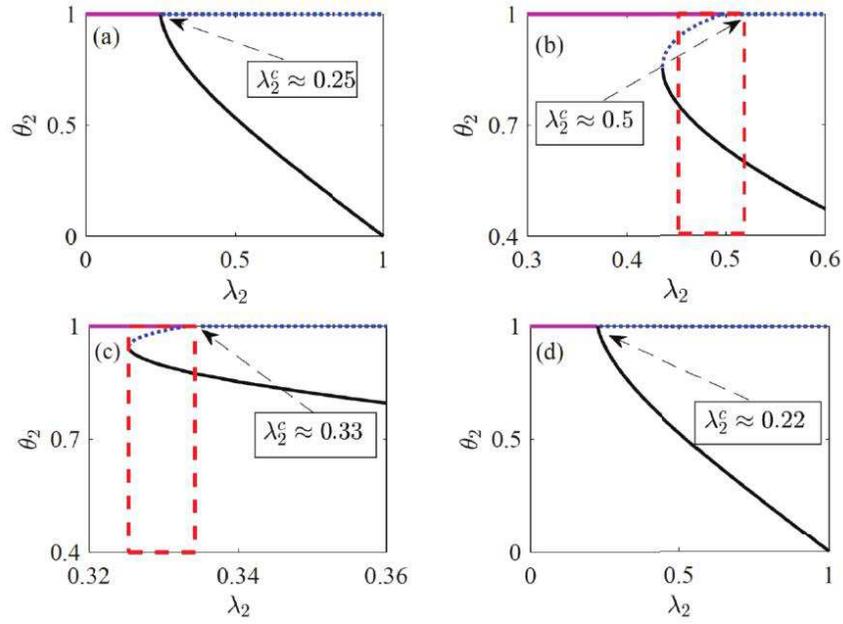}
\caption{(Color online) Graphical analysis of $\theta_2$ versus $\lambda_2$ with
inhibition and synergy effects on ER networks. (Top panel) $\theta_2$ versus
$\lambda_2$ with inhibition effect (i.e., $\alpha=0.5$). The
  parameters in (a) are $\lambda_1=0.2$, $\tau_1=\tau_2=0.4$, and $\gamma_1=\gamma_2=1.0$, and in (b)
  are $\lambda_1=0.8$, $\tau_1=\tau_2=0.4$, and $\gamma_1=\gamma_2=1.0$.
(Bottom panel) $\theta_2$ versus
  $\lambda_2$. The
  parameters in (c) are $\lambda_1=0.2$, $\tau_1=\tau_2=0.4$, and $\gamma_1=\gamma_2=1.0$, and in (d)
  are $\lambda_1=0.8$, $\tau_1=\tau_2=0.4$, and $\gamma_1=\gamma_2=1.0$.  Reproduced
  from Ref.~\cite{liu2018interactive}.}
\label{fig_seq_soc_inhit}
\end{center}
\end{figure}

The studies of two successively interacting epidemics were introduced in the previous section. In comparison, there are very few studies about two successively interacting social spreading processes. Liu \emph{et
al.}~\cite{liu2018interactive} proposed a successively interacting social contagion model, in which two behaviors, i.e., behavior 1 and behavior 2, spread successively on a network. The dynamical processes of both behaviors are depicted by a modified version of the SAR threshold model~\cite{wang2015dynamics}, with a different behavioral response function. During the spread of behavior 1, for a susceptible node of degree $k$ and with $m-1$ cumulative pieces of behavior information, when this node receives a new piece of information at this time step, the probability that it adopts the behavior is
$\pi(k,m)=1-{(1-\tau_1)}^m$,
where $\tau_1$ is the adopting probability for each reception of behavior information. At the termination of behavior 1, a small fraction of nodes are randomly chosen and set to be in the adopted state for behavior 2. The remaining nodes are in the susceptible state for behavior 2. The spreading dynamics of behavior 2 are mathematically identical to the dynamics of behavior 1, except that the transmission probability $\beta_2$ and recovery rate $\gamma_2$ are different. For simplicity, we denote the effective information transmission rate as $\lambda_2$, given by $\lambda_2=\beta_2/\gamma_2$. When a susceptible node $u$ of degree $k$ receives a piece of information about behavior 2 and the cumulative number of pieces of information about behavior 2 is $m$, then node $u$ adopts behavior 2 with a probability
\begin{equation}\label{e2}
\psi(k,m,X_u)=\left\{
\begin{array}{rcl}
{1-(1-\tau_2)^m,} & { { X_u=S},}\\
{1-(1-\alpha\tau_2)^m,} & { {X_u=R},}
\end{array} \right.
\end{equation}
{\noindent}where $\tau_2$ is the adopting probability for each reception of
behavior information 2, $X_u$ denotes the state of node $u$ for behavior 1,
and $\alpha$ quantifies the impact of behavior 1 adoption on adopting
behavior 2. When $\alpha>1$ ($\alpha<1$), the adopting probability for each
reception of behavior information 2 is increased (decreased), corresponding
to the cooperative (inhibitive) effects of behavior 1. The adoption of behavior
1 has no effect on adopting behavior 2 when $\alpha=1$, and $\alpha=0$
indicates that if a node has adopted behavior 1, it never adopts behavior 2,
which means that behavior 1 completely suppresses the adoption of behavior 2.
An edge-based compartmental method was established, and the equation for behavior
2 in the steady state is written as
\begin{equation}\label{e12}
\theta_2=\frac{\sum_{k^{\prime}}k^{\prime}P(k^{\prime})
\Phi_{2}(k^{\prime},\theta_1,\theta_2)}{\langle k \rangle}
+\frac{(1-\theta_2)(1-\gamma_2\lambda_2)}{\lambda_2},
\end{equation}
where $\theta_1$ ($\theta_2$) represents that in the final state, a random
individual does not successfully transmit the behavior information 1 (2)
to his/her neighbor along a random edge.
It is useful to define $g(\lambda_2,\theta_1,\theta_2)$ to be the right-hand
of Eq.~(\ref{e12}) minus the left-hand of Eq.~(\ref{e12}).
Thus, for a given $\theta_1$ (i.e., $\lambda_1=\beta_1/\gamma_1$), combining Eq.~(\ref{e12}) and
\begin{equation}\label{e12-d1}
\frac{dg(\lambda_2,\theta_1,\theta_2)}{d\theta_2}|_{\theta_2^c}=0,
\end{equation}
{\noindent}the critical transmission probability of behavior information 2 is determined and the type of phase transition can be analyzed by the bifurcation theory. To determine the critical condition for the change of phase transition types (from continuous to discontinuous in the presence of inhibition effects, and from discontinuous to continuous in the presence of synergy effects), Eqs.~(\ref{e12}),~(\ref{e12-d1}), and the second derivative of $g(\lambda_2,\theta_1,\theta_2)$ with respect to $\theta_2^c$ equals zero are solved numerically.
{\noindent}Liu \emph{et al.} showed that with the outbreak of
behavior 1, the inhibition effects of behavior 1 can cause the continuous phase transition of the spreading of behavior 2 to become discontinuous. Meanwhile, this discontinuous phase transition of behavior 2 can also become continuous when the effects of the adoption of behavior 1 become synergistic (see Fig.~\ref{fig_seq_soc_inhit}).

\subsubsection{Simultaneous contagions}

\begin{figure}
\begin{center}
\includegraphics[width=4.5in]{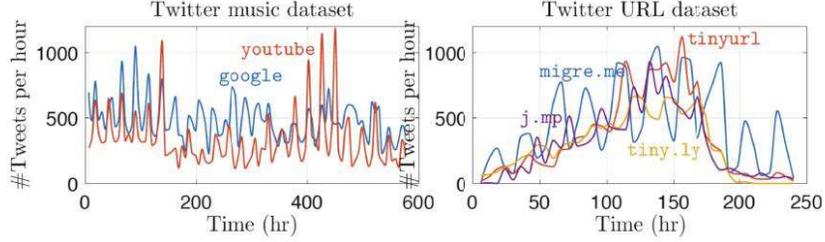}
\caption{(Color online) Visualization of correlated cascading behavior in real data.
(left) Tweets with terms Google and YouTube in Twitter music dataset are synchronized most of the time. (right) Different URL shortening services. $tiny.ly$ and $tinyurl$ are cooperating, whereas $migre.me$ and $j.mp$ are competing.
Reproduced from Ref.~\cite{zarezade2017correlated}.}
\label{fig_soc_sim_emp}
\end{center}
\end{figure}

In real social networks, two behaviors may spread simultaneously and interact with each other. By analyzing the number of tweets versus time in a Twitter music dataset, Zarezade \emph{et al.}~\cite{zarezade2017correlated} found that the usage of Google Play Music follows the same rhythm as that of YouTube when a new album arrives in both systems, and one URL link becomes popular as the other receives less attention (see Fig.~\ref{fig_soc_sim_emp}). The former implies cooperative contagions that promote the spreading adoption of both services, whereas the latter indicates the competitive contagions, where the adoption of one behavior inhibits that of the other. A correlated cascades model was also proposed to predict users' product adoption behavior. By evaluating the prediction accuracy, Zarezade \emph{et al.} claimed that the correlated model performs better than models that do not consider the interaction among spreading processes.

Liu \emph{et al.}~\cite{liu2018synergistic} proposed a synergistic behavior spreading model on two-layer networks, where the SAR model was adopted to describe the adoption process of each behavior. In this model, behavior 1 and behavior 2 are assumed to spread on layers $\mathcal{A}$ and $\mathcal{B}$, respectively, where the adoption thresholds of the two behaviors are respectively equal to $T_1$ and $T_2$. The synergistic mechanism is that the
adoption of one behavior by a node in one layer enhances its probability of adopting another behavior in the other layer. Mathematically speaking, once node $i$ has adopted behavior 1 (2), it generates an increase $\Delta T_2$ ($\Delta T_1$) in the number of pieces of information about behavior 2 (1). For example, if a node has adopted behavior 2 and has received a cumulative total of $m$ pieces of information about behavior 1 from distinct neighbors
in layer $\mathcal{A}$, this node will adopt behavior 1 if $\Delta T_1+m\geq T_1$.
An edge-based method was established to describe such model, 
In particular, when the dynamical parameters and network parameters for behaviors 1 and 2 are the same, one has
\begin{eqnarray}\label{sym_theta}
\theta=\frac{[1-\rho(0)]\sum_kkP(k)\Phi(k,\theta)}{\langle k \rangle }+\frac{(1-\theta)(1-\mu\lambda)}{\lambda}.
\end{eqnarray}
{\noindent}In the above equation, when $\lambda=\lambda_1=\lambda_2$ and $\gamma=\gamma_1=\gamma_2$, the bifurcation theory was used to analyze the above equations. One interesting result is that the synergistic interactions can greatly enhance the spreading of the behaviors in both layers. They also found that the synergy effects alter the nature of the phase transition of the behavior adoption processes, where a small (large) value of the transmission rate of behavior 1 (with a low adoption threshold) can lead to a discontinuous (continuous) phase transition in behavior 2 (with a high adoption threshold). Moreover, a two-stage spreading phenomenon is observed with the synergy effects, whereby nodes adopting the low-threshold behavior in one layer are more likely to adopt the high-threshold behavior and further stimulate the remaining nodes to quickly adopt the behavior in the other layer.

Chang and Fu~\cite{chang2018co} proposed a co-diffusion model for social contagion on a two-layer network, the layers of which are a periodic lattice and a random regular network. The model considers two aspects of the diffusion process. One is inclusive adoption, which allows individuals to adopt two behaviors (i.e., synergy adoption). At first, each naive individual (without adoption of any behaviors) becomes active with probability
\begin{eqnarray}
\beta(i\leftarrow~1~{\rm or}~2)=\frac{{(\frac{i_1}{K_1})
}^{\alpha}+{(\frac{i_2}{K_2})}^{\alpha}}{1+{
(\frac{i_1}{K_1})}^{\alpha}+{(\frac{i_2}{K_2})}^{\alpha}},
\end{eqnarray}
where $i_1$ and $i_2$ respectively represent the fractions of neighbors who have adopted behavior 1 and behavior 2, $K_1$ and $K_2$ are constants, and $\alpha$ is a free parameter. If it is activated, then with probability
\begin{eqnarray}
Pr(i\leftarrow~1)=\frac{{(\frac{i_1}{K_1})
}^{\alpha}}{{(\frac{i_1}{K_1})}^{\alpha}+{
(\frac{i_2}{K_2})}^{\alpha}},
\end{eqnarray}
it adopts behavior 1; otherwise, it adopts behavior 2. Once an individual adopts one behavior, it adopts another behavior with probability $\frac{{(\frac{i_2}{K_2})}^{\alpha}}{1+{(\frac{i_2}{K_2})}^{\alpha}}$. The other is stochastic dormancy, which means that an adopted individual becomes dormant at any given time step with a given rate. When an individual is considering adoption, if a neighbor is dormant then this neighbor's influence is discounted. Chang and Fu found that lower synergy makes contagions more susceptible to a global cascade, especially for those diffusing on lattices. Faster diffusion of one contagion with dormancy may block the diffusion of the other.

Czaplicka \emph{et al.}~\cite{czaplicka2016competition} explored the competition of the simple (SIS model) and complex adoption processes (Watts threshold model) on interdependent networks. Interdependent networks consist of a simple adoption layer and a complex adoption layer, respectively corresponding to the layers on which the SIS spreading and Watts threshold adoption processes occur. Interconnections between these two layers are added to couple the adoption processes. Czaplicka \emph{et al.} found that the transition points and the nature of transitions for both single dynamical processes are affected by the coupled dynamics. Specifically, the continuous transition can be observed in the complex adoption layer and the discontinuous transition occurs in the simple adoption layer, whereas in previous studies~\cite{pastor2015epidemic}, the SIS and Watts threshold models were found to exhibit continuous and discontinuous transitions, respectively. Srivastava \emph{et al.}~\cite{srivastava2016computing} investigated two competing cascades modeled by a generalized threshold model on a signed network, with trust and distrust edges. They developed a pairwise analytical method to approximately calculate the probability of a node being infected at any given time. Srivastava \emph{et al.} took the advantage of the derived solution to develop a heuristic method for the influence maximization problem, and showed that their proposed method significantly outperforms the state-of-the-art methods, particularly when the network is dominated by distrust relationships.

\begin{figure}
\centering
\includegraphics[width=1\linewidth]{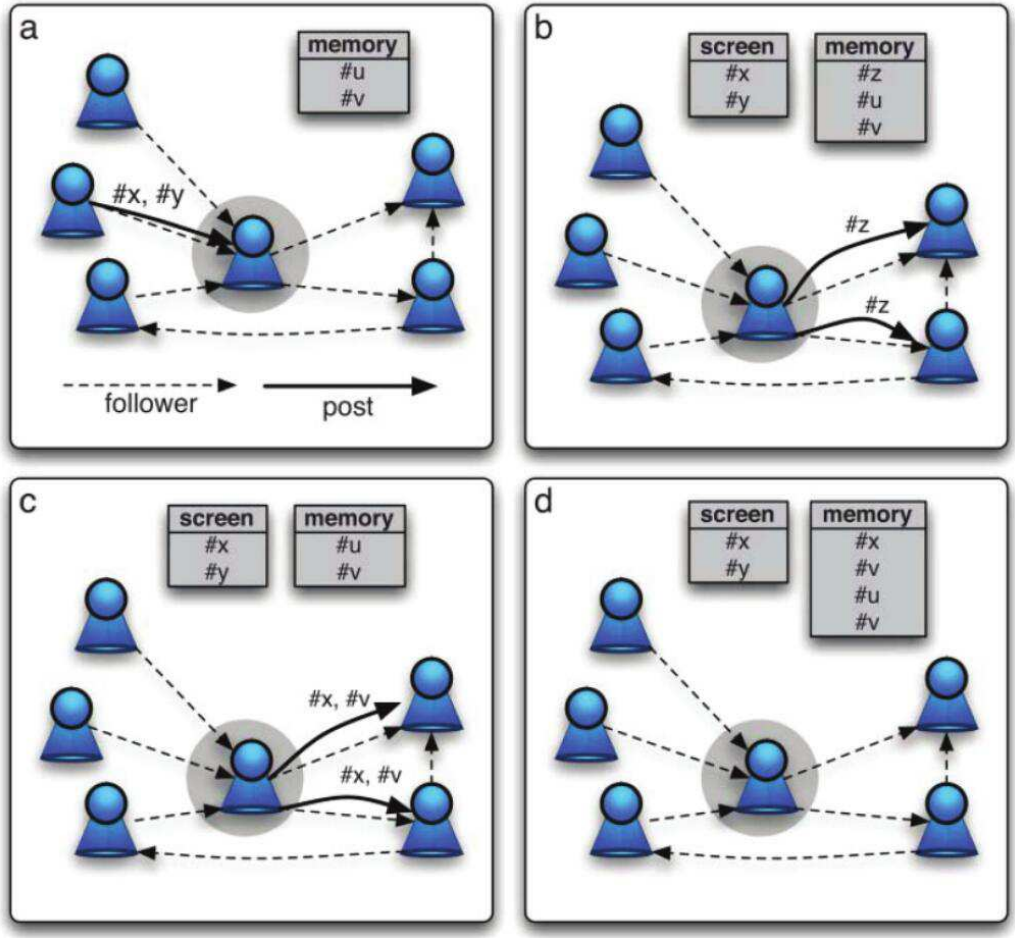}
\caption{Illustration of the meme diffusion model. Each user has a memory and a screen, both with limited sizes. (a) Memes are propagated along follower links. (b) The memes received by a user appear on their screen. With probability $\beta_n$, the user posts a new meme, which is stored in memory. (c) Otherwise, with probability $1-\beta_n$, the user scans the screen.
The user then either retweets a meme chosen from the screen with probability $1-\beta_m$, or tweets a meme chosen from memory with probability $\beta_m$. (d) All memes posted by the user are also stored in memory.
Reproduced from Ref.~\cite{weng2012competition}.}
\label{fig:meme_model}
\end{figure}

\subsection{Coevolution of multiple social contagions}

In the real world, multiple social contagion processes may compete or cooperate with each other and spread simultaneously, such that, with the wide adoption of social media and other socio-technical systems, the abundance of information to which we are exposed is exceeding our limited capacity~\cite{dunbar1998social,gonccalves2011modeling}. The information or ideas in different social media must compete for our limited attention to become popular. To understand the role of the limited attention of individual users in the diffusion of memes, Weng \emph{et al.}~\cite{weng2012competition} developed an agent-based model and investigated how they shape the spread of information (see Fig.~\ref{fig:meme_model} for the illustration of the meme diffusion model). By tuning the length of the time window during which posts are retained in an agent's screen or memory, they can change the extent of competition. For example, a shorter time window leads to less attention and thus enhances competition, whereas a longer time window allows for paying attention to more memes and thus decreases competition. By modeling the meme spreading on real social networks and random networks, they found that the combination of a social network structure and the competition for finite user attention is a sufficient condition for the emergence of broad diversity in meme popularity, lifetime, and user activity. Gleeson \emph{et al.}~\cite{gleeson2014competition} presented an analytically solvable model of selection behavior on a social network. They assumed each screen has capacity for only one meme, corresponding to fierce competition. During each time step (with time increment $\Delta t=1/N$), one node $i$ is chosen at random. Node $i$ then innovates with probability $\mu$ and generates a brand-new meme, which appears on its screen and is retweeted to all $i$'s followers. Otherwise, node $i$ retweets the meme currently on its screen. When a meme is retweeted, its popularity is incremented by 1 and the meme currently on the followers' screens are overwritten by this meme. They found that the competition among multiple memes for the limited user attention places the system at criticality. For a network of degree distribution $P(k) \propto k^{-\gamma_D}$, with $2<\gamma_D<3$, they derived the distribution of popularities at age $a$ by $q_n(a)$ (i.e., a meme has been retweeted $n$ times when its age is $a$) when $a \to \infty$ as
\begin{equation}\label{QNC}
q_n(\infty) \sim \left\{
\begin{array}{lr}
Bn^{-(\gamma/\gamma-1)}, & if~\mu=0, \\
Cn^{-\gamma}, & if~\mu>0 ,
\end{array}
\right.
\end{equation}
{\noindent}Here, $B$ and $C$ are the prefactors (see Ref.~\cite{gleeson2014competition} for details). From Eq.~(\ref{QNC}), they showed that the popularity growth of each meme is described by a critical branching process, and an asymptotic analysis predicts power-law distributions of popularities with very heavy tails in the zero-innovation limit ($\mu=0$). Gleeson \emph{et al.}~\cite{gleeson2016effects} further extended the above simplified model with the inclusion of memory times and heterogenous user activity rates. Even this generalized model is more complicated; they showed that it is analytically solvable and is able to reproduce several important characteristics of empirical microblogging data on hashtag usage, such as the time-dependent heavy-tailed distributions of meme popularity. In modeling the collective behavior of users' decisions on adopting different software applications, Gleeson \emph{et al.}~\cite{gleeson2014simple} proposed a model that incorporates two distinct mechanisms: one is associated with recent user decisions and the other reflects the cumulative popularity of each application. They found various combinations of the two mechanisms that can yield long-time behavior matched with the data. However, only the models that strongly emphasize the recent popularity of applications over their cumulative popularity can reproduce the observed temporal dynamics.

Myers \emph{et al.}~\cite{myers2012clash} studied URL diffusion on Twitter and proposed a statistical model that allows for the cooperation as well as competition of different contagions in the diffusion of retweeting behaviors. Competing contagions decrease each other¡¯s transmission rates, whereas cooperating contagions help each other. By evaluating the model on 18,000 contagions that are simultaneously spreading through the Twitter network, Myers \emph{et al.} estimated the probability of a user being infected, given a sequence of previously observed contagions, and also found that the interactions between simultaneous contagions cause a relative change in the spreading probability of a contagion by {71\%} on average. Valera \emph{et al.}~\cite{valera2015modeling} used continuous-time Hawkes processes to model the adoption of multiple products and conventions. An inference method was developed to estimate the parameters for both synthetic and real data. They found that this data-driven model predicts the adoption and the usage pattern of competing products and social conventions well. Meanwhile, the model reveals that the usage of more popular products and conventions is triggered by cooperation and using a less popular product or convention has a stronger inhibiting effect on future usage of a more popular product or convention than vice versa. Pathak \emph{et al.}~\cite{pathak2010generalized} proposed a generalized version of the linear threshold model to simulate multiple cascades on a network. In the model, nodes are allowed to switch their states back and forth. An algorithm was developed to estimate the most likely statistical properties of the cascades' spread and shown to be of high quality when tested with real data.

\subsection{Summary}

Social contagions are found in every aspect of the real world,
from online to offline behaviors. Experimentally, various types of
social contagions, ranging from health behaviors and new crops to Skype adoption,
exhibit the social reinforcement effect. That is to say, accumulative multiple contacts between
adopted and susceptible individuals are necessary to trigger the infection.
When including the social reinforcement effect in social contagions,
the order parameter, i.e., the cascade size or final fraction of adopted
(activated) individuals, changes
discontinuously or continuously, depending on the strength of the
social reinforcement. In addition, the network structures and the
underlying mechanism incorporated in social contagion models can alter the
cascade size and the type of phase transition.
For two successive and simultaneous social contagions, the inhibiting or
synergistic effects of one spreading behavior can render the discontinuous phase transition
of the other behavior as continuous. For the
coevolution of multiple competitive and multiple cooperative social contagion
processes, the competition induced by limited
attention or memory is a sufficient condition for the emergence of broad diversity in
meme popularity, and cooperative behaviors mutually promote each other's contagion.

\section{Coevolution of awareness diffusion and epidemic spreading} \label{ei}

When an epidemic spreads in a society,
information about the disease spreads through various
types of communication platforms, such as TV news, Facebook,
Twitter, text messages, phone calls, and WeChat. Once
healthy individuals obtain information about the epidemic, they will be aware of the epidemic and thus take certain actions
(e.g., wearing face masks or staying
at home) to protect themselves from being infected by the disease,
which will effectively suppress the outbreak of the infectious disease.
The search to understand how awareness spreading (also called information spreading
in the literature) can mitigate disease
outbreaks, and more broadly, the interplay between the two types of
spreading dynamics, has led to a
novel research domain in
network science~\cite{bauch2013social}. When investigating
the coevolution of awareness diffusion and epidemic spreading dynamics,
scientists assume that susceptible individuals will take certain nonpharmaceutical interventions (NPIs, e.g.,
hand-washing and social distancing) once they are aware of that they are
in dangerous circumstances, and thus
the susceptible nodes will have a lower probability of
being infected by their neighbors~\cite{zhang2016impact,
zhang2014suppression,liupnas}. Usually, the
disease outbreak threshold increases and the infected
size decreases when such interventions are implemented.
Vaccination is another effective measure to suppress the spreading
of infectious diseases~\cite{bauch2005imitation,
bauch2004vaccination}. In most cases,
the effects of vaccination
and NPIs on the epidemic spreading
have been investigated separately~\cite{zhang2013braess}.
Andrews and Bauch \cite{andrews2015disease} proposed a model
that includes both measures, and revealed that the practice of
NPIs decreases as vaccine coverage increases, and vice versa.
In the following, we discuss the recent progress
on the coevolution of awareness and epidemics
on single and multiplex networks.

\subsection{Empirical analyses}

Google is one the most popular
channels for persons to obtain
information about epidemics, and thus the volume
of related queries or the Google Flu Trends (GFT) should be highly correlated to
the number of infected persons~\cite{ginsberg2009detecting,
carneiro2009google,preis2014adaptive}. Therefore,
Ginsberg \emph{et al.}~\cite{ginsberg2009detecting} found a new way to improve the
early detection of influenza-like illness (ILI) by
using the data collected by GFT. They can accurately
estimate the current level of weekly influenza
activity in each region of the United States, with a
reporting lag of approximately one day. However, the
results reported by Ginsberg \emph{et al.}~\cite{ginsberg2009detecting}
were challenged until 2013,
as GFT predicted more than double
the fraction of doctor visits for ILI~\cite{butler2013google,olson}.
Lazer \emph{et al.} \cite{lazer2014} concluded
that two issues, big data hubris and the instability
of the algorithm, caused GFT's errors. On the
one hand, GFT developers report that the data are highly
correlated with the disease control and prevention
(CDC), but not related to the flu. As a result, the data quality is not very high, because certain detailed information
about the epidemic issues are ignored. On the other hand,
the Google search algorithm is time-varying, which
induces an unstable reflection of the prevalence
of the flu.

\begin{figure}
\begin{center}
\epsfig{file=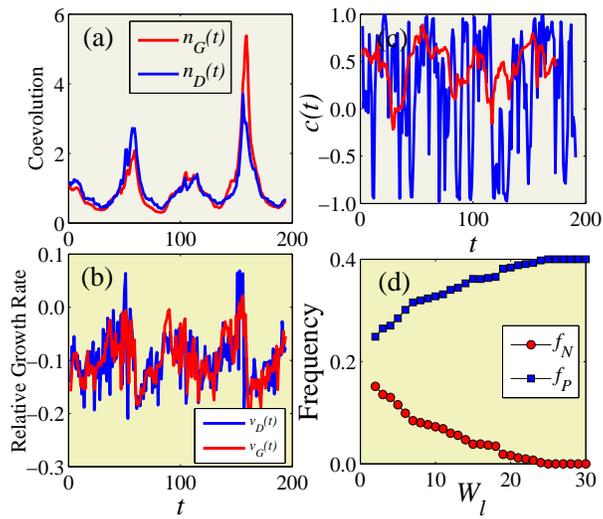,width=0.6\linewidth}
\caption{(Color online) Weekly outpatient visits
and Google Flu Trends (GFT) of influenza-like illness
(ILI) in the United States.
(a) Fractions of outpatient visits $n_D(t)$ (blue
line) and GFT $n_G(t)$ (red line) versus $t$. (b) Relative growth rate $v_D(t)$ (blue line) and
$v_G(t)$ (red line) of $n_D(t)$ and $n_G(t)$ versus $t$,
respectively. (c) Cross-correlation $c(t)$ between the
two time series of $v_G(t)$ and $v_D(t)$ for the given
window size $W_l=3$ (blue line) and $W_l=20$ (red line).
(d) Fraction of negative correlations
$f_P$ (blue squares) and positive correlations $f_N$
(red circles) as a function of $W_l$. In (a),
$n_G(t)$ and $n_D(t)$ are divided by their respective average values. Reproduced from
Ref.~\cite{wang2016suppressing}.}
\label{fig1_ie}
\end{center}
\end{figure}

To study the coevolution mechanism in real
data about information and epidemics,
Wang \emph{et al.}~\cite{wang2016suppressing}
investigated the weekly series of GFT and
ILI epidemics, and their synchronous evolution from 3 January 2010 to
10 December 2013 (nearly 200 weeks) in the United States~\cite{preis2014adaptive}.
Fig.~\ref{fig1_ie} shows the time series of
$n_G(t)$ (i.e., the number of ILIs) and
$n_D(t)$ (i.e., the volume of search queries).
From the macroscopic perspective,
ILI and GFT have similar trends, as shown in
Fig.~\ref{fig1_ie}(a). The time series
were further analyzed from a microscopic view. The relative growth rates
of $n_G(t)$ and $n_D(t)$ are defined as $v_G(t)$ and $v_D(t)$,
respectively. 
The growth trends
of $v_G(t)$ and $v_D(t)$ are the same, as shown in
 Fig.~\ref{fig1_ie}(b).
Furthermore, the cross-correlation $c(t)$ between the
$v_G(t)$ and $v_D(t)$ time series for a given window size $W_l$, as defined
in Ref.~\cite{podobnik2008detrended}, were investigated.
As shown in Fig.~\ref{fig1_ie}(c), positive and negative cross-correlations $c(t)$ are unveiled
for $W_l=3$ and $W_l=20$, respectively. Once the susceptible
individuals received information through GFT, they adopted
certain actions to protect themselves from being infected, and thus
the opposite growth was observed. The infected individuals
tend to search for information about the disease, which
leads to the same growth trends. Furthermore, Wang \emph{et al.}
found that the asymmetric interaction (i.e., the epidemic
spreading promotes information diffusion but information
diffusion suppresses the epidemic spreading) between information
diffusion and epidemic spreading only occurred for a short period,
as shown in Fig.~\ref{fig1_ie}(d).

Zhan \emph{et al.}~\cite{zhan2018coupling} analyzed the coevolution of information and two representative
diseases (i.e., H7N9 and Dengue fever). They collected
daily data about the two diseases from the Chinese Center for Disease Control
and Prevention, and crawled for information about the diseases from
Sina Weibo. Individuals who obtained information about the diseases were aware of the diseases, and may take some actions
to protect themselves from being infected. Through empirical analyses and
mathematical modeling, Zhan \emph{et al.} claimed that
the awareness and epidemic asymmetrically affect
each other, i.e., the information diffusion suppresses the
epidemic spreading, whereas the epidemic spreading promotes
the information diffusion. Similar results
are also reported in Refs.~\cite{wang2016suppressing,smith2016towards,gj}.

\subsection{Coevolution of awareness and epidemic on single networks}

\begin{figure}
\begin{center}
\epsfig{file=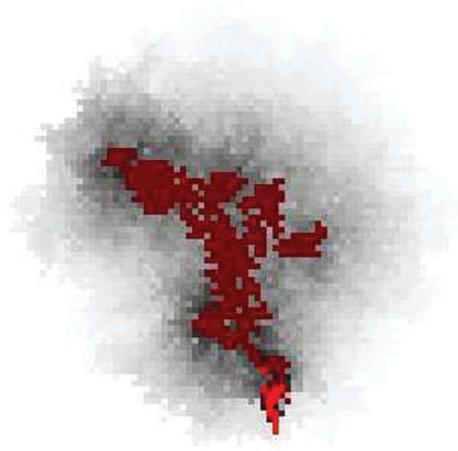,width=0.5\linewidth}
\caption{(Color online) Snapshot of awareness and epidemic
spreading on triangular lattice. Red represents nodes in
the recovered state, light red stands for nodes
in the infected state, and darker gray represents nodes with a higher
level of awareness. Reproduced from Ref.~\cite{funk2009spread}.}
\label{fig2_ie}
\end{center}
\end{figure}

Funk and his colleagues
investigated a variant SIR model that considers the coevolution of awareness and
diseases on well-mixed populations and lattices~\cite{funk2009spread}.
In their proposed model, information about
the disease is called awareness. An individual $i$
with awareness level $\ell_i$ represents that the awareness has
been through $\ell_i$ individuals before arriving at $i$, and
$\ell_i=0$ means that individual $i$ has first-hand awareness. The awareness level of individual $i$
is updated according to the following three rules. (i)
A new awareness $\ell_i=0$ is generated
with probability $\omega$ if individual $i$ is in the infected state.
(ii) If individual $j$ transmits a newer piece of awareness with
level $\ell_j$ to his/her neighbor, individual $i$, with rate $\alpha$, $\ell_i$ is updated as $\ell_i\leftarrow\ell_j+1$ when $\ell_j
<\ell_i$. (iii) At each time step, the awareness level
$\ell_i$ of node $i$ fades to $\ell_i+1$ with
probability $\epsilon$. For a susceptible
node with awareness level $\ell$, its susceptibility is $1-h^\ell$,
where $0<h<1$. That is, the older the
awareness is (i.e., the higher its level), the
weaker the protection power is.
In consideration of awareness, the
SIR epidemic evolves as follows. An infected node transmits the
infection to a susceptible neighbor of awareness level $\ell$
with probability $(1-h^\ell)\beta$, and recovers with
probability $\gamma$. Using a mean-field approach, the authors revealed that the outbreak threshold does not
change in a well-mixed population, although the epidemic size
is much lower than one without awareness. However,
the epidemic outbreak threshold increases on highly
structured networks, such as lattices.
As shown in Fig.~\ref{fig2_ie}, in the
snapshot, the infected nodes are surrounded
by a cloud of awareness, which greatly suppresses the
epidemic spread.

\begin{figure}
\begin{center}
\epsfig{file=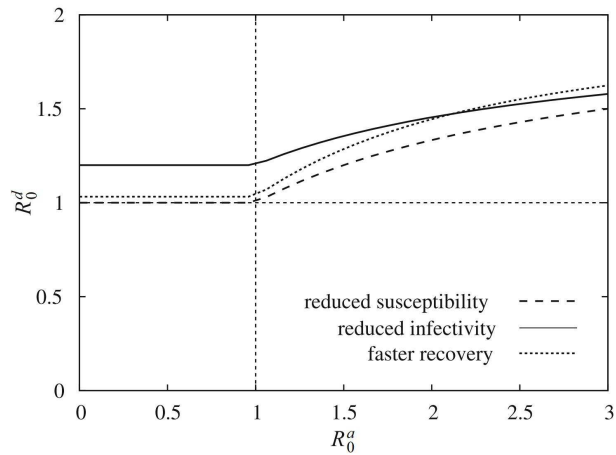,width=0.6\linewidth}
\caption{(Color online) Outbreak threshold under reduced susceptibility
($\sigma_S=0.5$), reduced infectivity ($\sigma_I=0.5$), and faster recovery ($\varepsilon
=2$). Reproduced from Ref.~\cite{funk2010endemic}.}
\label{fig4_ie}
\end{center}
\end{figure}

Funk \emph{et al.}~\cite{funk2010endemic}
further considered a more complicated case
in which the epidemic spreading is characterized
by a classical epidemic
SIRS model~\cite{anderson1992infectious} and the awareness spreading is described by
a generalized SIS model. In the former, the
infection, recovery, and loss of immunity
probabilities are $\beta$, $\gamma$, and $\delta$; in the latter, an infected node
becomes aware with probability $\omega$,
an aware node transmits the
awareness to unaware neighbors with probability $\alpha$,
and an aware node loses awareness with probability $\lambda$.
The dynamical parameters are dependent on
whether nodes are aware of the epidemic
information. (i) Reduced susceptibility and infectivity.
The awareness suppresses the infectious
transmission, as shown in Table \ref{tab1}.
(ii) Faster recovery. The recovery probability of
an aware infected node is $\varepsilon\gamma$. (iii) Longer preservation
of immunity. For an unaware recovered node, the probability of loss of immunity
is $\delta$, and for an aware recovered node, the probability of loss of
immunity is $\phi\delta$. With the above three modifications, the basic reproductive
number $R_0^d$ of the epidemic is different
for distinct situations. If the
susceptibility of a susceptible node is reduced, $R_0^d$ is only
dependent on the basic reproductive number $R_0^a$ of the awareness.
\textcolor[rgb]{1.00,0.00,0.00}{Funk \emph{et al.}~\cite{funk2010endemic} computed
the corresponding reproduction number for different situations.
By reducing infectivity of infected nodes and fixing other parameters,
i.e., $\sigma_S=\varepsilon=\phi=1$, the global epidemic
outbreak conditions are
$R_0^d>1+\frac{(1-\sigma_I)\omega}{\lambda+\gamma+\sigma_I
\omega}$
when $R_0^a<1$, and
$R_0^d>1+\frac{(1-\sigma_I)[R_0^a(1+\omega/
(\alpha+\gamma))-1]}{1+\sigma_I[R_0^a(1+\omega/(\alpha+\gamma))-1]}$
when $R_0^a>1$. When the faster recovery strategy is implemented,
the global epidemic  outbreak conditions are
$R_0^d>1+\frac{(\varepsilon-1)\omega}
{\lambda+\varepsilon\gamma+\omega}$
if $R_0^a<1$, and
$R_0^d>1+
\frac{R_0^a-1+\omega/(\alpha+\gamma+\omega)}{R_0^a+(\varepsilon-1)
\gamma/(\alpha+\gamma+\omega)}(\varepsilon-1)$
if $R_0^a>1$.}
The epidemic outbreak thresholds
for these situations are shown in Fig.~\ref{fig4_ie}. To sum up,
the epidemic threshold increases if the
infection and recovery rates are changed,
owing to the diffusion of awareness.
Agaba \emph{et al.}~\cite{agaba2017mathematical} proposed
a model in which the awareness is induced by two aspects:
direct contacts between aware and unaware
nodes and public information, and they
revealed distinct dynamical regimes.

\begin{table}
\scriptsize
\caption{Transmission probabilities from an
infected node to a susceptible node with different
states of awareness, where the two parameters are
$0<\sigma_s<1$ and $0<\sigma_I<1$.
} \label{tab1}
  \centering
  \begin{tabular}{ c|   c c| c }
    \hline
    \multirow{2}{*}{\textbf{Infected node}} &   \multicolumn{3}{ c }{\textbf{Susceptible node}}\\
    \cline{2-4}
    &  & Unaware & Aware  \\    \hline
     Unaware &   & $\beta$ & $\sigma_s\beta$  \\\hline
     Aware &   & $\sigma_I$ & $\sigma_s\sigma_I\beta$  \\
    \hline
  \end{tabular}
\end{table}

\begin{figure}
\begin{center}
\epsfig{file=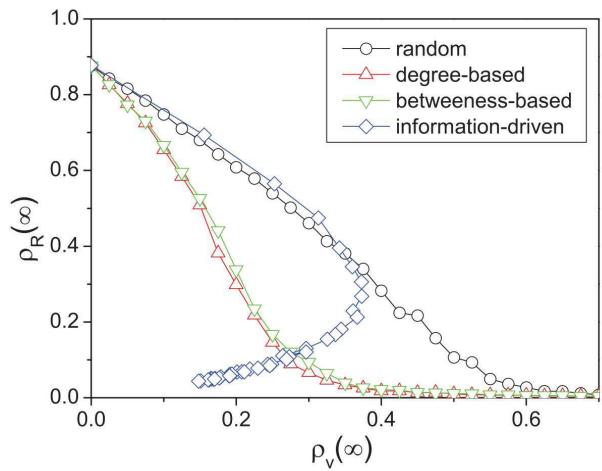,width=0.6\linewidth}
\caption{(Color online) Fraction of nodes $\rho_R(\infty)$ in the recovered
state versus the fraction of nodes $\rho_V(\infty)$ in the vaccinated state
under different immunization approaches. The network size is set as $N=2000$,
the average degree is $\langle k\rangle=6$, the infection probability is $\beta=0.06$,
and the recovery probability is $\mu=0.1$. Reproduced from Ref.~\cite{ruan2012epidemic}.}
\label{fig3_ie}
\end{center}
\end{figure}

Ruan \emph{et al.}~\cite{ruan2012epidemic} assumed
that the awareness can be generated by the
information transmission, which is described by
the delivery of information packets. A
susceptible-infected-recovered-vaccination
(SIRV) model was proposed to describe the epidemic spreading dynamics. Mathematically, the SIR component
of the spreading dynamics is identical to the classical SIR model, with
infection and recovery probabilities of $\beta$ and $\gamma$, respectively. A
susceptible node becomes vaccinated with probability $\kappa(t)$.
Nodes in the recovery and vaccinated states do not
participate in the spreading dynamics.
At each time
step, there are $\tau N$ newly generated packets with
randomly chosen origins and destinations,
following a shortest-path routing
protocol~\cite{yan2006efficient}. At time $t$, if
a node receives $m_I(t)$ packets from infected
neighbors, it becomes vaccinated with probability $\kappa
(t)=1-e^{-\eta\frac{m_I(t)}{k}}$, where $\eta$ is the strength of
the sensitivity to information, and $k$ is the degree of the node. Through
extensive simulations, Ruan \emph{et al.}~\cite{ruan2012epidemic} found that
the final epidemic size $\rho_R(\infty)$
monotonously changes with vaccination $\rho_V(\infty)$,
as shown in Fig.~\ref{fig3_ie}. Specifically,
the epidemic may outbreak globally or
locally for low vaccination prevalence, which is
markedly different from other classical immunization approaches,
e.g., random, degree-based, and betweenness-based
approaches.

Mass media, e.g., TV news broadcasts, is an important measure to distribute
information about an epidemic, and has observable effects on controlling
infectious epidemics~\cite{tchuenche2011impact,
rahman2007media}. Some studies considered the
coevolution of awareness and epidemics on well-mixed
populations~\cite{samanta2013effect},
in which the awareness is generated by mass media. Some interesting
phenomena are observed when awareness spreading is induced by mass media. For example,
sustained oscillations may occur if the awareness
is larger than a threshold~\cite{samanta2013effect}.
Wang \emph{et al.}~\cite{wang2013impact} investigated
the effects of mass media on the spread of epidemics on
complex networks, where the epidemic spreading is
described by the SIS model. The mass media density is a linear function of
the current epidemic prevalence. 
By using the approach developed
in Ref.~\cite{van2002reproduction}, 
Wang \emph{et al.} found that the basic reproductive number is
highly correlated with the volume of mass media.
Their results indicate that mass media is an effective tool in
controlling global epidemic outbreaks.

Another line about the awareness--epidemic spreading dynamics assumed
that the awareness is correlated with neighbors, such as the
number of infected neighbors~\cite{zhang2016impact,
zhang2014suppression}.
The nodes that are aware of the epidemic decrease their
infectivity probability~\cite{zhang2016impact} or activity~\cite{rizzo2014effect}.
In this case, the evolution of awareness is directly determined by the evolution of
the epidemic. Thus, we only need to write down the evolution equations
of the epidemic. Zhang \emph{et al.}~\cite{zhang2014suppression} proposed an awareness--epidemic model,
in which the awareness of a susceptible node is determined by the number
of its infected neighbors. In this model,
a susceptible node with awareness
has a smaller infectivity probability, i.e., the infection probability is
$\lambda(1-\alpha_r)^{n_I}$, where $\lambda$ is the basic infection probability,
$\alpha_r$ is the reduction factor, and $n_I$ is the number of infected
neighbors. Zhang \emph{et al.} used the heterogeneous mean-field theory
to describe the spreading dynamics, and found that the outbreak thresholds
of SIS and SIR models are the same, as follows:
\begin{equation}\label{sis_behavior}
\lambda_c =\frac{1}{1-\alpha_r}\frac{\langle k\rangle}{\langle k^2\rangle
-\langle k\rangle}.
\end{equation}
From Eq.~(\ref{sis_behavior}), one can observe
that the stronger awareness is, the larger the
outbreak threshold is.

To describe the belief and vaccination decision of individuals,
Xia and Liu proposed a novel model~\cite{xia2014belief}, which
assumes that an individual's belief about the severity and vaccine safety of
an epidemic is updated according to the awareness received from
their neighbors. Two key factors determine the adoption of vaccination.
The first factor is the fraction of nodes infected by the epidemic
and vaccinated, and the second factor is the fading of
awareness~\cite{funk2009spread}. The authors found that the
first factor affects the fraction of vaccinated nodes, and
the second factor influences the time when the nodes become vaccinated.

Time delays are pervasive real-world networked
dynamics and can radically alter the evolution
of dynamic processes in networks~\cite{wang2018effects,olfati2004consensus}.
Greenhalgh \emph{et al.} introduced time delays into awareness--epidemic coevolution
spreading dynamics~\cite{greenhalgh2015awareness}. On the one hand,
the awareness of individuals disappears after a certain number of
time steps $\tau_1$ have elapsed. On the other hand, the strength of
awareness is dependent on the number of infected individuals at time $t-\tau_2$,
because the policy-maker makes decisions based on previous reports about the
epidemic. Once time delays are included, the system exhibits limit-cycle
oscillation. Agaba \emph{et al.}~\cite{agaba2017dynamics}
assumed that the time delays are induced when
nodes became aware and modified their behavior.
They analytically studied the stability of
disease-free and endemic equilibria, and revealed that
the Hopf bifurcation is correlated to the dynamical
parameters and the time delay.

\subsection{Coevolution of awareness and epidemics on multiplex networks}

The previous subsection dealt with the case in which scientists studied the
the spreading of awareness and an epidemic
on the same network. However, a typical case is that in which the
epidemic spreads on contact networks, such
as sexual webs, and
the awareness spreads on various
types of social communication platforms,
such as Twitter and Facebook. Thus, using multiplex
networks to describe the coevolution spreading of
awareness and an epidemic is more
realistic. In this subsection, we will
introduce the progress of the coevolution of awareness
and epidemics on multiplex networks.

Granell \emph{et al.}~\cite{granell2013dynamical}
proposed an
unaware-aware-unaware+susceptible-infected-susceptible
(UAU+SIS) model on multiplex
networks to describe the coevolution of awareness
diffusion and epidemic spreading, where
nodes in the multiplex networks are
matched one-to-one randomly.
The awareness spreads through
virtual contact following an UAU model, in which each
node can be in an unaware or aware state.
The evolution of awareness follows three rules:
(i) Each unaware node is infected by
aware neighbors on the virtual network
with probability $\lambda_1$. (ii) Each
unaware node becomes aware if its
counterpart in the physical network is
infected by the epidemic. (iii) Each aware
node forgets the epidemic information
or relaxes its vigilance and becomes unaware
with probability $\gamma_1$. The epidemic
spreading through physical contacts is
described by an SIS model, following
two rules. (i) Each susceptible node
is infected by infected neighbors with probability
$\lambda_2=\lambda_2^U$ if its counterpart
in the virtual network is unaware,
and otherwise it is infected with
probability $\lambda_2^\prime=\delta\lambda_2^U$, where $0\leq\delta\leq1$. (ii) Each
infected node recovers with probability $\gamma_2$.

\begin{figure}
\begin{center}
\epsfig{file=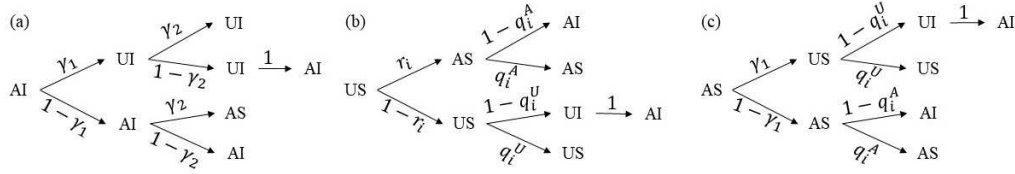,width=1\linewidth}
\caption{(Color online) Transition probabilities for the states
(a) AI, (b) US, and (c) AS at each time step.
Reproduced from Ref.~\cite{granell2013dynamical}.}
\label{add_ie}
\end{center}
\end{figure}

\begin{figure}
\begin{center}
\epsfig{file=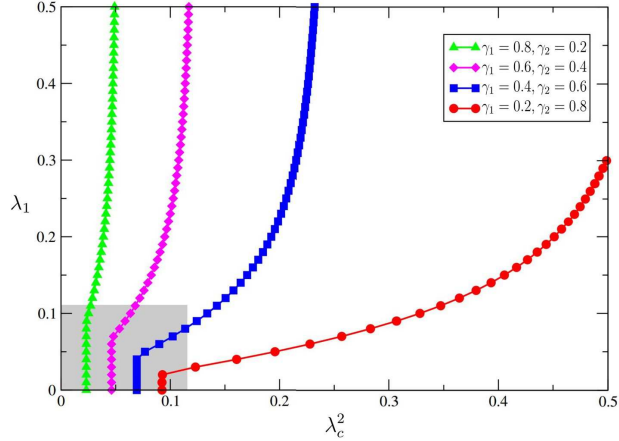,width=0.6\linewidth}
\caption{(Color online) Epidemic threshold $\lambda_c^2$ versus awareness
transmission probability $\lambda_1$ under different values of $\gamma_1$ and
$\gamma_2$. The shaded rectangle, which represents the metacritical points,
is bounded by $1/\Lambda_{\rm max}(A)$ and $1/\Lambda_{\rm max}(B)$.
Reproduced from Ref.~\cite{granell2013dynamical}.}
\label{fig5_ie}
\end{center}
\end{figure}

To analyze the model described above, Granell \emph{et al.}~\cite{granell2013dynamical}
applied a generalized microscopic Markov chain approach (MMCA)
\cite{gomez2010discrete}. $p_i^{AI}(t)$,
$p_i^{AS}(t)$, and $p_i^{US}(t)$ respectively denote the probability that node $i$ is in the
aware--infected, aware--susceptible, unaware--susceptible state at time $t$.
For node $i$, it does not get the awareness from neighbors at time $t$ with probability
$r_i(t)=\prod_j[1-A_{ji}\lambda_1 p_j^A(t)]$, where $A$ represents the adjacent matrix of the
virtual network, and $p_j^A=p_j^{AI}+p_j^{AS}$ is the probability that node $j$ is
in the aware state. If node $i$ is in the aware state,
it is not infected by the epidemic at time $t$ with probability $q_i^A(t)=\prod_j[1-
B_{ji}\lambda_2^\prime p_j^{AI}(t)]$, where $B$ is the adjacent
matrix of the physical network. Otherwise, if node $i$ is in the unaware state
in the virtual network, it is not infected by the epidemic with
probability $q_i^U=\prod_j[1-B_{ji}\lambda_2^U p_j^{AI}(t)]$. With the scheme
presented in Fig.~\ref{add_ie}, the generalized
MMCA equations of the coevolution dynamics are
\begin{equation} \label{mmca}
\begin{cases}
p_i^{US}(t+1)=p_i^{AI}(t)\gamma_1\gamma_2+p_i^{US}(t)r_i(t) q_i^U(t)
+p_i^{AS}\gamma_1 q_i^U(t),\\
p_i^{AS}(t+1)= p_i^{AI}(t)(1-\gamma_1)\gamma_2+p_i^{US}[1-r_i(t)]
q_i^A(t)+p_i^{AS}(t)(1-\gamma_1)q_i^A(t),\\
\begin{split}
p_i^{AI}(t+1)&=p_i^{AI}(t)(1-\gamma_2)+p_i^{US}(t)\{[1-r_i(t)][1-
q_i^A(t)]+r_i(t)[1-q_i^U(t)]\}\\
&+p_i^{AS}(t)\{\gamma_1[1-q_i^U(t)]
+(1-\gamma_1)[1-q_i^A(t)]\}.
\end{split}
\end{cases}
\end{equation}
In the steady state, linearizing Eq.~(\ref{mmca}), the epidemic
outbreak threshold is
\begin{equation}\label{SIS_AUA}
\lambda_c^2=\frac{\gamma_2}{\Lambda_{\rm max}(H)},
\end{equation}
where $\Lambda_{\rm max}(H)$ is the largest eigenvalue of
matrix $H$, whose elements are $H_{ji}=[1-(1-\delta)p_i^A]
B_{ji}$. From Eq.~(\ref{SIS_AUA}),
the epidemic threshold is dependent on the
topology of the physical network and the awareness outbreak
size. If the awareness cannot achieve an outbreak, the epidemic
threshold reduces to $\lambda_c^2=\gamma_2/\Lambda_{\rm max}
(B)$, which is the same as the classical epidemic
outbreak threshold~\cite{gomez2010discrete}. Granell \emph{et al.}
defined $(\lambda_c^1,\lambda_c^2)$ as
the metacritical point of epidemic spreading,
and the average accuracy of the approximation in the MMCA approach is approximately
98\%. The phase diagram of the coevolution dynamics is
shown in Fig.~\ref{fig5_ie} for different values of
$\gamma_1$ and $\gamma_2$. The metacritical point is bounded
by $[0,1/\Lambda_{\rm max}(A)]\times[0,1/\Lambda_
{\rm max}(B)]$, where $\Lambda_{\rm max}(A)$ is the largest
eigenvalue of matrix $A$.

\textcolor[rgb]{1.00,0.00,0.00}{Following the work of Granell \emph{et al.}~\cite{granell2013dynamical},
many interesting studies have been reported~\cite{granell2014competing,kan2017effects,
starnini2017effects}.}
The model proposed by Granell \emph{et al.}~\cite{granell2013dynamical} assumed
that (1) the infected nodes immediately become
aware, and (2) the aware nodes have a lower
susceptibility with a certain probability. What
happens when these two assumptions are removed? To
this end, Granell \emph{et al.} \cite{granell2014competing}
proposed a novel model without the two assumptions
but including a new aspect, i.e., that the mass media can
broadcast awareness.
A generalized MMCA approach is developed to describe
the proposed model, and it agrees markedly well with
numerical simulations. Granell \emph{et al.} revealed that the first assumption
has almost no effect on the epidemic spreading, whereas the second assumption
and mass media indeed alter the phase diagram. For instance, the metacritical
point disappears if the mass media effect is strong enough. Kan and Zhang
included this mechanism in the coevolution of awareness and epidemics,
and revealed that self-awareness generation (i.e.,
an unaware individual becoming aware spontaneously)
cannot alter the epidemic
outbreak threshold~\cite{kan2017effects}. Starnini
\emph{et al.}~\cite{starnini2017effects} found that
the temporal correlations
slow down the epidemic spreading, and slow down
(speed up) the awareness spreading for small (large)
values of epidemic transmission probability.

In reality, the awareness spreading is different from
the epidemic spreading. When an epidemic outbreaks,
the information about
the epidemic can spread on various channels such as Twitter,
Facebook, and text messages through mobile phones. However, the susceptible nodes may not become aware
immediately when they only receive one piece of information from their neighbors~\cite{centola2018behavior}.
Thus, the awareness spreading exhibits a social reinforcement or herd-like
feature~\cite{borge2013cascading}. Guo \emph{et al.} proposed a local awareness-controlled
contagion spreading (LACS) model to describe
the coevolution of awareness and
epidemics on multiplex networks~\cite{guo2015two}.
The Watts threshold model~\cite{watts2002simple} is used to describe
the awareness spreading, in which an unaware node becomes
aware if (1) the fraction of its aware neighbors exceeds a given threshold $\phi$,
or (2) it is infected by the epidemic. The aware node reverts to unaware
when it is susceptible or relaxes its vigilance.
For the epidemic spreading
dynamics, they adopted the model proposed in Ref.~\cite{granell2013dynamical}
with complete immunization, i.e., $\lambda_2^\prime=0$. Using a generalized MMCA
approach, Guo \emph{et al.} obtained a highly
accurate prediction of the epidemic threshold. Through
extensive numerical simulations and theoretical analyses,
the epidemic threshold $\lambda_c^2$ versus $\phi$ exhibits an abrupt
decrease at $\phi_c\approx0.5$. For the case of $\phi<\phi_c$,
the epidemic size and outbreak threshold remain nearly unchanged with $\phi$. When
$\phi>\phi_c$, the epidemic threshold does not vary with $\phi$, but the epidemic
size increases with $\phi$. Huang \emph{et al.}~\cite{huang2018hybrid} investigated the
coupled contagion of awareness and
an epidemic, where the former is depicted by the
unaware-aware-unaware (UAU) model on
the aware layer and the latter is represented by
the threshold model (susceptible-infected)
on the contact layer. The interacting mechanisms
are introduced such that an unaware individual immediately
changes to the aware state if it is infected
in the contact layer. A susceptible individual can be infected
by each of its infected neighbors with a certain probability if it is in the unaware state.
However, an aware and susceptible individual can only be infected by the epidemic when the
number of its infected neighbors is equal to or exceeds a threshold.
Huang \emph{et al.} found that such heterogenous interactions can induce hybrid phase transitions in which a continuous phase transition and bi-stable states coexist.

Node heterogeneity is widely observed in
many real-world systems, including different
degrees~\cite{albert2002statistical}, adoption
thresholds~\cite{wang2016dynamics,wang2018social}, and
waiting
times~\cite{holme2012temporal}.
Different nodes usually exhibit distinct
responses when they know information about the epidemic; therefore,
the awareness spreading dynamics should be described
by using a model with heterogeneous
thresholds. Pan and Yan~\cite{pan2018impact} proposed a coevolution model of awareness and epidemics,
in which the awareness of nodes comes from three types of information:
contact information as well as local and global prevalence of the epidemic.
The epidemic threshold increases with the
contact-based information, but does not change with the information
about the local and global prevalence of the epidemic. Furthermore,
the effects of the heterogeneity of individuals' responses,
and the structures of the virtual and contact networks are investigated, and
the existence
of two-stage effects on the epidemic threshold is demonstrated~\cite{pan2018impact}.
Zang considered that the awareness transmission probability equals the
fraction of nodes in the aware state, and found that the
epidemic spreading is greatly suppressed~\cite{zang2018effects}.

Because the transmission
probability of epidemic may be time-varying~\cite{stone2007seasonal,
liu2016hysteresis}, Sagar \emph{et al.}
\cite{sagar2018effect} proposed a generalized UAU+SIS
model, in which the transmission
probabilities of the awareness and epidemic vary with time.
The system can be in one of the sustained
oscillatory and damped dynamics. For the case of damped
dynamics (i.e., endemic state), the epidemic is greatly
suppressed if the awareness spreading is included.
Their results were further verified by using a generalized
MMCA approach. In addition, the evolution time scale is fatal for the coevolution
spreading dynamics~\cite{karrer2011competing}. Wang \emph{et al.}~\cite{wang2017epidemic}
investigated the effects of the time scale on the coevolution
of awareness and epidemics on multiplex networks.
Compared with the model in Ref.~\cite{granell2013dynamical},
Wang \emph{et al.} introduced a new parameter to adjust the
relative speed of awareness and epidemics, and assumed the infected
nodes in the contact network may not become aware instantly in the
virtual network. An individual-based mean-field
approximation method and MMCA approach were applied to
analyze the coevolution dynamics, and they found distinct accuracies
for different time scales, suggesting the existence of
an optimal time scale, at which the epidemic would be greatly suppressed.

Most previous studies about the coevolution of awareness
and epidemics on multiplex networks assumed that the time scale
of the network evolution is much longer than the spreading dynamics. Thus,
the network topology can be treated as static networks. However,
many experimental studies indicated that the network topology
varies during the spreading process~\cite{holme2012temporal}.
Guo \emph{et al.} \cite{guo2016epidemic} assumed that the virtual network varies faster than
the awareness spreading, and the contact network varies slower
than the epidemic spreading. Therefore, the activity-driven network
\cite{perra2012activity} is used to describe the virtual network, and
the contact network is still assumed to be static.
Then, the awareness spreads on the
temporal network according to the UAU model as proposed in
Ref.~\cite{granell2013dynamical}. The epidemic spreading dynamics on a contact network
evolves as the model in Ref.~\cite{perra2012activity}. Through
a generalized MMCA approach, Guo \emph{et al.} found that the metacritical
point is bounded by $[0,1/(m(\langle a\rangle+
\langle a^2\rangle))\times[0,1/\Lambda_{\rm max}(B)]$, where
$\langle a\rangle$ and $\langle a^2\rangle$ are the first and
second moments of the activity potential distribution. If the awareness spreading dynamics
follows a threshold model, the epidemic threshold
versus $\phi$ exhibits a sharp decrease at
$\phi\approx0.5$, which is similar to the results in
Ref.~\cite{guo2015two}. In addition, the time-variation of the
virtual network promotes the epidemic spreading on the contact
network.

\begin{figure}
\begin{center}
\epsfig{file=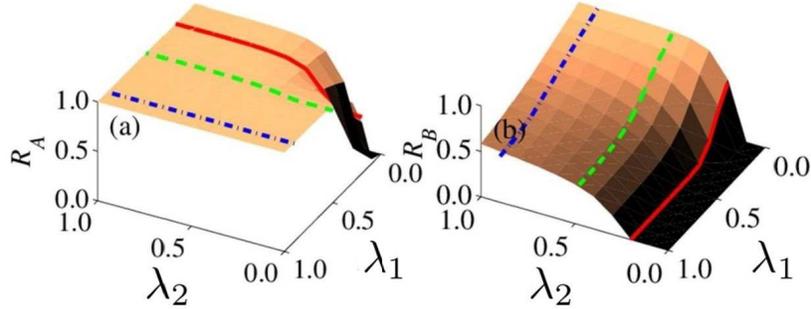,width=0.8\linewidth}
\caption{Coevolution of awareness and epidemics on SF-ER networks.
(a) The final awareness outbreak size $R_A$ and (b) the
final epidemic outbreak size $R_B$ versus $\lambda_1$ and
$\lambda_2$. The other parameter
is set as $p=0.5$.
The lines are theoretical predictions with $\lambda_1=0.2,0.5$, and
0.9 in (a) and with $\lambda_2=0.2,0.5$, and 0.9 in (b).
Reproduced from Ref.~\cite{wang2014asymmetrically}.}
\label{fig7_ie}
\end{center}
\end{figure}

Different from the reversible SIS model, certain epidemics
such as measles and chickenpox should be described by using the
irreversible SIR model~\cite{keeling2008modeling}.
Wang \emph{et al.}~\cite{wang2014asymmetrically} assumed that
the awareness spreads
on a communication network (layer $\mathcal{A}$), and the irreversible epidemic transmits
through a contact network (layer $\mathcal{B}$). 
For the awareness spreading dynamics, they used an SIR-like
model. Specifically, each aware (or infected) node
transmits the awareness to unaware (i.e., susceptible) neighbors
with probability $\lambda_1$, and recovers with probability $\gamma_1=1$.
The unaware node becomes aware about the epidemic
when its counterpart in the contact network is infected by the epidemic.
For the epidemic spreading dynamics, an SIRV model is adopted. The infected nodes transmit the
infection to susceptible neighbors with probability $\lambda_2$,
and recover with probability $\gamma_2=1$. A susceptible node
$i_B$ in layer $\mathcal{B}$ becomes vaccinated if its counterpart in layer $\mathcal{A}$ becomes aware. Wang \emph{et al.} developed a generalized heterogeneous mean-field theory to
quantitatively describe the coevolution spreading dynamics.
To obtain the outbreak threshold $\lambda_c^1$ of the awareness spreading, Wang \emph{et al.} used a linear approximation method, and got
\begin{equation} \label{Athreshold}
\lambda_c^1 = \left\{\begin{array}{l}\langle k_1\rangle/({\langle k_1^2\rangle
-\langle k_1\rangle}),
\quad {\rm for} \quad \lambda_2\leq \langle k_2\rangle/
(\langle k_2^2\rangle-\langle k_2\rangle) \\
0, \quad \quad {\rm others}.
\end{array} \right.
\end{equation}
When computing the outbreak threshold of the epidemic in the
thermodynamic limit, the competing percolation theory~\cite{karrer2011competing}
is used. If the awareness does
not outbreak, i.e., $\lambda_1\leq\lambda_c^1$, the epidemic
outbreak threshold is $\lambda_c^2=\langle k_2\rangle/
(\langle k_2^2\rangle-\langle k_2\rangle)$, which is
the same as that when there is no awareness spreading
\cite{newman2002spread}. When $\lambda_1>\lambda_c^1$,
there are two different situations: (i)
If the epidemic spreads faster than the awareness,
i.e., $ \lambda_2\langle k_1\rangle/({\langle k_1^2\rangle
-\langle k_1\rangle})>\lambda_1\langle k_2\rangle/({\langle k_2^2\rangle
-\langle k_2\rangle}) $, the threshold is still
$\lambda_c^2=\langle k_2\rangle/(\langle k_2^2\rangle
-\langle k_2\rangle)$. (ii) If the awareness spreads
faster than the epidemic, i.e., $ \lambda_2\langle k_1\rangle/({\langle k_1^2\rangle
-\langle k_1\rangle})<\lambda_1\langle k_2\rangle/({\langle k_2^2\rangle
-\langle k_2\rangle}) $, the threshold
is
\begin{equation} \label{betaBc}
\lambda_c^2 = \frac{\langle k_{B}\rangle}{(1-pS_A)
(\langle k_{B}^{2}\rangle-\langle k_{B}\rangle)},
\end{equation}
where $S_A$ is the fraction of the aware nodes in layer
$A$, which can be obtained by using the bond percolation theory~\cite{newman2002spread}. In the simulations,
the communication network is generated by using the
SF uncorrelated configuration model~\cite{catanzaro2005generation}, and
the contact network is an ER network.
This coupled network is hereinafter called a SF-ER network (i.e., the communication network is SF and the contact network is an ER network). As shown
in Fig.~\ref{fig7_ie},
the theoretical predictions agree well with the simulations.
One can observe that the
final awareness outbreak size $R_A$ increases with $\lambda_1$ and
$\lambda_2$, whereas the final
epidemic outbreak size $R_B$ decreases with $\lambda_1$.

Juher and Salda\~{n}a investigated the effects of the overlap of two
layers on the coevolution of awareness and epidemics~\cite{juher2018tuning}.
They first proposed an approach to adjust the overlap and the cross-layer
correlations in two-layer networks. Juher and Salda\~{n}a assumed that
the awareness spreads on layer $\mathcal{A}$ and the epidemic spreads
on layer $\mathcal{B}$. The coevolution mechanisms of the awareness and epidemic
are similar to the model in Ref.~\cite{wang2014asymmetrically}.
By contrast, for a pair of nodes with susceptible
and infected states, the susceptible node reduces its
infection probability to $\lambda_o$.
\textcolor[rgb]{1.00,0.00,0.00}{They proposed a variant mean-field theory, where edges are
divided into two categories: the ones existing in both
networks are called common edges while the ones only in one
network are called private edges.} The epidemic
outbreak threshold fulfills the following condition
\begin{equation}
\frac{\langle k_2^2\rangle}{\langle k_2\rangle}\lambda_2(\alpha)-\gamma_2=0,
\end{equation}
where $\alpha$ is the overlap coefficient between the two layers, and
$\lambda_2(\alpha)=\frac{1}{1+\alpha}[\lambda_2(1-\frac{\langle k_1\rangle}
{\langle k_2\rangle}\alpha)+ \lambda_o
(1+\frac{\langle k_1\rangle}{\langle k_2\rangle})\alpha]$. For the
case of $\alpha=0$, the epidemic threshold is the same as the
classical epidemic threshold. When $\alpha>0$, the epidemic threshold
decreases with $\alpha$, that is, overlap edges promote the epidemic outbreak.


\begin{figure}
\begin{center}
\epsfig{file=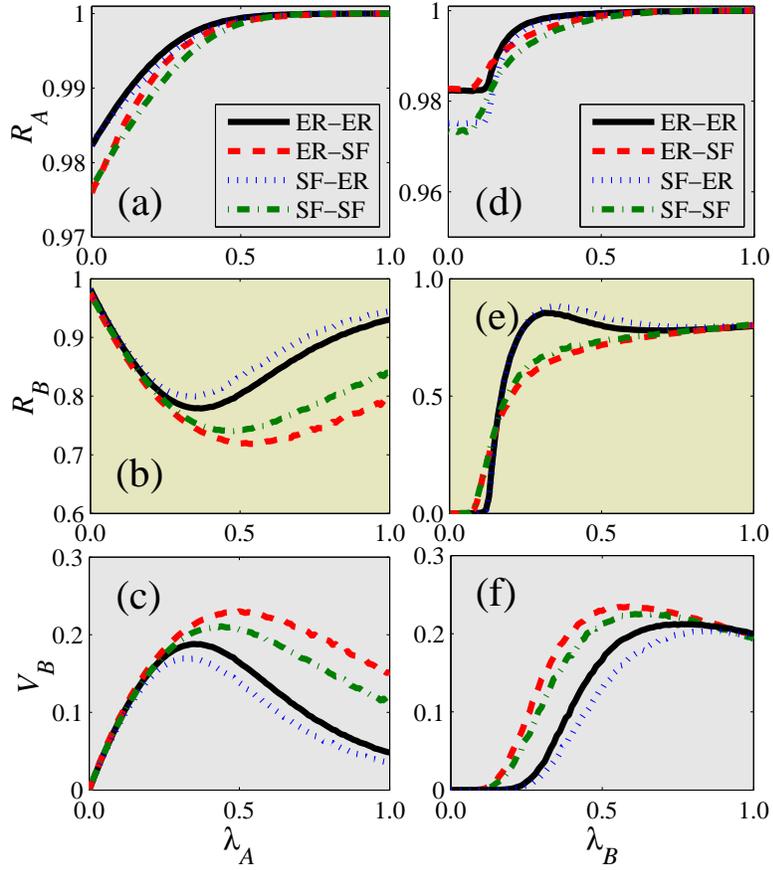,width=0.8\linewidth}
\caption{(Color online) Coevolution of awareness and epidemic
spreading dynamics on different types of multiplex networks.
(a) Awareness outbreak size $R_{A}$, (b) epidemic
outbreak size $R_{B}$ and (c) vaccination size $V_{B}$
versus effective awareness transmission probability
$\lambda_{A}=\lambda_1/\gamma_1$ on ER-ER, ER-SF, SF-ER, and SF-SF
multiplex networks with $\lambda_{B}=0.5$. (d) $R_{A}$,
(e) $R_{B}$, and (f) $V_{B}$ as functions of $\lambda_{B}
=\lambda_2/\gamma_2$ with $\lambda_{A}=0.5$. Other parameters
are set as $\phi=2$, $p=0.8$, and $\langle k_1\rangle =\langle
k_2\rangle=8$. Reproduced from Ref.~\cite{wang2016suppressing}.}
\label{fig8_ie}
\end{center}
\end{figure}

In reality, immunization is always expensive and risky~\cite{
altarelli2014containing}. For rational
individuals, they verify the necessity before adopting immunization.
Along this line, Wang \emph{et al.}~\cite{wang2016suppressing} generalized the
model proposed in Ref.~\cite{wang2014asymmetrically} and assumed that a
susceptible node adopting immunization should consider two aspects:
(i) whether its counterpart in the communication
network is aware of the epidemic, and (ii) whether
the number of its infected neighbors in the contact network
is larger than a threshold $\phi$. Using the heterogeneous
mean-field theory, Wang \emph{et al.} showed that the awareness outbreak threshold is identical to that in
Eq.~(\ref{Athreshold}). However, the epidemic outbreak
threshold is the same as that where there is no awareness spreading on
the communication network when $\phi\geq1$,
whereas if $\phi=0$,
the epidemic threshold is identical to that in
Eq.~(\ref{betaBc}). Through extensive
numerical simulations and theoretical analyses on
different types of artificial multiplex networks,
Wang \emph{et al.} found that there is
an optimal awareness transmission probability $\lambda_1^o$, at
which the epidemic is greatly suppressed, as shown in Fig.
\ref{fig8_ie}.

\begin{figure}
\begin{center}
\epsfig{file=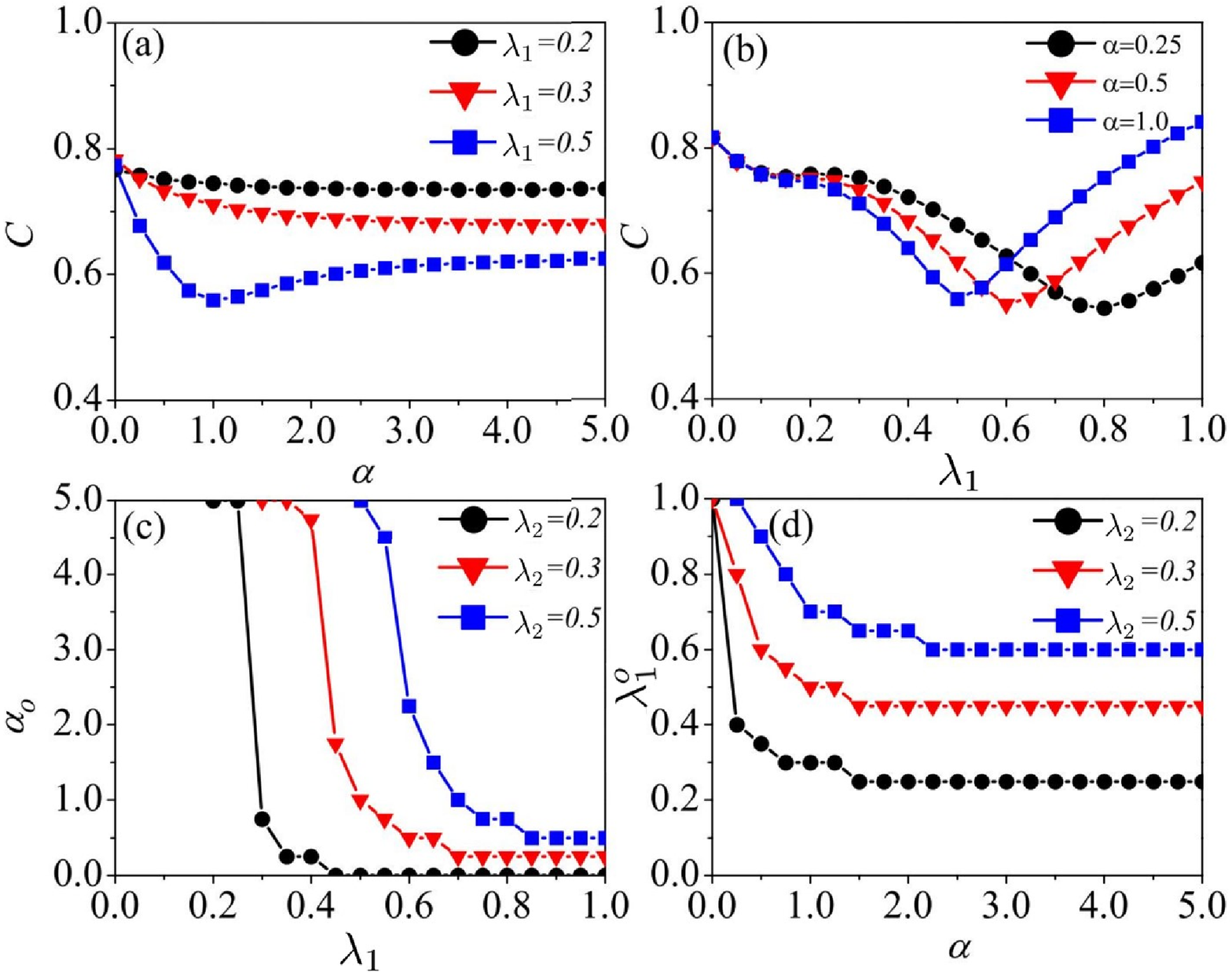,width=0.8\linewidth}
\caption{(Color online) Social cost of the coevolution of awareness
and epidemic spreading on SF-ER multiplex networks. The social cost
$C$ versus (a) the strength of the social reinforcement effect $\alpha$,
and (b) awareness transmission probability $\lambda_1$ (b). (c) The optimal
social reinforcement effect $\alpha_o$ versus $\lambda_1$, and (d)
the optimal awareness transmission probability $\lambda_1^o$ versus
$\alpha$. Reproduced from Ref.~\cite{liu2016impacts}.}
\label{fig9_ie}
\end{center}
\end{figure}

Most of the previous studies assumed that the adoption by individuals of
vaccination or immunization behaviors only depends on their
current perceptive awareness of the epidemic. Liu \emph{et al.}~\cite{liu2016impacts}
proposed a non-Markovian model that assumes
the probability of adopting immunization behavior is
dependent on the cumulative awareness, say
$\xi_M=\xi_1+(1-\xi_1)[1-e^{-\alpha(M-1)}]$, where $\xi_1$ is
the basic
vaccination probability when a node receives the first piece
of awareness on the communication network, and $\alpha>0$ is used
to reflect the strength of social reinforcement:
a larger $\alpha$ corresponds to
stronger social reinforcement. By using the
heterogeneous mean-field theory, Liu \emph{et al.} showed
that the awareness
outbreak threshold is identical to that in
Eq.~(\ref{Athreshold}), namely the social reinforcement
has no effect on the threshold of awareness spreading. For the epidemic outbreak threshold, there are
two different situations. (i) If the awareness cannot
outbreak on the communication network, the epidemic outbreak
threshold is $\lambda_c^2=\langle k_2\rangle/(\langle k_2^2\rangle
-\langle k_2\rangle)$. (ii) If the awareness breaks out and the epidemic spreads
faster than the awareness, the epidemic outbreak threshold
does not change; otherwise, the epidemic outbreak threshold
is
\begin{equation}\label{beta_hui}
\lambda_c^2=\frac{\langle k_2\rangle}{[1-v_B(\infty)](\langle
k_2^2\rangle-\langle k_2\rangle)},
\end{equation}
where $v_B(\infty)$ is the fraction of vaccination nodes with
$\lambda_2=0$. Because
vaccination and infection incur some social cost,
Liu \emph{et al.} defined the social cost of the system
as $C=\frac{1}{N}\sum_{i=1}^N(c_VV_{B,i}+c_RR_{B,i})$, where $V_{B,i}=1$ means
that node $i$ is in the vaccination state, otherwise, $V_{B,i}=0$.
Analogously, $R_{B,i}=1$ means that node $i$ is in the
recovered state, otherwise, $R_{B,i}=0$. The parameters
$c_V$ and $c_R$ denote the social unit cost of vaccination and treatment,
respectively. There exists
an optimal strength of the social reinforcement effect $\alpha_o$,
at which the social cost $C$ is minimized when $\lambda_1>\lambda_2$,
as shown in Fig.~\ref{fig9_ie}(a). Similarly, in Fig.~\ref{fig9_ie}(b),
$\lambda_1$ also exhibits a minimum value at $\lambda_1^o$.
The value of $\alpha_o$ decreases with $\lambda_1$,
and $\lambda_1^o$ decreases
with $\lambda_1$, as shown in Figs.~\ref{fig9_ie}(c) and (d).

\subsection{Summary}

In this section, we reported
the progress of the coevolution of awareness and epidemics. Empirically,
the coevolution of awareness and epidemics exhibits an asymmetric coupling
between the two dynamics: the awareness suppresses the epidemic spreading,
whereas the epidemic spreading promotes the awareness diffusion. For the
coevolution on single networks, the network topology (e.g., degree
distribution and clustering) remarkably affects the coevolution spreading
dynamics. For the coevolution of awareness and epidemics on multiplex
networks, the epidemic threshold of the UAU+SIS model has a critical value
(i.e., the metacritical point), which is correlated with the awareness
diffusion and the topology of the underlying network. In the SIR+SIR model,
there exists an optimal awareness diffusion probability at which the
epidemic spreading is maximally suppressed.

\section{Coevolution of resource diffusion and epidemic spreading} \label{res}

The treatment and control of epidemics require human intervention,
which is infeasible without the resource input from the government or other
institutions. Thus, the effect on resource input in coping with the epidemic propagation
is an issue of great social significance. A traditional  research domain named
economic epidemiology~\cite{philipson2000economic}, is mainly interested in the occurrence of infectious epidemics and the effects
of public health interventions designed to control them. The possibility
of eradicating epidemics, as well as the welfare
loss induced by epidemics are highlighted.
In the area of public health, there are some outstanding studies on the government resource
input in coping with epidemic spreading \cite{colizza2007modeling}. It is found
that sharing the resource of antiviral drugs among countries helps to
contain the epidemic outbreak at the global level, and the more cooperative the resources,
the more effective are the containments in all regions of the world.
It has been pointed out that the lack of resource allocation can arouse
critical financial crises. For example, the scarcity of a resource is closely related to the
susceptibility of the trade network with respect to cascading shocks~\cite{klimek2015systemic}.
However, the study of the field under the framework of network coevolution dynamics
has come to the fore only recently. Most importantly, the effects of resources are often
critical on this epidemic--resource coevolution process.

\subsection{Epidemic spreading dynamics under constant resources}

The resources (e.g., vaccines, funding, and human beings) for curing an infectious epidemic
are always limited and expensive. Given limited resources, determining how to effectively
allocate them is extremely important \cite{nagel1995resource}.
For instance, we could randomly allocate resources to infected individuals,
or prefer to allocate resources to certain important individuals and areas;
in the latter case, how to find the important individuals to allocate the
resources becomes a related optimization problem. Moreover, the critical phenomena of epidemic spreading
dynamics could be different under different strategies for resource allocation. In this section, we first introduce the progress on
the influences of constant resources on the evolution of epidemic spreading
dynamics, and then discuss the critical phenomena induced by constant
resources.

\subsubsection{Optimal allocation of resources}

From the perspective of macroscopic resource cost in terms of vaccinations and the
social welfare loss associated with epidemics, Francis \emph{et al.} \cite{francis2004optimal}
solved an optimal control problem of suppressing the epidemic
through vaccination while minimizing the total cost.
The model is a standard SIR model incorporating an extra vaccination rate $r(t)$,
such that the susceptible can turn into the recovered state at this rate.
Then, Francis \emph{et al.} solved an optimal control problem to minimize the total cost during
the disease process, where the total cost is a weighted summation of the
cost of vaccination and the cost from loss of utility when people are infected.
The resource regulation effects from both government policy and the market were discussed.
In a more realistic scene, the resource allocation is performed on spatial
districts. For example, the epidemic outbreaks may occur in different but
interconnected regions. Under this setting, Mbah \emph{et al.} \cite{mbah2011resource} studied
an optimal control model to minimize the discounted number of infected
individuals during the course of an epidemic with economic constraints,
and investigated preferential treatment strategies.
Their model is a two-region SIRS model in which the infected can be cured
with a cost. The discounted number has the form $\int_0^\infty e^{-rt}\rho(t)dt$,
where the discount rate $r$ aims to give greater emphasis to control in
the short rather than the long term. Mbah \emph{et al.} discuss the optimal control problem
to minimize the discounted numbers of the two regions with total budget constraints.
Their results indicated that when faced with the dilemma of choosing between socially equitable and purely
efficient strategies, the optimal control strategy is not apparent but
should be determined by many epidemiological factors, such as the basic
reproductive number and the efficiency of the treatment measure. Apart from multiple populations from different
regions, the resource allocation can occur over multiple time periods. Zaric \emph{et al.} ~\cite{zaric2001resource,zaric2002dynamic} studied a dynamic resource
allocation model in which a limited budget is allocated over multiple time periods to interventions.
The epidemic model is a generalized epidemic model with multiple compartments and the interventions
turn to the changes in the model parameters.
Through a heuristic numerical study, Zaric \emph{et al.} found
that allowing for some reallocation of resources or funds
over the time horizon, rather than allocating resources just once
at the beginning of the time horizon, can lead to significant increases in health benefits.

Because the allocation to districts is preliminary and rough, to better
quantify the effect of resources, one can further consider the case in which
resources are allocated to individuals in networks (such as distributing
vaccines and antidotes to individuals throughout the network).
A number of studies focus on studying the algorithm of the network distribution problem.
Preciado \emph{et al.} \cite{preciado2013optimal} proposed a model
in which the infection rate of each
individual can be reduced by distributing the vaccination resources to them (such that they become less infective).
Thus, individuals in the SIS epidemic model
in the network present different levels of susceptibility, depending on how many resources they gain.
Considering how to minimize the total cost of the corresponding
vaccination and the corresponding outbreak asymptotic exponential decaying rate,
a convex framework to find the optimal
distribution of vaccinations to contain the epidemics in
arbitrary contact networks is proposed.
In a similar study, Enyioha \emph{et al.} \cite{enyioha2015distributed} considered a
linearized SIS epidemic spreading model with the setting that
the resources can be used to reduce individual infection rates or/and to increase the
recovery rates. Enyioha \emph{et al.} proposed a distributed
alternating direction method of multipliers algorithm to solve this
problem, and obtained a distributed solution in which agents in the
network were able to locally compute their optimal allocations.

In general, the curing rate of each node is positively correlated to its medical
resources, that is, the more resources given, the higher the curing rate.
In a more realistic case, the total amount of medical resources is limited and the average
curing rate is thus fixed.
Chen \emph{et al.} \cite{chen2017optimal} analyzed how to best allocate limited
resources to each node so as to minimize the epidemic prevalence. They formulated the
SIS model by the mean-field theory and solved the corresponding optimal control problem
by the Lagrange multiplier method, and found counterintuitively that in the strong infection region, the low-degree
nodes should be allocated more medical resources than the high-degree nodes to minimize prevalence.
The above models assume that resources can be distributed to reduce the infectiousness
or increase the recovery of a node. A similar idea can also be conducted on the edges.
Nowzari \emph{et al.} \cite{nowzari2015optimal} assumed that the epidemic spreading can be
suppressed by decreasing the strength or weight of edges when allocating resources.
For example, the government might be
able to decrease the edge weight by reducing interactions between two nodes,
such as by limiting the amount of transportation between two cities. Nowzari \emph{et al.} considered an SIS epidemic model on time-varying networks and studied
how to optimally allocate the budget to best combat the undesired epidemic within a given budget.
They showed that this problem can be formulated as a geometric programming and solved in polynomial time.

The optimal allocation can also be formulated from a mathematical perspective.
Ogura \emph{et al.} \cite{ogura2017optimal} focused on the mathematical problem
of finding the optimal allocation of containment resources to eradicate
epidemic outbreaks over models of temporal and adaptive networks, including
 Markovian temporal networks, aggregated-Markovian temporal networks,
and stochastically adaptive networks. For each model,
a rigorous and tractable mathematical framework to efficiently find the
optimal distribution of control resources to eliminate the epidemic
is developed.

\subsubsection{Resource-induced critical phenomena}

\begin{figure}
\centering\includegraphics[width=1\linewidth]{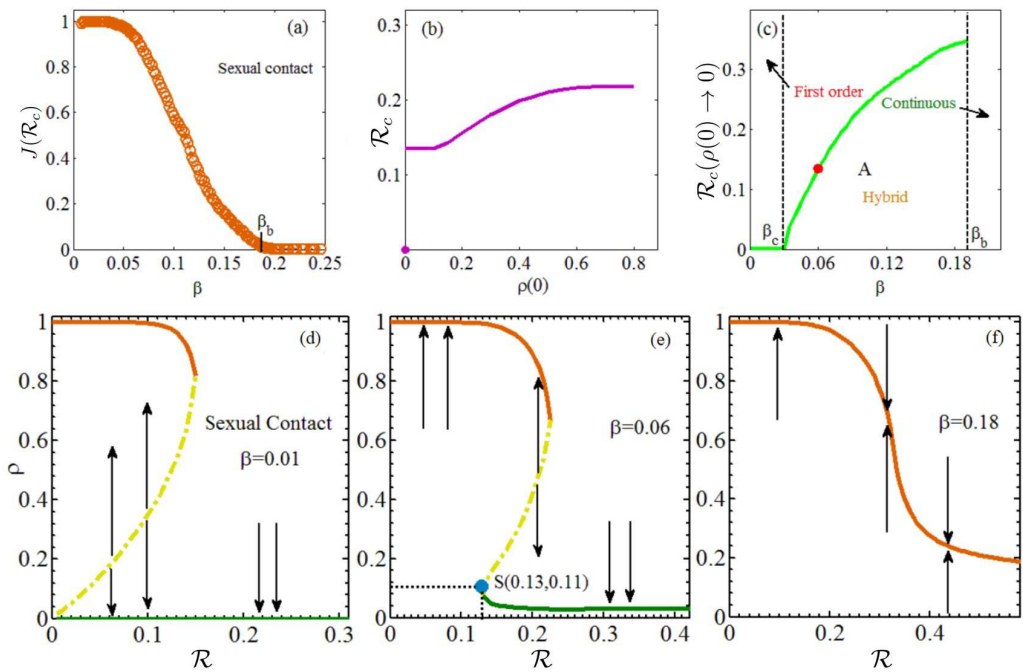}
\caption{Nontrivial critical resource amount and multiphase behaviors in real sexual
contact network. (a) The size of jump $J(\mathcal{R}_c)$ as a function of $\beta$. When $\beta>\beta_b$, the catastrophic
jump behavior disappears, switching to
a continuous phase. (b) $\mathcal{R}_c$ versus $\rho(0)$ with $\beta=0.06$. (c) The critical
resource $\mathcal{R}_c$ vs $\beta$ when $\rho(0)\to 0$.
$\rho$ as a function of $\mathcal{R}$ with $\beta=0.01$ (d), $\beta=0.06$ (e), and $\beta=0.18$ (f).
Reproduced from Ref.~\cite{chen2016critical}.}
\label{FIG:Chen2016Critical}
\end{figure}

Chen \emph{et al.} \cite{chen2016critical} studied an SIS model that incorporates
the relationship between the devoted resource $\mathcal{R}$ and the recovery rate $\gamma$.
This is motivated by their finding from the empirical cured rate of cholera
that the recovery rate $\gamma(t)$ can be formulated in terms of $
\gamma(\mathcal{R},\rho) = e^{-\rho/\mathcal{R}}$,
where $\mathcal{R}$ indicates the amount of the resource.
The infected proportion $\rho(t)$ is governed by the equation
\begin{equation}
\frac{d\rho(t)}{dt}=k \beta \rho(t) (1-\rho(t)) - e^{-\rho(t)/\mathcal{R}} \rho(t),
\label{En:RRdynamics}
\end{equation}
where $\beta$ is the infection rate. In the steady state,
Eq.~(\ref{En:RRdynamics}) could have multiple stable points,
which depend on the initial infected population $\rho(0)$. Chen \emph{et al.}
found that there is a critical resource value $\mathcal{R}_c$, such
that only when $R< \mathcal{R}_c$ will the epidemic be widespread.
In real-world networks and artificial networks, Chen \emph{et al.}
found three types of phase transitions: discontinuous,
hybrid, and continuous, as shown in Fig.~\ref{FIG:Chen2016Critical}.
The regions for the three distinct phase
transitions are determined by the value of $\beta$. The three
phase regimes are separated by $\beta_c$ and $\beta_b$. When
$\beta<\beta_c$, increasing resource $\mathcal{R}$ can always eradicate the
epidemic spreading.
When $\beta>\beta_b$, the infected fraction changes continuously
with $\mathcal{R}$, and this region is the continuous phase transition. When
$\beta_c<\beta<\beta_b$, there is a hybrid phase transition. In this region,
because $\beta>\beta_c$, the epidemic can never be totally removed.

Jiang \emph{et al.}~\cite{jiang2018resource} generalized the above model to multilayer networks.
For a two-layer network, through a discrete-time Markov chain approach,
the spreading dynamics on network $\mathcal{A}$ can be described by the equation
\begin{equation}
\begin{split}
\rho_{A,i}(t+1)&=[1-\rho_{A,i}(t)][1-\prod_{j=1}^N[1-\beta_AA_{ij}p_{1,j}(t)]]\\
&+[1-\gamma_A(t)]\rho_{A,i}(t)+\beta_A^\prime\rho_{B,i}(t)[1-\rho_{A,i}(t)],\\
\end{split}
\label{En:Jiangmultipayer}
\end{equation}
where $\rho_{A,i}(t)$ and $\rho_{B,i}(t)$ respectively represent the probabilities that node $i$ in
the networks $\mathcal{A}$ and $\mathcal{B}$ is infected at time $t$. The epidemic is described
by the SIS dynamics in each layer, with $\beta_A$ ($\beta_B$) and $\gamma_A$ ($\gamma_B$) being the
infectious and recovery probability, respectively, in network $\mathcal{A}$ ($\mathcal{B}$).
Note that the recovery process is the same as for the model in Ref.~\cite{chen2016critical}.
The parameter $\beta^\prime_A$ ($\beta^\prime_B$) represents the infectious probability
from network $\mathcal{B}$ ($\mathcal{A}$) to network $\mathcal{A}$ ($\mathcal{B}$). The
evolution of epidemic on network $\mathcal{B}$ can be expressed similarly.
Jiang \emph{et al.} found that a significant fraction of the total population
may be infected if the resource amount is below the resource threshold. Moreover,
the resource threshold is dependent upon both the inter- and intra-connections
of the two networks. It is found that the inter-layer infectious
strength can lead to a phase transition from the discontinuous to
the continuous phase or vice versa, whereas the internal infectious strength
can result in a hybrid phase transition. Thus, the links between the two networks and the edges within a
network play different roles in the resource-induced critical phenomena.

Time delay is another important factor that affects the epidemic spread dynamics.
In the spreading on multilayer networks, the time delay can arise from the
difference in physical properties between the layers.
Jiang \emph{et al.} ~\cite{jiang2018influence} further studied the effects of
time delays between the layers. This model modifies Eq.~(\ref{En:Jiangmultipayer})
such that $\gamma_1\rho_{B,i}(t)\rightarrow \gamma_1\rho_{B,i}(t-\tau^2)$ and
$\gamma_2\rho_{A,i}(t)\rightarrow \gamma_2\rho_{A,i}(t-\tau^1)$. Here, $\tau^1$ and $\tau^2$
represent the time delays involved of transmitting between the layers.
Interestingly, it is found that
the time delay can induce discontinuous, continuous, and hybrid phase transitions
among them, depending on the resource amount, the infectious
strength between the layers, and their internal structures. In addition, there is
a critical threshold of the time delay, such that even a small resource amount
can effectively control the epidemic spreading if the delay is beyond this
threshold, whereas a huge amount of the resource is needed otherwise. Thus, the effect
of time delays in the presence of a limited resource is more important.

The resource discussed above is not relevant to the network topology and
thus it is a global resource. One can further consider a resource that
is provided by the neighboring individuals in the network. Such local
resources can induce more abundant effects on the epidemic spreading process.
In an SIS model proposed by Bottcher \emph{et al.} \cite{bottcher2016connectivity}, the infected nodes
can only recover when they remain connected to a predefined central node, which provides the resources.
In other words, there is a single central node in the system that controls healing.
An infected node can heal if, and only if, it is connected to the central node via
a path involving only healthy nodes.
Interestingly, through numerical simulation, a two-phase behavior is observed such that the
system converges to only one of two stationary states: either the whole
population is healthy or it becomes completely infected. This gives rise
to discontinuous jumps of different sizes in the infected population and
larger jumps tend to emerge at lower infection rates.
The discontinuous jumps can be understood by the fact that
at some point, the central node may become surrounded
by infected nodes, such that nodes outside cannot be cured, which
leads to a sudden jump to a fully infected absorbing state.

\subsection{Coevolution of epidemic--resource spreading dynamics}

The coevolution process of an epidemic and resources is more realistic in
social networks. In the view of global (government) resources, the outbreak
of an epidemic will reduce the labor force, as sick workers lose their productivity,
and thus decrease the resource output. In the view of individual
resources, social individuals may distribute their resources to their sick
friends. In all these cases, the amount of resources should be affected by the
epidemic process. In other words, the evolution of epidemic spreading is affected by the amount of resources, and at the same
time, the amount of resources is also influenced by the epidemic spreading.
Unlike the case of constant resources, the dynamical evolution of resources can introduce
nontrivial effects on the epidemic spreading. Research on the
epidemic--resource coevolution processes has only recently come to the fore.
In this coevolution process, the critical outbreak of an epidemic
depends on the evolution properties of epidemics and resources, the complex
interplay among them, as well as the network topology.

\subsubsection{Effects of global resources}

\begin{figure}
\centering
\includegraphics[width=0.6\linewidth]{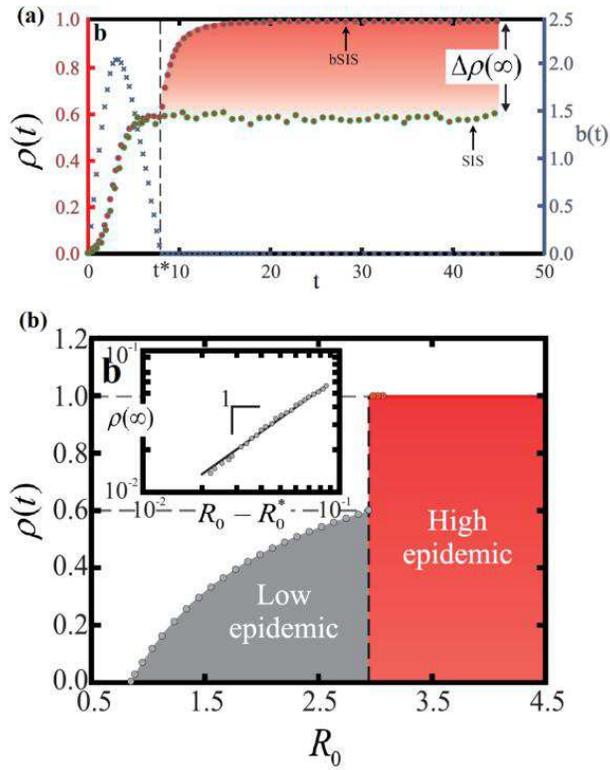}
\caption{Phase behavior of the bSIS model. (a) Coevolution of the proportion
of infected individuals $\rho(t)$ and the budget $b(t)$ on the friendship network.
Other parameters are set to be $\gamma_0=0$,
$\gamma_b=0.8$ and $\beta=0.285$.
(b) Asymptotic fraction of the infected
individuals $\rho(\infty)$ on the friendship network, for $c =
0.833$ and $\gamma_b = 0.8$, as a function of $R_0$.
The inset shows that in the epidemic regime $\rho(\infty) \sim (R_0 - R_0^*)$,
where $R_0^*= 1.6488\pm0.0001$.
Reproduced from Ref. \cite{bottcher2015disease}.}
\label{FIG:BottcherSR}
\end{figure}

If an epidemic becomes more prevalent, it can limit the availability of
the resources needed to effectively treat those who have fallen ill,
because the recovery of sick individuals may depend on the availability
of healing resources that are generated by the healthy population.
Bottcher \emph{et al.} \cite{bottcher2015disease} proposed a budget-constrained
SIS (bSIS) model, in which healthy individuals produce resources and infected
individuals consumes resources to recover. The bSIS model modifies the classical
SIS dynamic by introducing a global budget $b$ that can be increased by the
number of healthy individuals per time step. There is a basic recovery process
that is independent of the available budget, occurring at a rate $\gamma_0$.
Furthermore, each individual can recover through treatment and this requires
a cost $c$. This resource-mediated recovery requires the budget.
However, the rate of recovery through treatment is $\gamma_b
f(b)$, where $f(b)$ is a function of the budget, satisfying $f(b) = 0$
for $ b \leq 0$ and $0 < f(b) \leq 1$ for $b > 0$. With the mean-field assumption,
the dynamics of the fraction of infected individuals is given by
\begin{equation}
\frac{d \rho(t)}{dt}=k\beta\rho(t)s(t)-[\gamma_0+\gamma_b f(b)]\rho(t).
\label{bSISmeanfield}
\end{equation}
Fig.~\ref{FIG:BottcherSR}(a) shows
an explosive epidemic where the entire population is infected in the steady
state [$\rho(\infty) = 1$]. Furthermore, there is a discontinuous transition
in the model parameters. For example, as illustrated in Fig.~\ref{FIG:BottcherSR}(b),
the transition is discontinuous in the basic reproduction number.
Therefore, there is a jump, $\Delta \rho(\infty)$, in the infection level
(see Fig.~\ref{FIG:BottcherSR}(a)). The discontinuity of this jump in the final
infected fraction $\rho(\infty)$ has important implications for the resilience
of the healthy system and the control of the epidemic. It implies that small changes
in the properties of the epidemic can abruptly increase the infection
level, such that the epidemic is harder to control.
In short, an insufficient resource production rate would induce the outbreak
of the epidemic. Equivalently, the epidemic spreads out explosively if the cost
of recovery is above a critical value.


Another way to contain the spread of an epidemic is to make the population
aware of the epidemic and possible self-protection methods. For this purpose,
the government can allocate funds to make the public aware through mass media,
print media, pamphlets, etc. In this case, Misra \emph{et al.} \cite{misra2018modeling}
proposed an SIS model in which the infectious rate $\beta$ can be decreased
by the resource input. The available resource $\mathcal{R}(t)$ itself evolves
in a logistic growth dynamical process. They used a compartmental model to
describe this coevolution process,
and found that although
 increasing the funds reduces the number of infected individuals,
the delay in providing the funds can destabilizes the interior equilibrium and
may cause stability switches, resulting in epidemic outbreaks.

The effect of global resources can also be seen from the strategy of the usage of resources.
Assume that each individual can choose to use the resource to vaccinate or otherwise not
to vaccinate. Without vaccination, there is a possibility that they become infected and
this would also result in a cost to treat the epidemic. In this situation, the evolution
of the resource is reflected in the evolution of the strategy about how individuals use
the resource. Human behavior and the networking-constrained interactions among
individuals significantly impact the coevolution of the epidemic and the resource
strategy.

\begin{figure}
\centering
\includegraphics[width=1\linewidth]{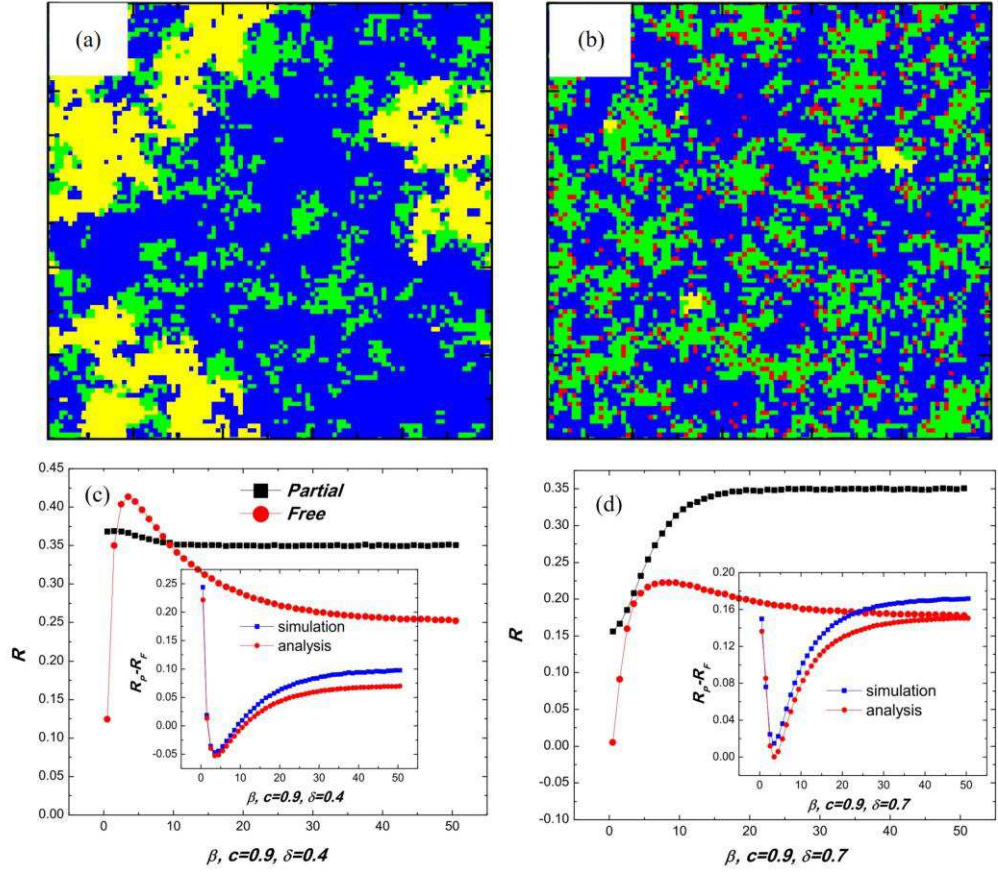}
\caption{Epidemic dynamics with incentive vaccination strategy.
Snapshots of the stationary state configuration in square lattices with
partial-subsidy (a) and free-subsidy (b) policies. Other parameters are
$N=10^4$, $c=0.5$, $\sigma=1.0$, and $\delta=0.4$. The vaccination coverage $V$ (c)
and epidemic size $R$ (d) as a function of $\sigma$ in
homogeneous small-world network. The insets show the differences in the
vaccination coverage and epidemic size between the partial-subsidy and free-subsidy
policies, $V_P - V_F$ and $R_P - R_F$, respectively. Reproduced from Ref.~\cite{zhang2014effects}. }
\label{FIG:ZhangSR14}
\end{figure}

Zhang \emph{et al.} \cite{zhang2013impacts,zhang2014effects} proposed a
theoretical framework to study the coevolution of epidemic and the resources
game. Taking into account the periodic outbreaks of flu-like epidemics
and the limited effectiveness of vaccines, they studied models with seasonal
updating and pre-emptive vaccination, in which individuals decide whether
or not to get vaccinated before each epidemic season. The strategy for individuals
is to determine whether or not they vaccinate. Although it requires a cost to vaccinate, individuals
also bear the cost to treat the epidemic if they are infected when they do not receive
vaccination. Once the epidemic ends, individuals update their vaccination decisions for the next season by imitating
the strategy of their neighbors. The neighbor with a higher payoff has higher probability
for their strategy to be learned, which is described by the famous Fermi rule in game theory \cite{perc2010coevolutionary}.
By this rule, an individual, say $i$, updates his/her vaccination strategy by
randomly choosing one of its immediate neighbors, say $j$,
and adopts the strategy of $j$ with the following probability determined by
the payoff difference
\begin{equation}
y(v_i \leftarrow v_j)=\frac{1}{1+{\rm exp}[-\sigma(\mathcal{P}_j-\mathcal{P}_i)]},
\label{imitate}
\end{equation}
where $v_i=1$ or $0$ denotes the vaccination choice for individual $i$: either
vaccinated or not, and $\mathcal{P}_i$ is the current payoff of individual $i$ at the current season.
Without any subsidy, according to the costs of vaccination and infection,
we have
\begin{equation}
\mathcal{P}_i =\left\{
             \begin{array}{lr}
             -c, \quad {\rm vaccination},  \\
             -1, \quad {\rm infected},\\
             0, \quad {\rm free rider},
             \end{array}
\right.
\label{imitatecost}
\end{equation}
where $\sigma$ is a parameter characterizing the rationality of the individuals:
higher $\sigma$ implies more rational.
By incorporating evolutionary games into epidemic dynamics, Zhang \emph{et al.}
found that under the partial-subsidy policy, the
vaccination coverage depends monotonically on the sensitivity of individuals
to payoff difference, $\sigma$, but the dependence is nonmonotonic for the
free-subsidy policy, referring to Figs.~\ref{FIG:ZhangSR14}(c) and ~\ref{FIG:ZhangSR14}(d).
For the case of irrational individuals where $\sigma=1$, the free-subsidy
policy can in general lead to higher vaccination coverage, referring to
Figs.~\ref{FIG:ZhangSR14}(a)--(b).

In \cite{zhang2017preferential}, Zhang \emph{et al.} further explored the effect of a preferential imitation rule,
where individuals choose their imitating neighbors with a tendency rather than randomly selecting them.
It is found that the targeted subsidy policy is only advantageous when
individuals prefer to imitate the subsidized individuals' strategy.
Otherwise, the effectiveness of the targeted policy is worse. More
importantly, under the targeted subsidy policy, preferential imitation
causes a nontrivial phenomenon whereby the final epidemic size increases
with the increase in the proportion of subsidized individuals. In
summary, epidemic-control policy through resource input depends on the
complex interplay among the intrinsic mathematical rules of epidemic
spreading, governmental policies, and the behavioral responses of individuals.
Moreover, this complexity may introduce the so-called Braess's Paradox \cite{zhang2013braess},
such that increasing the effectiveness of the strategy may in contrary lead to worse strategy outcomes
(more resource to be consumed).

The increasing level of vaccination in the population
helps to inhibit the epidemic spreading, which in turn, however, discourages
people from participating in vaccination owing to the cost.
Cai \emph{et al.} \cite{cai2014effect} studied the
impact of some other vaccination strategies on the epidemic spreading.
The epidemic and game dynamics are similar to the above process in
Eq.~(\ref{imitate}) and Eq.~(\ref{imitatecost}). Individuals
randomly choose a neighbor to perform the Fermi rule.
Consider three strategies: (i) deterministic (individuals choose
to vaccinate or not before the epidemic season),
(ii) probabilistic (individuals choose to vaccinate with
certain probabilities), and (iii) probabilistic with random mutation (after choosing the
vaccinating probability, there is still a random change in the probability).
Cai \emph{et al.} showed that there is a critical value of $c$ in Eq.~(\ref{imitatecost}),
below which, the lower the mutation probability, the
higher the vaccination level, and above which the opposite effect takes place.
Both the final vaccination level and epidemic size in the continuous-strategy case are less than
those in the pure-strategy case when vaccination is cheap.

In terms of the resource strategy game, Wang \emph{et al.} \cite{wang2009decentralized}
examined how two countries would allocate resources at the onset of an
epidemic when they seek to protect their own populations.
They build a two-region SIR model in which infected individuals can transmit
between the two regions (countries). Each country can distribute the total resource
to itself or to the other country. The effect of the resource is to decrease
the initial number of susceptibles. As a game between selfish countries,
each country aims to minimize its own outbreak size over the entire time horizon.
Wang \emph{et al.} mathematically analyze the model and show that the best strategy for selfish
countries is to allocate all their resources to themselves to decrease their
own effective reproduction ratio. Moreover, they further identify the mathematical
conditions under which the total number of infected in the whole population is
minimized by their the best strategy. In this case, even though each country
selfishly seeks their own optimal strategy without communication, the global optimal
situation can be achieved such that a major global outbreak may still be avoided.

\subsubsection{Effect of individual resources}

Long \emph{et al.} \cite{long2018malicious} considered an SIS model
in which the recovery rate of a node $i$ is positively related to the resource provided by its healthy neighbors.
That is, $\gamma_i=1-(1-\gamma_0)^{\omega \mathcal{R}_i}$, with $\gamma_0$ being the basic
recovery rate and $\omega$ a scale factor, where
$\mathcal{R}_i$ is the resource provided by its healthy neighbors and takes the form
\begin{equation}
\mathcal{R}_i= \sum_{j=1}^N A_{ij}s_j\frac{k_j^\alpha}{\rm max(k_{min}^\alpha,k_{\rm max}^\alpha)}.
\label{En:Longresource}
\end{equation}
Here, $A_{ij}$ is the connectivity matrix of the networks,
$k_{\rm min},k_{\rm max}$ are its minimum and maximum degrees,
$s_j$ indicates the state of node $j$, which equals 1 if
it is healthy and 0 if infected, and
$\alpha$ is a preference factor.
Consider the cases in which the recovery rate of infected nodes is
heterogeneous in the sense that nodes with large degree tend to
receive more (or less) resources, corresponding to $\alpha>0$ and $\alpha<0$, respectively.
Through analysis and numerical simulations, Long \emph{et al.} find that the virus spreading
can be optimally suppressed if there is no such relation between
the node degrees and resource amounts they received.
In other words, if each healthy node contributes equal resources to the infected nodes,
the virus can be optimally suppressed and there will be a maximum outbreak threshold
and minimum fraction of infected nodes. Moreover, they find that in a homogeneous network,
the uneven distribution of resources (if the recovery resources of infected nodes
mainly rely on nodes with large or small degrees) would lead to a discontinuous
phase transition, but the phase transition is continuous under even distribution of resources.
However, heterogeneous networks always go through continuous transitions. In a similar scenario, Chen \emph{et al.}~\cite{chen2018controlling} studied
another model with a preferential effect.
When the transmission rate is small, the resources of the healthy
nodes should be allocated preferentially to the highly infectious nodes (nodes with more infected neighbors).
When the transmission rate is large, in the early stage, resources should be allocated preferentially
to the highly infectious nodes, whereas after the early stage,
resources should be allocated to the less infectious nodes.
With the individual resource diffusion framework,
Chen \emph{et al.} found that the allocation strategy can adaptively change with the current fraction of
infected nodes and the epidemic can be maximally suppressed under the proposed strategy, which gives
a novel viewpoint to the optimal epidemic control problem under resource constraints.

\begin{figure}
\centering
\includegraphics[width=0.8\linewidth]{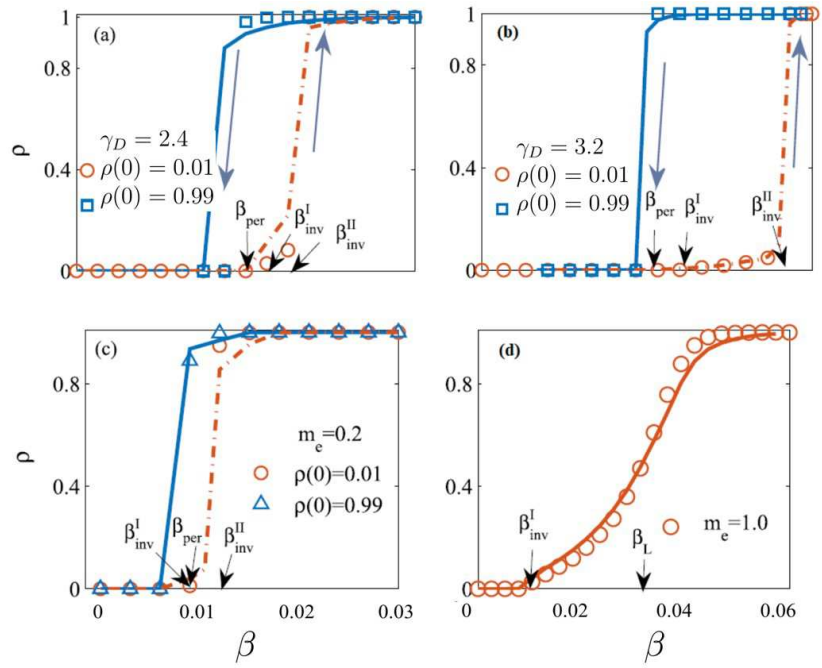}
\caption{Epidemic spreading on multiplex networks with resource allocation.
The infected density $\rho$ as a function of epidemic transmission rate $\beta$.
Results in scale-free network with exponent $\gamma_D = 2.4$ (a) and $\gamma_D = 3.2$ (b).
Results in scale-free network with exponent $\gamma_D = 2.2$ with edge overlapping fraction
$m_e=0.2$ (c) and $m_e=1.0$ (d). Reproduced from Ref. \cite{chen2018suppressing}.}
\label{FIG:ChenNJP18}
\end{figure}

In the above model, the epidemic process and resource diffusion occur on the same network.
Chen \emph{et al.}~\cite{chen2018suppressing} proposed a coevolution
spreading model in multilayer networks, where the epidemic and resources
separately spread in the contact and social layers, respectively.
At each time step, each healthy individual generates one resource unit,
which are distributed equally to infected neighbors through the social layer
and infected nodes consume all of the received resources to improve their recovery rate.
The recovery rate of node $i$ is $\gamma_i(t)=\mu_r \mathcal{R}_i(t)/ k_i$, where
$\mathcal{R}_i(t)$ are the resources that node $i$ receives from its healthy neighbors and
$\mu_r \in [0,1]$ is an efficiency coefficient. Mathematically, the epidemic threshold can be approximately computed
by a generalized dynamical message-passing approach
based on the nonbacktracking matrix $\mathbf{B}$ of the contact layer.
Under this approach, the epidemic threshold is given by
\begin{equation}
\beta_c = \frac{1}{\Lambda_J},
\label{EN:threshold_ChenNJP18}
\end{equation}
where $\Lambda_J$ is the largest eigenvalue of $J$, defined as $J_{j \rightarrow i,l \rightarrow
h} = -\delta_{lj} \delta_{ih} \mu_r + \beta \mathbf{B}_{j \rightarrow i,l \rightarrow
h}$, with $\delta$ being the Dirac delta function.
Chen \emph{et al.} found that
a hybrid phase transition can be observed in SF networks, as shown
in Figs.~\ref{FIG:ChenNJP18}(a) and (b).
The final infected proportion $\rho$ increases continuously
with infection probability $\beta$
at $\beta^{\rm I}_{\rm inv}$; a small increase
in $\beta$ would induce a sudden jump in $\rho$ at $\beta^{\rm II}_{\rm inv}$. Here,
$\beta^{\rm I}_{\rm inv}$ and $\beta^{\rm II}_{\rm inv}$ are the first and
second thresholds, respectively.
In addition, there are hysteresis loops in the phase transition. When the seed
density is initially low, the epidemic breaks out at
the invasion threshold $\beta^{\rm I}_{\rm inv}$, but at the persistence threshold
$\beta_{\rm per}$ if it is initially high.
When the fraction of edge overlap of the two layers is decreased (e.g., when $m_e=0.2$),
a hybrid phase transition appears (see Fig. \ref{FIG:ChenNJP18}(c)).
Note that when the edges of the two layers overlap completely ($m_e=1.0$) the infected density
$\rho$ smoothly increases from $0$ to $1$ (see Fig. \ref{FIG:ChenNJP18}(d)).

In this multilayer network framework, Chen \emph{et al.}~\cite{chen2018optimal}
further considered the case in which the diffusion of resources has a preferential tendency
that the resource tends to spread to the nodes with higher ($\alpha_p>0$) or lower
($\alpha_p<0$) degrees, determined by a bias parameter $\alpha_p$.
Specifically, the resource transfer probability
from node $i$ to $j$, $\phi_{i \to j}$, is
\begin{equation}
\phi_{i \to j}=\frac{(a_{ij}+\delta_{ij})k^{\alpha_p}_j}{\sum_l a_{li}k^{\alpha_p}_l+k^{\alpha_p}_i},
\end{equation}
where $\delta_{ij}=1$ only when $i=j$.
The model exhibits different types of phase transition,
depending on the preference value $\alpha_p$.
This dependence relation is nontrivial and determined by the interlayer degree correlation of the network.
Moreover, there is an optimal strategy at any given strength of
interlayer correlation, where the threshold reaches
a maximum and under which the epidemic can be maximally suppressed.


The coevolution of resources and the network topology is also
an interesting problem. Aoki \emph{et al.}~\cite{aoki2012scale}
proposed an adaptive model in which resources diffuse over a weighted network
with edge weights adaptively varying depending on the resource distribution.
The dynamics of node resources are governed by a
reaction--diffusion process in which nodes are coupled
through the weighted links of the network.
Aoki \emph{et al.} showed that under feasible conditions,
the weights of the network robustly acquire SF distribution in the asymptotic state,
even when the underlying topology of the network is not a SF degree distribution.
Interestingly, in the case in which the system includes dissipation, it asymptotically
realizes a dynamical phase characterized by an organized SF network, in which
the ranking of each node with respect to the quantity of the resource it possesses changes ceaselessly. Regarding the epidemic dynamics, the weight of edges of a
network may indicate the infectious strength, which can adaptively change during
the epidemic process. Because the coevolution of resources and the network
weight exhibits abundant and various properties, the coevolution of resources, network weight,
and epidemic deserves further study.

\subsection{Summary}

 In this section, we reviewed the progress in
the study of the effects of resource diffusion on epidemic spreading
processes. Resource consumption arises from the resource input to reduce
the spreading rate and/or from the medical cost of curing to improve the
recovery rate. The optimal solution of resource allocation can be found under
constant resources. If the amount of total resources is limited, the epidemic
may exhibit abrupt phase transitions at certain critical points, which is different
from the traditional understanding that the optimal allocation strategy and
its consequential outbreak size changes continuously with the total amount of
resources. This critical phenomenon indicates that a slight lack of resource
input may result in a catastrophic epidemic outbreak. In the case of coevolution,
an inadequate resource production rate may cause an epidemic outbreak through abrupt
phase transitions, similar to the case of inadequate constant resources. Individual
behavior can induce complex coevolution dynamics with different types of critical
phase transitions. This would depend on the complicated interplay among the
intrinsic rules of epidemic spreading and behavioral responses of individuals,
such as resource allocation strategies and the interlayer degree correlations between two networks.

\section{Conclusions and outlooks} \label{con}

Studies of coevolution spreading dynamics are very popular
in the field of network science, especially because of
recent empirical observations, which indicate
that some collective phenomena can hardly be understood
by using single contagion models and the ignorance of the
coevolving nature may yield misleading results.
For instance, why a global epidemic can be
contained once the awareness about the epidemic is induced,
why a virus grows discontinuously in biology, and why an epidemic
outbreak size exhibits bistable states for limited resources.
Correct answers to these questions cannot be obtained by analyzing a single spreading process.

\begin{figure} \label{framework}
  \begin{adjustbox}{addcode={\begin{minipage}{\width}}{\caption{%
      Landscape of coevolution spreading dynamics.
      }\end{minipage}},rotate=90,center}
      \includegraphics[scale=1.2]{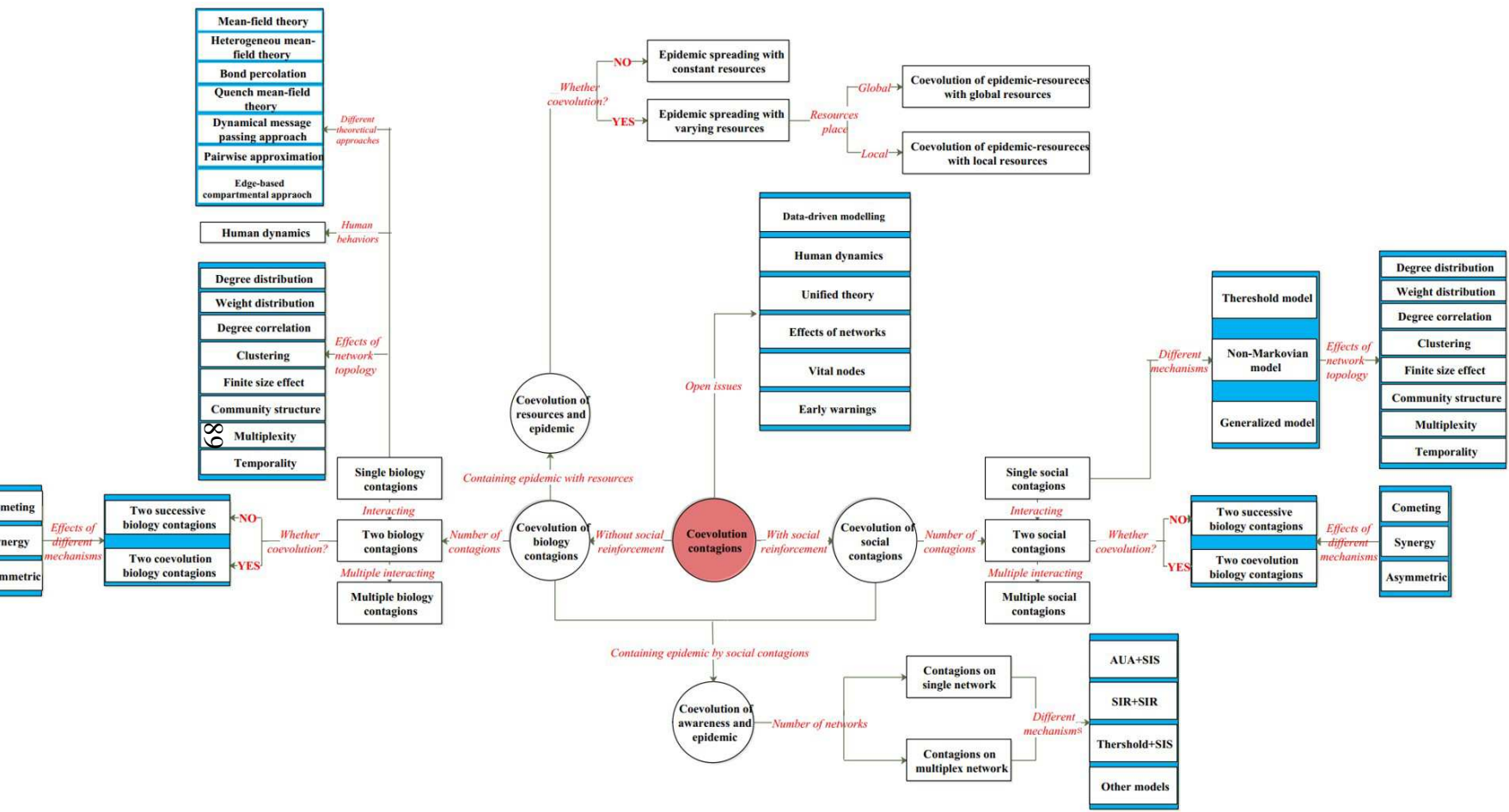}%
  \end{adjustbox}
\end{figure}

In this review, we presented the state of the art
of the progress of coevolution spreading dynamics on complex networks.
The landscape of the studies on coveolution spreading dynamics is presented in
Fig.~\ref{framework}. The two main classes of spreading dynamics are
biological and social contagions, and the two types of coevolution spreading processes related to the control strategies are epidemic--awareness spreading and epidemic--resource spreading. As in the early stage of the progress of evolution spreading studies, many efforts have been made in analyzing various factors that are known to play significant roles in single spreading dynamics. In addition, most theoretical approaches introduced in this review were first developed to analyze single spreading dynamics. This situation will likely persist for years and then completely novel viewpoints and approaches may arise.

In the so-called big data era, the increasingly
available empirical data on
social platforms (such as Twitter and Facebook) and biological systems
enables us to build more realistic data-driven coevolution
spreading models.
For the existing works, most of the known interacting
mechanisms between two spreading contagions, such as synergy, competition, and
asymmetrical, have not been verified by
empirical data. Whether those coevolving mechanisms
are sufficient to capture the main properties of real
dynamics, or whether we still need more intricate interacting
models is not yet clear to us. Therefore, we should
first collect as much related data as possible. For
example, one could crawl the forwarding time series,
network topology, and personal
information about news and products, and then analyze
the interacting mechanisms between the
news and products. To reveal the
representative features of real coevolution
spreading processes from empirical data (or, more valuably but with more
difficulty, from designed experiments) is
the most significant task facing the advancement of credible
studies on mathematical models.

Another important aspect is that the coevolution spreading
dynamics are largely affected by the network topology.
To get analytical solutions of the coevolution spreading
dynamics, most previous studies made some assumptions
about the network topology, such as that the network is large,
sparse, local-tree-like, or static. Studies on
single spreading dynamics have already indicated that each assumption
markedly alters the spreading dynamics; therefore, we should
systematically study the effects of network topology on coevolution spreading dynamics from
the microcosmic, mesoscopic, and macroscopic views.
Furthermore, the spreading on adaptive networks~\cite{gross2008adaptive}
that also incorporates the coevolution of network topology
should be investigated.

In addition, studies on human dynamics
have revealed some inherent
regularities of human behaviors, such as memory and
burstiness in temporal activities
\cite{barabasi2005origin,zha2016unfolding} and
heterogeneity in mobility patterns~\cite{Brockmann2006,Yan2018}, which remarkably
influence single spreading dynamics.
Once the human dynamics are included, the interaction
patterns and transmissibility among nodes will be affected,
and thus the coevolution spreading will also be affected.
Temporal and spatial network
representations~\cite{barthelemy2011spatial},
as well as Monte Carlo simulations may be
useful tools in this direction.

The resilience of coevolution spreading is a potentially interesting
topic. The resilience of a system is its ability to maintain its functions
when some errors and attacks occur, or some environmental and dynamical
parameters are changed~\cite{gao2016universal}. Previous studies
mainly focused on the resilience of
single dynamics, such as epidemic spreading~\cite{fu2017center}, biological
dynamics~\cite{carpenter2011early}, climate changes~\cite{
lenton2008tipping}, and financial dynamics
\cite{may2009systemic}. Pananos \emph{et al.} \cite{pananos2017critical} studied the
critical dynamics in a population with vaccinating behavior
by extracting data on measles-related tweets and Google
searches. They revealed the critical slowing down
and critical speeding up for coevolution dynamics, which are
markedly different from single spreading. Because the resilience
depends on the dynamical process, do there exist some
common characteristics for the resilience of coevolution spreading
dynamics? For a given coevolution spreading process, how might
its resilience be estimated?

To control the coevolution spreading dynamics,
we should identify the most influential nodes, such that we can promote the spreading by letting these influential nodes be infected seeds or suppress the spreading by immunizing these nodes. For
single spreading dynamics, some effective measures
(e.g., $H$-index~\cite{lu2016h} and $k$-core~\cite{kitsak2010identification}) and algorithms (e.g., PageRank~\cite{Brin1998}, LeaderRank~\cite{lu2011leaders},
collective influence~\cite{morone2015influence},
and some heuristic algorithms~\cite{chen2009efficient})
have been proposed. More detailed progress on vital
node identification is presented in a recent review
\cite{lu2016vital}. We should note that an effective
vital node identification algorithm may not
work for coevolution spreading dynamics. For instance,
the nodes with high $k$-shell values are more likely
to be the influential nodes; however, those nodes may
inhibit the spreading when there exist asymmetrical
interactions between two dynamics. Therefore, the
vital node identification problem should be redefined and reanalyzed for
coevolution spreading dynamics.

Lastly, to our best knowledge, the existing theoretical
approaches were originally designed for single spreading dynamics,
and can only deal with certain specific situations in coevolution
spreading dynamics. Generally speaking, the dynamical correlations
in coevolution dynamics are stronger and more complicated than
those in single dynamics, which demand novel theoretical
approaches that may be based on but beyond the message passing
approaches~\cite{karrer2010message}. According to our intuition, the
nonbacktracking matrix~\cite{Martin2014} and the Hankel
matrix~\cite{Zhang2019} may be useful tools these theoretical analyses.

\section*{Acknowledgements}
This work was partially supported by the China Postdoctoral Science Foundation
(No.~2018M631073), China Postdoctoral Science Special Foundation (No.~2019T120829), and
Fundamental Research Funds for the Central Universities.
Y.H. and J.L. are supported by the National Natural
Science Foundation of China Grants (Nos. 61773412 and U1711265), the Program for
Guangdong Introducing Innovative and Entrepreneurial Teams (No.2017ZT07X355), Guangzhou Science and Technology Project (No. 201804010473), Guangdong Research and Development Program in Key Fields under Grant 2019B020214002 and Three Big Constructions-Supercomputing Application Cultivation Projects sponsored by National Supercomputer Center in Guangzhou. T.Z. acknowledges the National Natural
Science Foundation of China (No. 61433014) and the
Science Promotion Program of UESTC (No. Y03111023901014006).
We would like to thank J.C. Miller for his meaningful
comments.











\begin{spacing}{0}

\end{spacing}

\end{document}